\documentclass[useAMS,usenatbib]{mn2e}
\pdfoutput=1
\usepackage{times}
\usepackage{graphicx}
\usepackage[space]{grffile}
\usepackage{latexsym}
\usepackage{amsmath}
\usepackage{amssymb}
\usepackage{url}
\usepackage{fancyref}
\usepackage{colortbl}
\usepackage{xcolor}
\usepackage[caption = false]{subfig}
\usepackage{array,booktabs}
\usepackage{tabu}
\usepackage{pdflscape}
\usepackage{afterpage}

\begin{document}

\title[BVI CATALOGUE OF STAR CLUSTERS IN 5 HCGs]{A COMPREHENSIVE HST BVI CATALOGUE OF STAR CLUSTERS IN FIVE HICKSON COMPACT GROUPS OF GALAXIES.}

\author[K. Fedotov et al.]
{K.~Fedotov,$^{1, 2}$
S.~C.~Gallagher,$^1$
P.~R.~Durrell,$^3$
N.~Bastian,$^4$
\newauthor
I.~S.~Konstantopoulos,$^5$
J.~Charlton,$^6$
K.~E.~Johnson,$^7$
R.~Chandar$^8$\\
$^1$University of Western Ontario, London, ON N6A 3K7, Canada\\
$^2$Herzberg Institute of Astrophysics, National Research Council of Canada, Victoria, BC V9E 2E7, Canada\\
$^3$Department of Physics \& Astronomy, Youngstown State University, Youngstown, OH 44555, USA\\
$^4$Astrophysics Research Institute, Liverpool John Moores University, Liverpool L3 5RF, UK\\
$^5$Australian Astronomical Observatory, North Ryde, NSW 1670, Australia\\
$^6$Pennsylvania State University, University Park, PA 16802, USA\\
$^7$Department of Astronomy, University of Virginia, Charlottesville, VA 22904, USA\\
$^8$Department of Physics \& Astronomy, University of Toledo, Toledo, OH 43606, USA}
\maketitle
\begin{abstract}
We present a photometric catalogue of star cluster candidates in Hickson
compact groups (HCGs) 7, 31, 42, 59, and 92, based on observations with 
the Advanced Camera for Surveys and the Wide Field Camera 3 on the 
\textit{Hubble Space Telescope}.  The catalogue contains precise cluster 
positions (right ascension and declination), magnitudes, and colours in 
the $BVI$ filters.  The number of detected sources ranges from 2200 to 
5600 per group, from which we construct the high-confidence sample by 
applying a number of criteria designed to reduce foreground and 
background contaminants. Furthermore, the high-confidence cluster 
candidates for each of the 16 galaxies in our sample are split into two 
sub-populations: one that may contain young star clusters and one that is 
dominated by globular older clusters.  The ratio of young star cluster to 
globular cluster candidates varies from group to group, from equal 
numbers to the extreme of HCG 31 which has a ratio of 8 to 1, due to a 
recent starburst induced by interactions in the group.  We find that the 
number of blue clusters with $M_{\rm V} < -9$ correlates well with the 
current star formation rate in an individual galaxy, while the number of 
globular cluster candidates with $M_{\rm V}<-7.8$ correlates well (though 
with large scatter) with the stellar mass.  Analyses of the high-
confidence sample presented in this paper show that star clusters can be 
successfully used to infer the gross star formation history of the host 
groups and therefore determine their placement in a proposed 
evolutionary sequence for compact galaxy groups. 
\end{abstract}


\bibliographystyle{apj}
\newcolumntype{C}{@{}>{\kern\tabcolsep}c<{\kern\tabcolsep}}
\newcolumntype{L}{@{}>{\kern\tabcolsep}l<{\kern\tabcolsep}}
\newcolumntype{R}{@{}>{\kern\tabcolsep}r<{\kern\tabcolsep}}
\newcommand{\gr}{\rowcolor{black!10}[0pt][0pt]}
\newcommand\one{\,{\sc i}}
\newcommand\two{\,{\sc ii}}
\newcommand{\OIII}{[O~{\sc iii}]}
\newcommand{\hst}{{\emph{HST}}}
\newcommand\ub{$U_{336}$}
\newcommand\bb{$B_{438}$}
\newcommand\vmb{$V_{547}$}
\newcommand\vb{$V_{606}$}
\newcommand\hb{$H\alpha_{665}$}
\newcommand\ib{$I_{814}$}
\newcommand\bvi{\textit{BVI}}
\newcommand\ubvi{\textit{UBVI}}
\newcommand\ubvmvi{\textit{UB$V_{\text m}$VI}}
\def\msun{\hbox{$\hbox{M}_{\odot}$}}
\def\Msun{\hbox{$\hbox{M}_{\odot}$}}
\newcommand\Rsun{\hbox{R$_\odot$}}
\newcommand\Lsun{\hbox{L$_\odot$}}
\newcommand\Zsun{\hbox{Z$_\odot$}}
\newcommand\kms{\hbox{$\,$km$\,$s$^{-1}$}}
\newcommand\ie{\textit{i.\,e.}}
\newcommand\eg{e.\,g.}
\def\farcs{\hbox{$.\!\!^{\prime\prime}$}}
\def\fs{\hbox{$.\!\!^{\rm{s}}$}}
\DeclareRobustCommand{\rchi}{{\mathpalette\irchi\relax}}
\newcommand{\irchi}[2]{\raisebox{\depth}{$#1\chi$}} 
\let\newcommand=\providecommand
\def\arcsec{\hbox{$^{\hbox{\rlap{\hbox{\lower4pt\hbox{$\,\prime\prime$}}
          }\hbox{$\frown$}}}$}}

\def\aj{AJ}                   
\def\araa{ARA\&A}             
\def\apj{ApJ}                 
\def\apjl{ApJ}                
\def\apjs{ApJS}               
\def\ao{Appl.Optics}          
\def\apss{Ap\&SS}             
\def\aap{A\&A}                
\def\aapr{A\&A~Rev.}          
\def\aaps{A\&AS}              
\def\azh{AZh}                 
\def\baas{BAAS}               
\def\jrasc{JRASC}             
\def\memras{MmRAS}            
\def\mnras{MNRAS}             
\def\pra{Phys.Rev.A}          
\def\prb{Phys.Rev.B}          
\def\prc{Phys.Rev.C}          
\def\prd{Phys.Rev.D}          
\def\prl{Phys.Rev.Lett}       
\def\pasp{PASP}               
\def\pasj{PASJ}               
\def\qjras{QJRAS}             
\def\skytel{S\&T}             
\def\solphys{Solar~Phys.}     
\def\sovast{Soviet~Ast.}      
\def\ssr{Space~Sci.Rev.}      
\def\zap{ZAp}                 
\let\astap=\aap
\let\apjlett=\apjl
\let\apjsupp=\apjs

\section{Introduction}
\setcounter{figure}{0}
\setcounter{table}{0}

It is widely accepted that interactions and mergers between gas-rich
galaxies lead to star formation \citep[e.g.,][]{Kennicutt1987, Mihos1996, 
BartonGillespie2003, Springel2005}, and that the majority of stars form 
in clusters and associations \citep{Lada2003, Bressert2010}.  It therefore 
follows that a detailed analysis of star cluster populations in a galaxy 
can reveal its history of interaction events \citep[e.g.,][]
{Whitmore1999, Gallagher2001, Bastian2005, Wilson2006}.

In that light, star clusters are a powerful tool for studying star 
formation events triggered by mergers and tidal interactions between 
galaxies.  In particular, star clusters could prove useful for  
studying compact groups (CGs), specifically Hickson Compact Groups 
\citep[HCGs;][]{Hickson1982, Hickson1997, Hickson1989, Hickson1992}. 
By virtue of their selection criteria (low velocity dispersions and high 
galaxy number densities), HCGs represent an environment with frequent 
and prolonged interactions, that can trigger the formation of star 
cluster populations associated with specific events. 

Initially motivated by the work of \citet{Verdes-Montenegro2001},
\citet{Johnson2007} proposed an evolutionary sequence of HCGs by
separating them into three types based on the ratio of their H~{\sc i}
content (a proxy for the available reservoir of cool gas for star
formation) and the dynamical mass\footnote{Alternative dynamical 
classifications exist that determine the state of the group through 
means other than mass ratios. For example, \citet{Bitsakis2011} use 
early-type galaxy fractions to ascertain the dynamical state of a 
group.}, with Type I being the gas-rich groups and Type 
III the gas-poor ones. \citet{Johnson2007} also report -- based on 
{\em Spitzer} mid-infrared colours -- that galaxies in Type I groups 
are more actively star-forming than galaxies in Type II groups while 
galaxies in Type III groups are relatively quiescent. \citet{Iraklis2010} 
expanded on this classification by splitting group types into two 
parallel sequences according to their gas distributions: Sequence~A 
groups maintain the bulk of their cold gas inside galaxies, whereas 
Sequence~B groups have gas dispersed throughout the intra-group medium 
(IGM) \citep[their fig. 1]{Iraklis2010}.  The gas distribution of 
Sequence~B groups likely results from strong interactions that occur 
while disk galaxies are still gas-rich. The initial conditions of the 
positions and relative velocities of Sequence~A group galaxies are such 
that only softer interactions occur, and while secular evolution may be 
enhanced and lead to a boost in star formation rates in individual 
galaxies, the bulk of the cold gas is not pulled into the IGM.  As a 
consequence, the groups in Sequence~A are expected to ultimately lead 
to the formation of a single elliptical galaxy with little to no X-ray
envelope, as gas is consumed within galaxies before late-stage dry
mergers.  Groups in Sequence~B -- where galaxies interact strongly
before gas is consumed -- would be more likely to form ellipticals
with a strong X-ray envelope (heated by star-formation triggered by
one or more gas-rich mergers), as can be seen around some massive
elliptical galaxies or so-called `fossil' groups \citep{Jones2003}.
The differences in star-forming histories, which vary depending on gas
content and distribution and advance along the evolutionary path, must
be reflected in the star cluster populations of the groups. Thus, star
cluster populations can potentially be used to infer their hosts'
placement on the CG evolutionary sequence proposed by
\citet{Iraklis2010}.

In this paper, we consolidate the information on star clusters in
compact groups of galaxies that has been presented in a number of
projects \citep{Gallagher2010, Iraklis2010, Fedotov2011, Iraklis2012,
Iraklis2013} and present it in a consistent, coherent catalogue,
with the goal of further assisting researchers in star cluster-related
studies.  We also take this opportunity to compare the basic
properties between cluster populations in compact groups at distinct
evolutionary stages.

This paper is organized in the following way: in Section 2, we
describe the samples and data sets. We outline the procedure for
constructing the catalogue in Section 3, and present our results and
discuss them in Section 4. Lastly, we summarize the main conclusions
in Section 5.  Throughout, we use the cosmology H$_0 = 73.0$
km\,s$^{-1}$\,Mpc$^{-1}$, $\Omega_{\text{matter}} = 0.27$, and
$\Omega_{\text{vacuum}} = 0.73$ to determine distances and physical
sizes.

\section{Data}

The data for this project were obtained with the {\em Hubble Space 
Telescope} ({\em HST}) Advanced Camera for Surveys (ACS) and Wide 
Field Camera 3 (WFC3). These observations are part of two programs: 
ID 10787 (PI J. Charlton) and ID 11502 (PI K. Noll). The observations 
were carried out in the F435W (F438W for WFC3), F606W, and F814W 
filters, which are similar to the Johnson $BVI$-bands. Hereafter, 
we refer to the {\em HST} filters as $B_{435}$, $B_{438}$, $V_{606}$, 
and $I_{814}$, although we did not make transformations to the 
Johnson-Cousins system.

\afterpage{
\begin{table*}
\setcounter{subfigure}{0}
\setcounter{table}{0}
\caption{Details of observations.}
\label{tab:observations}
\begin{minipage}{130mm}
\begin{center}
\begin{tabu}{@{} cCCCCCCCL @{}}
\toprule
Group & Galaxy & Instrument & Program ID &\multicolumn{3}{c}{$t_{exp}$ (s)} & Date & References \\
\cmidrule{5-7}
\rowfont{\small}
 &  &  & & F435W$^*$ & F606W & F814W & & \\
\midrule
HCG 07 & A, B, D    & ACS/WFC & 10787 & 1710 & 1230 & 1065 & Sept 2006 & K10\\
\gr    & C          & ACS/WFC &       & 1710 & 1230 & 1065 & Sept 2006 & \\
HCG 31 & A--C, E--H & ACS/WFC & 10787 & 1710 & 1230 & 1065 & Aug 2006  & G10\\
\gr HCG 42 & A, C   & ACS/WFC & 10787 & 1710 & 1230 & 1080 & Dec 2007  & K13\\
HCG 59 & A, C       & ACS/WFC & 10787 & 1710 & 1230 & 1065 & Dec 2007  & K12\\
\gr    & B, I       & ACS/WFC &       & 1710 & 1230 & 1065 & Nov 2006  & \\
HCG 92 & B, D       & WFC3    & 11502 & 3410 & 1395 & 1860 & Aug 2009  & F11, T12\\
\gr    & C, B       & WFC3    &       & 3410 & 1395 & 1860 & Aug 2009  & \\
       & E          & WFC3    &       & 3410 & 1395 & 1860 & July 2009 & \\
\bottomrule
\end{tabu}\vspace{0.1cm}
\end{center}
\textbf{Notes.}  $^*$ For HCG 92 B filter is F438W of WCF3 camera.  In the last 
column we reference star cluster studies that used the associated observations.\\
G10: \citet{Gallagher2010}\\
F11: \citet{Fedotov2011}\\
K10: \citet{Iraklis2010}\\
K12: \citet{Iraklis2012}\\
K13: \citet{Iraklis2013}\\
T12: \citet{Trancho2012}\\
\end{minipage}
\end{table*}
}

Table~\ref{tab:observations} contains an observation log. The last
column in the table lists publications related to those
observations. Table~\ref{tab:HCG_info} presents properties of the
16 individual galaxies within the five compact groups included
in this sample.

\afterpage{
\begin{table*}
\caption{HCG Galaxies Information.}
\label{tab:HCG_info}
\begin{minipage}{130mm}
\begin{center}
\begin{tabu}{@{}CCCCCCCC@{}}
\toprule
Galaxy & Name & RA  & Dec  & Type & $v_r$ & $M_*$ & SFR \\
\rowfont{\scriptsize}
 &  & (2000) & (2000) & & km\,s$^{-1}$ & $\log(M_{\odot})$ & $(M_{\odot}\,$yr$^{-1})$\\
\midrule
07A & NGC0192  & 00h 39m 13.4s & + 00d 51m 52s  & Sb   & 4133 & 11.28 & 3.88$\pm$0.47\\
\gr 07B & NGC0196  & 00h 39m 17.8s & + 00d 54m 46s  & SB0  & 4255 & 10.76 & 0.23$\pm$0.02\\
07C & NGC0201  & 00h 39m 34.8s & + 00d 51m 36s  & SBc  & 4415 & 10.89 & 2.06$\pm$0.17\\
\gr 07D & NGC0197  & 00h 39m 18.8s & + 00d 53m 31s  & SBc$^a$  & 4121 & 10.15 & 0.43$\pm$0.04\\
31A & NGC1741  & 05h 01m 38.7s & -- 04d 15m 34s & Sdm  & 4074$^b$ & 10.32$^c$ & 8.11$\pm$0.74$^c$\\
\gr 31B &          & 05h 01m 36.2s & -- 04d 15m 43s & Sm   & 4136$^b$ & 9.51 & 0.78$\pm$0.07\\  
31C & M1089    & 05h 01m 37.7s & -- 04d 15m 28s & Im   & 4019$^b$ & $^c$ & $^d$\\
\gr 31E &          & 05h 01m 37.5s & -- 04d 15m 57s & Sdm  & 4009$^e$ & ... & $^d$\\
31F &          & 05h 01m 40.0s & -- 04d 16m 22s & Sbc  & 3969$^e$ & ... & 0.19$\pm$0.02\\
\gr 31G & IC 0399  & 05h 01m 44.0s & -- 04d 17m 20s & Sbc  & 3991 & 10.04 & 1.47$\pm$0.12\\
42A & NGC3091  & 10h 00m 14.3s & -- 19d 38m 13s & E3   & 3964 & 11.53 & 0.44$\pm$0.04\\
\gr 42C &          & 10h 00m 10.3s & -- 19d 37m 19s & E2   & 4005 & 10.70 & 0.09$\pm$0.02\\
59A & IC 0737  & 11h 48m 27.5s & + 12d 43m 39s  & Sa$^f$   & 4109 & 10.19 & 4.99$\pm$0.67\\
\gr 59B & IC 0736  & 11h 48m 20.1s & + 12d 43m 00s  & E0$^g$   & 4004 & 10.14 & 0.02$\pm$0.01\\
59C & KUG 1145+129 & 11h 48m 32.4s & + 12d 42m 19s & Sb   & 4394 & 9.82 & 0.16$\pm$0.03\\
\gr 59D & KUG 1145+130 & 11h 48m 30.6s & + 12d 43m 47s & Im   & 3635 & 9.38 & 0.48$\pm$0.04\\
92B & NGC7318B & 22h 35m 58.4s & + 33d 57m 57s & Sbc   & 5774 & 10.89 & 0.52$\pm$0.01$^h$\\
\gr 92C & NGC7319  & 22h 36m 03.5s & + 33d 58m 33s & SBc   & 6747 & 11.35 & 0.08$\pm$0.05$^h$\\
92D & NGC7318A & 22h 35m 56.7s & + 33d 57m 56s & E2    & 6630 & 11.19 & 0.05$\pm$0.01$^h$\\
\gr 92E & NGC7317  & 22h 35m 51.9s & + 33d 56m 42s & E4    & 6599 & 10.83 & 0.03$\pm$0.01$^h$\\
\bottomrule
\end{tabu}\vspace{0.1cm}
\end{center}
{\bf Notes.} Unless indicated otherwise stellar masses taken from 
\citet{Tyler2014}, SFR values listed are from \citet{Tzanavaris2010}, 
velocities values from \citet{1991rc3..book.....D}, and morphology 
types from \citealt{Hickson1989}.\\
$^a$ In RC3 galaxy listed as SB0.\\
$^b$ \citet{2000AJ....120.1691N}.\\
$^c$ Stellar mass value listed is measured as the total mass of galaxies 31A and -C.\\
$^d$ SFR value listed is combined value of galaxies 31A, -C, and -E.\\
$^e$ \citet{2006AJ....132..570M}.\\
$^f$ In RC3 galaxy listed as E?.\\
$^g$ In RC3 galaxy listed as S0.\\
$^h$ \citet{2014A&A...565A..25B}.\\
\end{minipage}
\end{table*}
}

\section{Data Analysis}

Before we go into the detailed description of selecting and sorting
the detected sources, here are a few words about our terminology to
clarify the differences between samples.

Our catalogue consists of all detected point sources that passed the
criteria described below.  However, the high-confidence portion of the
catalogue is divided into two subcategories that we denote as star
cluster candidates (SCCs) and globular cluster candidates (GCCs).  The
major difference between these two subcategories is that the selection
criteria for the SCCs do not discriminate against sources with
significant nebular emission.  In contrast, the selection criteria for
GCCs filter out objects with nebular emission (both in terms of their
colours and their spatial extent), while being less strict about the
lower luminosity limit (see Fig.~\ref{fig:GCC_selection}).  These
two subcategories are not mutually exclusive, i.e., the same cluster
may be present in both categories. Indeed, the selection criteria 
(described in detail in the section below) for the globular cluster 
candidates are fine-tuned to filter out potentially young star clusters 
in the dynamically young groups with active star formation (e.g., HCG 
31) where we detect many young SCs. Thus, we expect to see a small 
fraction of SCCs being classified as GCCs in such groups (for HCG 31 
out of 338 SCCs only 13 of them, or less than 4\%, are also GCCs).  On 
the other hand, in a group such as HCG 42, which in our sample is 
represented by the giant elliptical NGC~3091 and where the majority of 
the detected sources are expected to be old star clusters, we anticipate 
a majority of SCCs to be also classified as GCCs (out of 356 SCCs for 
HCG 42, 331 of them, or $\sim 93\%$ are classified as GCCs).  Hence, 
we use the SCC sample primarily to study the young star cluster 
populations and GCCs to study globular cluster populations. 

Throughout the paper we use the terms young, intermediate, and old
when we talk about star clusters. These are general terms without
standard definitions in the literature. For this paper, when we
talk about young star clusters we mean clusters that are younger
(according to their location in $BVI$ colour space) than 10 Myr. For
intermediate clusters the age range is between a low hundreds Myrs and 
a few Gyrs, and for the old clusters the range is from $\sim 5$ 
Gyr to 14 Gyr. The intervals not covered by our definitions are grey 
areas, and star clusters in those intervals are labelled according to
context. For example, if we are talking about intermediate age
clusters (750 Myr old), clusters less than 50 Myr old could be
referred to as young/younger SCs. For most of the cases, whenever we
are using these terms (young, intermediate, old) we are specifying the
time intervals within the parentheses following the term.

\subsection{Star Cluster Selection}\label{subsec:SC_select}

For the ACS observations, the closest group is HCG 07 at a distance
of 56.6~Mpc (from the initial radial velocity measurement by
\citet{Hickson1982} modified based on the velocity field model of
\citet{Mould2000}), equivalent to a distance modulus of $(m-M)_0=33.76$ 
mag. At this distance, the $0\farcs049$ pixel size of the ACS corresponds 
to 13.7 pc. The distance to Stephan's Quintet which was observed with 
the WFC3, is adopted to be 88.6 Mpc \citep{Hickson1982,Mould2000}, 
equivalent to $(m-M)_0=34.74$~mag. Thus, one $0 \farcs 04$ pixel on 
WFC3 corresponds to $\sim$ 17.2~pc. With the average star cluster 
half-light radius of $\sim 4$~pc \citep[e.g.,][]{Larsen2004, Barmby2006, 
Scheepmaker2007}, the majority of the detected clusters in all 
observations are expected to be 
unresolved or marginally resolved in the case of the closest groups 
\citep[e.g.,][]{Harris2009}. Therefore, for this catalogue we used 
point spread function (PSF) fitting as the preferred method of 
obtaining photometry. In the next section we give a more detailed 
explanation for favouring PSF photometry.

For point source detection we used the $V_{606}$ images, as they offer
the faintest limiting magnitudes at 50\% completeness level among the
observations (see Table~\ref{tab:completeness}). To detect point
sources, we run the {\tt DAOFIND} \citep{Stetson1987} task in {\tt 
IRAF}\footnote{{\tt IRAF} is distributed by the National Optical 
Astronomy Observatory, which is operated by the Association of 
Universities for Research in Astronomy (AURA) under cooperative 
agreement with the National Science Foundation.} on median-divided
images, obtained by division of the original images by median
boxcar-smoothed ones (with a 13$\times$13 pixel smoothing window). The
coordinates of point sources detected in $V_{606}$ were transformed
into the $B_{435/438}$ and $I_{814}$ coordinate systems. PSF
photometry was performed on each image independently, and the results
were cross-matched (with matching radius of 1.5 pixels) to yield point
sources that were present in all three frames.

PSF models for each filter of each group were constructed from bright,
isolated, and unsaturated stars with smooth radial curves of
growth. The aperture correction was calculated as an average of the
difference between magnitudes obtained from aperture photometry (with
a 10-pixel aperture) and PSF-magnitudes (calculated at 3 pixels). The
aperture corrections between 10 pixels and infinity in the case of ACS
observations were taken from \citet{Sirianni2005}, and in the case of
WFC3 observation were calculated from the enclosed energy curves as
the difference between unity and the enclosed energy in the given
aperture and wavelength (B. Whitmore, priv. comm.). The foreground 
extinctions were obtained from \citet{Schlafly2011}, published in the 
NASA/IPAC Extragalactic Database (NED)\footnote{http://ned.ipac.caltech.
edu/ngi/}.  Table~\ref{tab:PSF_info} gives overall information on the 
number of stars used in the PSF constructions, the aperture correction 
values, and foreground extinctions for different filters and targets.

\afterpage{
\begin{table*}
\begin{minipage}{\textwidth}
\caption{Information on number of stars used to create a PSF model, aperture corrections, and foreground extinctions in $BVI$ filters for every pointing.}
\label{tab:PSF_info}
\begin{center}
\begin{tabu}{CC@{}|CCC@{}|CCC@{}|C@{}|C@{}|C@{}|C@{}|C@{}|C@{}}
\toprule
Group & Pointing & \multicolumn{3}{C|}{\# of PSF stars} & \multicolumn{3}{C|}{Foregroung extinction} & \multicolumn{6}{c}{Aperture correction} \\
\cmidrule{3-5} \cmidrule{6-8} \cmidrule{9-14}
\rowfont{\small}
 &  & B$_{435}$ & V$_{606}$ & I$_{814}$ & \multicolumn{1}{C}{B$_{435}\,\,$} & \multicolumn{1}{C}{V$_{606}\,\,$} & \multicolumn{1}{C}{I$_{814}$} & \multicolumn{2}{|C|}{B$_{435}$} & \multicolumn{2}{C|}{V$_{606}$} & \multicolumn{2}{C}{I$_{814}$}\\
\cmidrule{9-14}
\rowfont{\scriptsize}
 & & & & & mag & mag & mag & $3\,\rightarrow\,10$ & $10\,\rightarrow\,\infty$ & $3 \rightarrow 10$ & $10 \rightarrow \infty$ & $3 \rightarrow 10$ & $10 \rightarrow \infty$ \\      
\rowfont{\scriptsize}
 & & & & & & & & mag & mag & mag & mag & mag & mag \\
\midrule

\gr HCG 07 & 1 & 8 & 8 & 6 & 0.081 & 0.056 & 0.036 & 0.144 & 0.107 & 0.209 & 0.088 & 0.213 & 0.087\\
           & 2 & 10 & 12 & 12 & 0.081 & 0.056 & 0.036 & 0.223 & 0.107 & 0.139 & 0.088 & 0.224 & 0.087\\
\gr HCG 31 & 1 & 13 & 13 & 18 & 0.210 & 0.144 & 0.093 & 0.161 & 0.107 & 0.121 & 0.088 & 0.211 & 0.087\\
    HCG 42 & 1 & 24 & 41 & 55 & 0.174 & 0.119 & 0.078 & 0.107 & 0.107 & 0.113 & 0.088 & 0.030 & 0.087\\
\gr HCG 59 & 1 & 8 & 11 & 9 & 0.151 & 0.104 & 0.066 & 0.137 & 0.107 & 0.162 & 0.088 & 0.203 & 0.087\\
           & 2 & 11 & 8 & 16 & 0.151 & 0.104 & 0.066 & 0.160 & 0.107 & 0.174 & 0.088 & 0.221 & 0.087\\
\gr HCG 92 & 1 & 27 & 14 & 48 & 0.32 & 0.22 & 0.14 & 0.207 & 0.110 & 0.258 & 0.103 & 0.342 & 0.108 \\
\bottomrule
\end{tabu}\vspace{0.1cm}
\end{center}
\textbf{Notes.} Foreground extinction coefficients in this table were obtained with \citealt{Schlegel1998}, published on NED. $3\,\rightarrow\,10$ aperture corrections were calculated as the mean brightness difference of the stars we used to construct the PSF, between the 3 pixel PSF photometry and brightness measured in a 10 pixel aperture. $10 \rightarrow \infty$ aperture corrections for ACS observations were taken from \citealt{Sirianni2005}. $10 \rightarrow \infty$ aperture corrections for WFC3 observations were calculated as the difference between unity and the enclosed energy in the given aperture.
\end{minipage}
\end{table*}
}

At this stage, we had an unfiltered point source list for each group,
which we refer to as the ``extended sample''.

\subsection{PSF photometry justification}
Quite often in studies of extragalactic star clusters, the question of
which type of photometry to use (aperture or PSF) arises.  If star
clusters can be resolved, then, logically, aperture photometry would
produce the most accurate results. On the other hand, for unresolved
objects PSF photometry is preferred. What about sources that are
marginally resolved?  \citet{Harris2009} have shown that if a SC is
larger than $\sim 10\%$ of the FWHM of the stellar PSF then it can be
treated as marginally resolved and aperture photometry should be
applied.  This result is certainly important for studies that look at
SC sizes, but what about work that focussing on the colours (and
colour-based properties such as ages) of these SCs?

Colours are differences between the magnitudes in different filters
and as such are less dependent on the accuracy of the total magnitude
measurements than the relative magnitudes. That is, if measurements
are carried out consistently for all of the filters, the magnitude
differences between filters (i.e., the colours) should be similar for
both PSF and aperture photometry. For our study, most of the compact
groups are at distances where their SCs may be marginally resolved,
except, perhaps HCG~92 (at 88.6~Mpc). However, most of our groups have
areas with highly variable backgrounds, and in some cases, albeit
rarely, the issue of crowding arises. The above concerns lead us to
prefer PSF photometry. Additionally, if a PSF for a given filter is
generated and applied correctly, PSF photometry can potentially give
more accurate magnitude measurements than aperture photometry, and with
much smaller uncertainties.

We have compared the aperture photometry with our PSF photometry and 
found that the difference between the two are within the accepted error 
level for this study ($< 0.17$ mag).  Moreover, as was mentioned 
previously, the differences in the photometries becoming even less 
noticeable in colour indices ($< 0.09$ mag).  Thus, we conclude that on 
average the two photometries produce comparable results, and we prefer 
PSF photometry for the reasons given above.

\subsection{Star Cluster Candidate Selection}\label{sec:SCC_selection}
To create a catalogue of high confidence star cluster candidates we
applied to our extended samples the following selection criteria:

\begin{enumerate}
\renewcommand{\theenumi}{\arabic{enumi}.}
\item \textbf{Magnitude cut at $\bmath{M_{V_{606}} < -9}$~mag (S1)}\\
To eliminate the contamination from individual luminous supergiants, 
which can reach $M_{V_{606}} \simeq -8$ mag, we only considered 
sources with $M_{V_{606}} < -9$ mag.  This roughly corresponds to 
a cluster mass of a few $\times10^4\, \msun$ at 10 Myr and  
$\sim10^6\, \msun$ at 10 Gyr (depending on metallicity and distance 
modulus).\\

\item \textbf{Photometric error $\bmath{\sigma < 0.3}$~mag in 
all three filters (S2)}\\
To maintain photometric quality of magnitudes and colours, all 
point sources that had photometric errors larger than 0.3 mag 
in any of the $B_{435/438}$, $V_{606}$, or $I_{814}$ filters were 
discarded.\\
 
\item \textbf{Sharpness between $-$2 and 2 in all bands (S3)}\\
To further minimize contamination from cosmic rays and background 
galaxies, we applied a sharpness filter which, essentially, is a 
constraint on the intrinsic angular size of the detected objects.  
Sharpness is measured as the difference between the square of the 
width of the object and the square of the width of the PSF, and 
for our purposes should be between $-2$ and 2: large positive 
values of sharpness are indicative of blended sources and 
partially resolved galaxies, whereas large negative numbers are 
flags for cosmic rays and blemishes. For a well-matched width 
the sharpness is zero.\\

\item  \textbf{$\bmath{\rchi<3\ \text{in the } I_{814}}$-band (S4)}\\
The {\tt DAOPHOT} goodness of fit factor $\chi$ from PSF-fitting in 
the $I_{814}$-band should be less than 3. The use of $I_{814}$-band 
for the $\chi$ parameter is twofold: (a) the PSF model is typically 
best determined in that band because we have the most PSF stars; and 
(b) no contamination from nebular emission lines (e.g., H$\alpha$) 
is expected (nebular emission around a young star cluster may cause 
them to be marginally resolved in the $V_{\rm 606}$-band images).\\

\item \textbf{Colour cuts (S5)}\\
In order to minimize the contamination from foreground Galactic 
stars, we applied colour cuts, where all sources that had colours 
$B_{435/438}-V_{606} > 1.5$ mag or $V_{606}-I_{814} > 1.0$ mag were 
discarded.  As can be seen on fig. 8 of \citet{Trancho2012}, applying 
these colour cuts removes the majority of spectroscopically confirmed 
contaminants.\\
\end{enumerate}

\subsection{Globular Cluster Candidate Selection}\label{sec:GCC_selection}
\begin{figure*}
\centering
\includegraphics[width=0.9\textwidth]{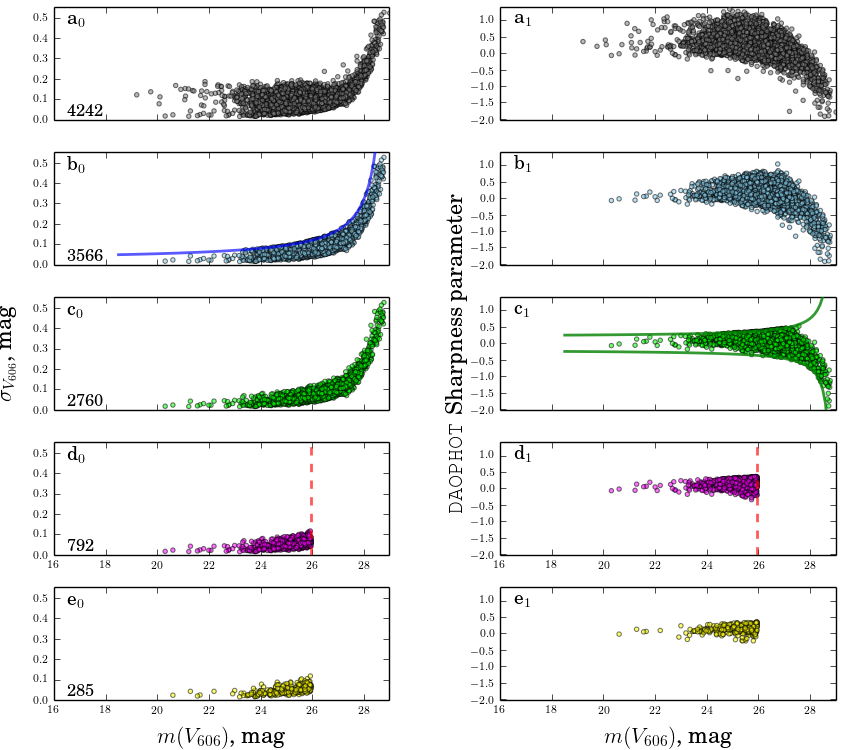}
\caption[The effect of selection criteria applied to point source catalogue]
{Illustration of the criteria applied to the HCG 07 extended
point source catalogue to produce the GCC sample for that group.  The
rows in this figure show the same point sources being plotted in
error, $\sigma_{V_{606}}$, vs. $V_{606}$ magnitude (left-hand-side
panels with subscript 0) and in {\tt DAOPHOT} sharpness parameter
vs. $V_{606}$ magnitude space (right-hand-side panels with subscript
1).  The upper row of panels shows the initial point sources. We apply
the hyperbolic filter in magnitude error, as observed in panel b$_0$,
and panel b$_1$ represents the point sources that satisfy that
criterion in sharpness parameter space.  Panel c$_1$ shows the
hyperbolic filter that was applied in the sharpness parameter space,
and c$_0$ displays point sources in magnitude error space that pass
that filtration. We apply a magnitude cut ($M_{V_{606}} < -7.8$) which
is illustrated by panels d$_0$ and d$_1$. To obtain the final GCC
sample for HCG~07, we select only those sources that have colours
similar to those of the Milky Way globulars (see Fig.~\ref{fig:GC07}, 
panel (a) for reference). Resulting objects are presented in panels 
e$_0$ and e$_1$.  The number in the lower left corner of the 
left-hand-side panels indicates the number of point sources that remain 
after application of each criterion.  (A colour version of this figure 
is available in the online journal.)}
\label{fig:GCC_selection}
\end{figure*}

In a similar manner we created a catalogue of Globular Cluster Candidates, 
with the following (generally stricter) criteria (based on \citet{Rejkuba2005}):

\begin{enumerate}
\renewcommand{\theenumi}{\arabic{enumi}.}
\item  \textbf{Hyperbolic filters in photometric error (G1)}\\
The hyperbolic filter in the photometric error was defined in such a 
manner as to retain $\sim 97\%$ of all recovered artificial sources 
(these are the sources that were used in determining the completeness 
level, see Section~\ref{sec:Completeness}). The application effect of 
this filter on detected sources can be observed in 
Fig.~\ref{fig:GCC_selection}, panels b$_0$ and b$_1$.\\

\item  \textbf{Hyperbolic filters in {\tt DAOPHOT} sharpness 
parameter (G2)}\\
We defined the hyperbolic filter in the $V_{606}$ {\tt DAOPHOT} 
sharpness parameter in the same manner as the photometric error filter.
That is, the sharpness filter is tuned so that it retains $\sim 97\%$ 
of all recovered artificial sources.  However, there is a possibility 
that for the closest groups (e.g., HCG 07, HCG 42) some of the GCCs 
might be marginally resolved.  In these cases we relax the upper envelope 
of the sharpness filter to make sure that we do not discard larger 
GCCs.  The example of the application and effect of the sharpness filter on 
detected sources can be seen in Fig.~\ref{fig:GCC_selection}, panels 
c$_0$ and c$_1$.\\

\item \textbf{Magnitude cut at $\bmath{M_{V_{606}} < -7.8}$ mag (G3)}\\
Because contamination from supergiants is less of a problem for objects 
with GC-like colours, we relaxed our luminosity cut to $M_{V_{606}} < -7.8$ 
mag to get closer to the expected peak of the GC luminosity function $M_V 
\sim -7.4$ \citep[e.g.,][]{Ashman1998, Harris2001}.  The $-7.8$ magnitude 
cutoff roughly corresponds to a globular cluster mass of $\sim5\times10^5\,
\msun$ (depending on metallicity and distance modulus).  The effect of 
application of this filter is shown on Fig.~\ref{fig:GCC_selection}, 
panels d$_0$ and d$_1$.\\

\item  \textbf{Milky Way Globular Cluster selection parallelogram (G4)}\\
We use dereddened GCs from the updated \citet{Harris1996} Milky Way 
Globular Cluster catalogue to derive a selection parallelogram in 
$B-V$ vs. $V-I$ colour space. We convert the vertices of the parallelogram 
from Johnson's $B$, $V$, $I$ magnitudes to $B_{435}$, $V_{606}$, $I_{814}$ 
magnitudes via transformations derived from \citet{Sirianni2005}. We keep 
all sources that would fall into the selection parallelogram in the 
$B_{435} - V_{606}$ vs $V_{606} - I_{814}$ colour space or that would 
overlap the selection region with their 1$\sigma$ error bars.  HCG 92 
was observed with WFC3 and there is (to our knowledge) currently no 
equivalent to the \citet{Sirianni2005} calibration paper for this 
instrument.  Therefore, we apply the same selection box for globular 
clusters to these data as for the ACS data.  Given the similarities 
between these two instruments and their filter sets, we would not 
expect a significant change in the number of globular cluster candidates 
as the result of that action.
\end{enumerate}

Table~\ref{tab:criteria} specifies what fraction (in percent) 
of the initial number of point sources (rightmost column) remains 
after applying each criterion individually. For example, after 
applying the S1 criterion (absolute magnitude cut) on 4243 
sources detected in HCG 07 only 11.6\% (or 493 sources) of 
the original sample remains, whereas after applying the S2 
criterion (photometric error $<$ 0.3 mag in all three filters) 
on the initial list of 4243 sources 91.3\% of the extended 
sample (or 3874 sources) remain, and so on.

\afterpage{
\begin{table*}
\caption{Percentage remaining of the initial sources after applying particular criteria.}
\label{tab:criteria}
\begin{center}
\begin{minipage}{144mm}
\begin{tabu}{@{} CC@{}|CCCCC@{}|C@{}| CCCC@{}| C@{}|C @{}}
\toprule
Group  & Number of & S1  & S2  & S3  & S4  & S5  & SCC & G1  & G2  & G3  & G4 & GCC & N$_{\text{common}}$\\
	& initial sources & \%  & \%  & \%  & \%  & \%  & \% & \%  & \%  & \%  & \%  & \% & \\
\midrule
\gr HCG 07 & 4243&	11.6&	91.3&	100.0&	92.3&	88.3&	6.9&	84.0&	66.1&	40.6&	28.2&	6.7&	86\\
    HCG 31 & 2670&	25.5&	93.7&	99.9&	84.1&	88.8&	12.7&	64.4&	44.7&	58.6&	22.3&	2.9&	13\\
\gr HCG 42 & 2262&	22.1&	94.7&	99.7&	94.2&	82.2&	15.7&	92.1&	85.2&	58.5&	67.8&	39.7&	331\\
    HCG 59 & 3445&	12.9&	76.3&	99.9&	93.5&	84.7&	7.5&	78.8&	79.6&	41.1&	39.1&	10.9&	121\\
\gr HCG 92 & 5493&	52.9&	94.5&	99.3&	82.9&	78.7&	29.4&	92.4&	81.4&	92.6&	23.5&	17.2&	341\\
\bottomrule
\end{tabu}\vspace{0.1cm}
\begin{flushleft}
\textbf{Notes.} Criteria S1 through S5 are for star clusters, criteria G1 through G4 are for 
globular clusters; S1 -- Magnitude cut at $M_{V_{606}} < -9$ mag, S2 -- Photometric error 
less than $0.3$ mag in all three frames, S3 -- Sharpness between $-2$ and $2$ in all bands, 
S4 -- $\chi$ in I-filter less than $3$, S5 -- Color cuts,  G1 -- Hyperbolic filter in the 
magnitude errors, G2 -- Hyperbolic filter in the sharpness, G3 -- Magnitude cut at 
$M_{V_{606}} < -7.8$ mag, G4 -- Colours similar to those of the dereddened Milky Way globular 
clusters.  SCC and GCC columns show the resulting percentage of high confidence star and globular 
cluster candidates, respectively. N$_{\text{common}}$ column shows the number of clusters that apear 
in both SCC and GCC catalogues.
\end{flushleft}
\end{minipage}
\end{center}
\end{table*}}

\subsection{Completeness Levels}\label{sec:Completeness}

To determine completeness of our catalogue we carried out the
following routine for each group in our sample. For groups HCG 07, 42,
31 and 59 we used {\tt ADDSTAR} to add 3000 artificial stars to the
images (over the entire field, including the galaxies) in the apparent
magnitude range 24{\textendash}28 mag, i.e., absolute magnitudes
ranging between $-9.99$~mag and $-5.76$~mag (taking into consideration
that these groups are located in the range of distances from 56.6 and
62.8 Mpc).  For HCG 92 we used {\tt ADDSTAR} to add 5000 artificial
stars to the image as the image covers a larger field of view. The
apparent magnitude range of artificial stars was the same, i.e.,
24{\textendash}28 mag, which translates into absolute magnitudes range
between $-10.74$~mag and $-6.74$~mag, given that the distance modulus
for HCG 92 is 34.74 mag. After artificial stars were added to the
images, we applied the same algorithm for point source detection to
determine the recovery rates. The average limiting magnitudes for the
50\% and 90\% recovery rates for each group are presented in
Table~\ref{tab:completeness}.  As can be seen, we operate at the
slightly higher than 90\% completeness level (in V$_{606}$ filter) for
GCCs and at even higher completeness level for the SCCs, except for
HCG 92 which is the most distant group in our sample. In that case we
are at $\gtrsim 50\%$ and $\sim 90\%$ completeness level for the GCCs
and SCCs, respectively.  We point out that the method described gives
the values for the average completeness level over the entire image.
Because of the random distribution of artificial sources and because
galaxies (with elevated surface brightness) typically take up a
smaller fraction of the field-of-view, our actual completeness levels
will generally be lower as star clusters tend to be found in galaxies
(with HCG~92 as a notable exception).

\begin{figure*}
\centering
\includegraphics[width=0.496\textwidth]{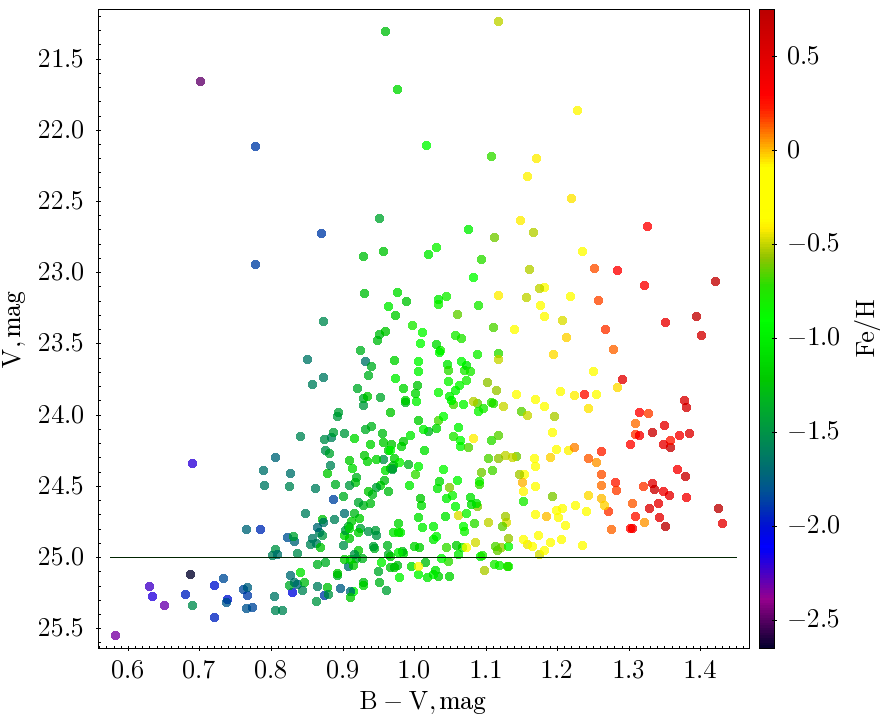}
\includegraphics[width=0.496\textwidth]{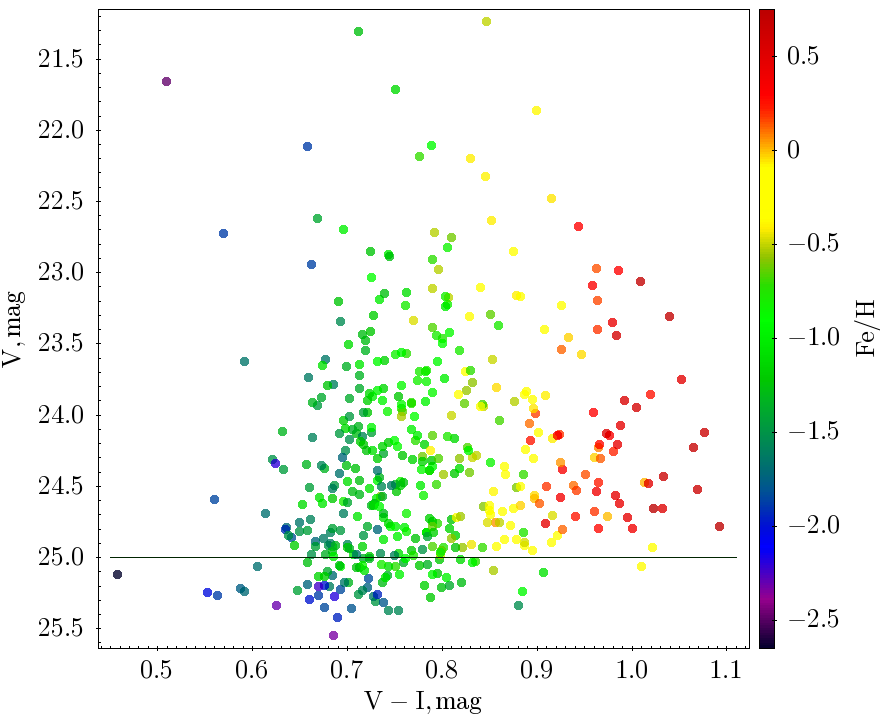}
\caption[$B-V$ and $V-I$ colour-magnitude diagrams for HCG 42]
{$B-V$ and $V-I$ colour-magnitude diagrams for HCG 42.  The 
displayed sources are GCCs at or above 90\% completeness level in all 
three filters.  The sources themselves are colour coded according to 
their metallicities (see \S\ref{sec:metal}), represented with colour 
bars on the right hand side of each plot.  As can be seen from the 
plots, even at the 90\% completeness level we are still missing some 
of the metal enriched GCCs.  To minimise this effect, we are making a 
magnitude cut at $\text{V}=25$~mag for this group, and at 
$\text{V}=25.5$~mag for HCG 07, 31, and 59.  Thus, for obtaining 
values of the expected total number of GCs in a galaxy, $N_{\text{total}}$, 
the value of specific frequency, $S_N$, or for the analysis of the 
metallicity distribution of GCCs we will be using a sub-catalogue 
with aforementioned magnitude cut. (A colour version of this figure is available in the online journal.)}
\label{fig:CMD}
\end{figure*}

The level of completeness becomes especially important when we are
dealing with the specific frequencies and metallicity distributions
for globular cluster candidates.  The $B-V$ colour for GCCs is, on
average, about 1 although our selection parallelogram (the G4
criterion, \S\ref{sec:GCC_selection}) goes down to $B-V=1.38$.  As an
example, consider a hypothetical GCC source in HCG 42.  If this source
has m$_V = 26$~mag (just above the 90\% completeness level in the
$V$ filter; Table~\ref{tab:completeness}), its magnitude in
$B$, according to our G4 criterion, will be between 26.68 and
27.38.  However, the 90\% completeness level in $B$ is $26.2$~mag.
Thus, if we force our sources to have 90\% completeness level only in
the $V$ filter, we will be missing some of the red sources at the
faint end.  Moreover, even forcing 90\% completeness in the 
limiting filter (in our case it is $B$ filter), we are still risking 
missing some objects (Fig.~\ref{fig:CMD}).  
So, for the calculation of the total number of clusters in the GC 
system of a host galaxy, the globular cluster specific frequency, 
and the metallicity distributions, we will be using a portion of our 
catalogue, in which all sources are at or above 90\% completeness 
level in all three filters and which also minimises the loss of the 
faint metal rich GCCs.  For our sample of HCGs, that corresponds to 
$V$-filter magnitude cutoffs at 25.0 mag for HCG 42, and 25.5 mag for 
HCG 07, 31, and 59.  The only exception is HCG 92, the farthest 
group in our sample.  For this group we went down to 50\% 
completeness level, to maximise the number of GCCs  to strengthen 
the statistical conclusions validity.   Unfortunately, even at that 
completeness level, we are still sampling only about 10\% of the 
GCLF making the derived values of the total number in the GC populations 
($N_{\text{total}}$) and specific frequencies ($S_N$) of HCG 92 
galaxies highly uncertain ($\sim$ factor of 5).  Although we forego 
determining $N_{\text{total}}$ and $S_N$ for HCG 92 (for aforementioned 
reasons), we are still attempting a GCC population analysis, based on a 
``face value'' GCC catalogue containing sources that are at or 
above the 50\% completeness level with a cutoff at $V=25.5$~mag 
(to minimise the loss of the faint metal rich GCCs).

\afterpage{
\begin{table*}
\caption{Completeness levels for our Hickson Compact Group sample.}
\label{tab:completeness}
\begin{minipage}{144mm}
\begin{center}
\begin{tabu}{CCCCCCCCCC@{}}
\toprule
Group & \multicolumn{2}{C}{B$_{435}$} & \multicolumn{2}{C}{V$_{606}$} & \multicolumn{2}{C}{I$_{814}$} & Distance & GCC cutoff & SCC cutoff\\
\cmidrule{2-3} \cmidrule{4-5} \cmidrule{6-7}
\rowfont{\small}
      & 50\% & 90\% & 50\% & 90\% & 50\% & 90\% & modulus & $M_{V_{606}}<-7.8$ mag & $M_{V_{606}}<-9$ mag\\
\midrule
\gr HCG 07 & 27.4 & 26.7 & 27.5 & 26.3 & 27.1 & 26.4 & 33.76 & 25.96 & 24.76 \\
HCG 31 & 27.3 & 26.7 & 27.4 & 26.4 & 27.2 & 26.5 & 33.68 & 25.88 & 24.68 \\
\gr HCG 42 & 27.2 & 26.2 & 27.2 & 26.1 & 27.1 & 26.0 & 33.86 & 26.06 & 24.86 \\
HCG 59 & 27.2 & 26.6 & 27.2 & 26.5 & 27.1 & 26.5 & 33.99 & 26.19 & 24.99 \\
\gr HCG 92 & 26.9$^*$ & 25.9$^*$ & 27.1 & 26.0 & 27.0 & 25.9 & 34.74 & 26.94 & 25.74 \\
\bottomrule
\end{tabu}\vspace{0.1cm}
\end{center}
The values for distance moduli were taken from NED with the following cosmology parameters: 
H$_0 = 73.0$ Mpc\,km\,s$^{-1}$, $\Omega_{\text{matter}} = 0.27$, and $\Omega_{\text{vacuum}} = 0.73$.\\
$^*$ For HCG 92 B filter is B$_{438}$ of WCF3 camera
\end{minipage}
\end{table*}
}

\subsection{Physical Extent of the SCC and GCC Systems}

\begin{figure}
\centering
\includegraphics[width=1.0\columnwidth]{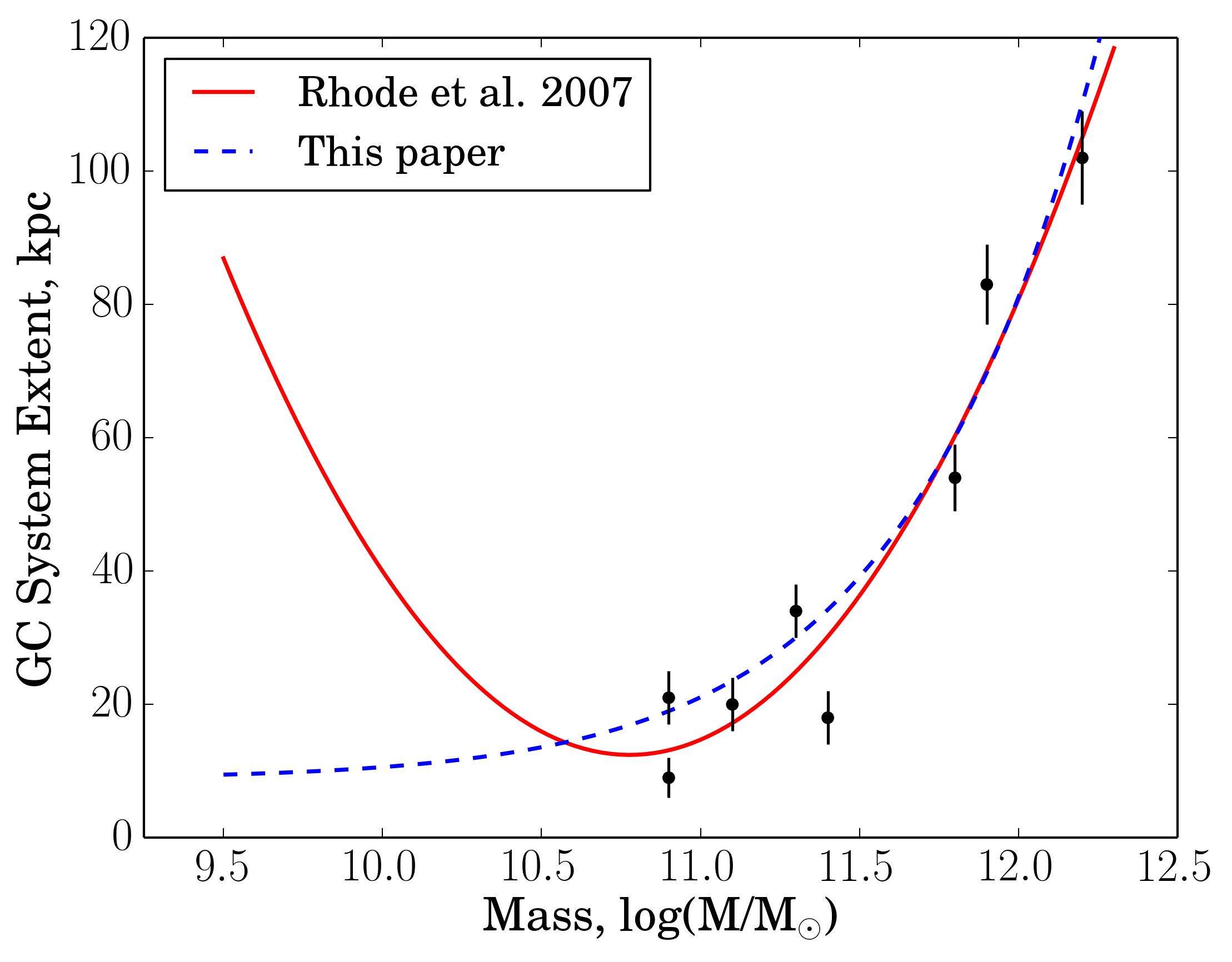}\\
\caption[Modified relation for GC system expected extent]
{The relation between the expected extent of a globular 
cluster system from \citet{Rhode2007} (solid line) and the modified 
version used in this paper (dashed line).  The modified relation was 
meant to be used only on the mass range presented in this plot (i.e., 
from $10^9$ to $\sim 10^{12.5}$ \Msun).  The black dots are from the 
sample of \citet{Rhode2007}, which consists of the radial extent of 
the GC system and the galaxy mass in solar masses for nine elliptical, 
S0, and spiral  galaxies from their wide-field GC system survey.  
(A colour version of this figure is available in the online journal.)}
\label{fig:GC_ext}
\end{figure}

For each group, we present an image with regions that define the 
expected extent of the star cluster and globular cluster systems 
for each galaxy in that group (e.g., panel (b) of Fig.~\ref{fig:CC07}, 
panel (c) of Fig.~\ref{fig:GC07}).  We define the extent of the 
star cluster systems as a brightness contour of $\sim 1.25\sigma$ 
above the background level in the $V_{606}$ image to trace the 
stellar light in each galaxy.  The value of $1.25\sigma$ was identified 
after experimenting with different values to both cover the optical 
extent of the galaxies and include the apparent star cluster populations.  
We successfully applied this criterion in our previous studies, and we 
enclose areas large enough to include sources that are likely associated 
with the host galaxies.  
To define the expected extent for the globular cluster systems 
(which populate much fainter galaxy halos),  we initially use the 
formula outlined in \citet{Rhode2007}: \\
\begin{equation}
y = [\,(45.7 \pm 9.5)\,x^2\,] - [\,(985 \pm 217)\,x\,] + (5320 \pm 1240),
\end{equation}
where $x$ is the mass of a host galaxy in $\log(M/M_{\odot})$ and 
$y$ is the expected radial extent of a system in kpc. However, given 
the quadratic nature of the above equation and the low mass of some 
galaxies in our sample, we needed to modify that expression to get 
more realistic numbers for the lower range of stellar masses.  We 
made an assumption that for a small galaxy with a mass of $10^9 M_{\odot}$, 
the expected radial extent of the GC system would be 9 kpc.  This 
assumption is supported by observed GCC distributions in galaxies 
in our HCG sample.  Combining that with data of the GC systems extent 
from \citet{Rhode2007} and fitting a power-law function, we obtain 
the following relation:\\
\begin{equation}
y = 8.85 + (6.6\times10^{-21})\,x^{20.42},
\end{equation}
with $x$ and $y$ defined as above.  This relation is valid for the 
range of log(mass) values from 9 to 12.5 \Msun. The result (dashed
line) can be seen in Fig.~\ref{fig:GC_ext}.

\subsection{Photometric Estimates of Metallicity}\label{sec:metal}

We use the \citet{Sirianni2005} synthetic transformations to convert
$(B-I)$ colours for GCs to the Johnson photometric system $(B-I)_0$
colours.  We estimate the globular cluster metallicities using the
$(B-I)_0$ to [Fe/H] transformation from \citet{Harris2006}:\\
\begin{equation}
(B - I)_0 = 2.158 + 0.375 [\text{Fe/H}].
\end{equation}
For the GC population of our sample galaxies with a sufficient number
of clusters (we use populations with 40 or more GCs), we plot the
$(B-I)_0$ colour distribution and measure the specific frequency. We
also plot $(B-I)_0$ colour distributions of GCs for each galaxy group
in our sample.  We use the GMM (Gaussian Mixture Modeling) code of
\citet{Muratov2010} to probe the bimodality and to determine the peaks
and dispersions of these distributions. Because mixture modeling codes 
are generally sensitive to extended tails we use GMM on the distribution 
between $-2.5 < [\text{Fe/H}] < 1.0$. The GMM results are recorded in 
Table~\ref{tab:unibi} similarly to table~2 of \citet{Blakeslee2012}, 
where one can find a detailed guide to interpreting the GMM results.  The 
plots of the metallicity distributions for groups and galaxies from 
Table~\ref{tab:unibi} are presented in Figures~\ref{fig:GC07} -- 
\ref{fig:GC92}, and described on a group-by-group basis in \S4 below.

\afterpage{
\begin{table*}
\caption{GMM results of probing modality of metallicity distributions.}
\label{tab:unibi}
\begin{minipage}{\textwidth}
\begin{center}
\begin{tabu}{@{} CCCCCCCCC @{}}
\toprule
Group/Galaxy  & N  & kurt & p-val & p1 & p2 & D & Frac(2) & Bimodal\\
(1) & (2) & (3) & (4) & (5) & (6) & (7) & (8) & (9)\\
\midrule
\gr HCG 07 & 165 & $-0.748$ & 0.115 & $-1.21 \pm 0.23$ & $-0.08 \pm 0.26$ & $2.46 \pm 0.32$ & $0.21 \pm 0.19$ & No\\
HCG 07BD & 83 & $-0.576$ & 0.442 & $-0.86 \pm 0.23$ & $0.31 \pm 0.27$ & $2.73 \pm 0.57$ & $0.07 \pm 0.18$ & No\\
\gr HCG 42 & 400 & $-0.598$ & 0.001 & $-1.05 \pm 0.07$ & $0.15 \pm 0.14$ & $3.14 \pm 0.33$ & $0.25 \pm 0.08$ & Yes\\
HCG 42A & 393 & $-0.611$ & 0.001 & $-1.04 \pm 0.07$ & $0.16 \pm 0.15$ & $3.14 \pm 0.35$ & $0.26 \pm 0.08$ & Yes\\
\gr HCG 59 & 198 & $-0.475$ & 0.592 & $-1.20 \pm 0.46$ & $-0.37 \pm 0.33$ & $1.69 \pm 0.63$ & $0.20 \pm 0.33$ & No\\
HCG 59B & 112 & $-0.306$ & 0.803 & $-1.05 \pm 0.36$ & $-0.43 \pm 0.28$ & $1.36 \pm 0.80$ & $0.38 \pm 0.29$ & No\\
\gr HCG 92 & 290 & $-0.575$ & 0.004 & $-0.79 \pm 0.62$ & $0.63 \pm 0.53$ & $2.65 \pm 0.39$ & $0.09 \pm 0.36$ & Yes\\
HCG 92BD & 67 & $-0.541$ & 0.766 & $-0.46 \pm 0.20$ & $0.68 \pm 0.32$ & $2.76 \pm 0.76$ & $0.07 \pm 0.19$ & No\\
\gr HCG 92C & 46 & $-0.814$ & 0.010 & $-0.79 \pm 0.34$ & $0.59 \pm 0.34$ & $3.24 \pm 0.61$ & $0.18 \pm 0.22$ & Yes?\\
HCG 92E & 52 & $-0.608$ & 0.738 & $-0.93 \pm 0.28$ & $0.09 \pm 0.29$ & $2.36 \pm 0.67$ & $0.29 \pm 0.20$ & No\\
\bottomrule
\end{tabu}\vspace{0.1cm}
\end{center}
\textbf{Notes.} Columns list: (1) group, galaxy or region for which modality is being determined; 
(2) number of analysed GCs in given distribution; (3) kurtosis of the input distribution; (4) an 
indicator of the significance of a bimodal Gaussian over a unimodal Gaussian distribution (lower 
p-values are more significant); (5) mean of the first peak of the proposed bimodal Gaussian 
distribution; (6) mean of the second peak of the proposed bimodal Gaussian distribution; (7) 
separation of the peaks; (8) the fraction of GCCs that was assigned to the second peak of bimodal 
Gaussian distribution; (9) our expectation on bimodality of the given distribution based on the 
evidences displayed in this table.  For the purpose of statistic significance only galaxies/regions 
with over 40 GCCs were checked for bimodality.  The GCCs that were used for this purpose were all at 
or above 90\% completeness level (except for HCG 92, where sources were at or above 50\% completeness 
level).  To minimise the loss of the faint metal enriched GCCs, we further apply a $V=25.5$ magnitude 
cut ($V=25.0$~mag for HCG 42).  All these manipulations reduces the GCC numbers available for analysis, 
but at the same time, brings our completeness to 95\% level for HCG 07, 31, 42, and 59.   
\end{minipage}
\end{table*}
}

\subsection{Empirical Estimate of GC System Population}\label{sec:Predictor}

In recent work, \citet{Harris2013} have determined an empirical 
predictor of the total number of GCs for galaxies of all 
luminosities as a function of effective radius ($R_{\text{e}}$) and 
velocity dispersion ($\sigma_{\text{e}}$), given by equation:\\
\begin{equation}\label{eq:Harris}
N_{GC} = (600 \pm 35)\,\left[\left( \frac{R_{\text{e}}}{10~\text{kpc}}\right)\,\left(\frac{\sigma_{\text{e}}}{100~\text{km}\,\text{s}^{-1}}\right)\right]^{1.29\pm0.03}.
\end{equation}

For those galaxies in our sample for which we were able to find the 
values of $R_{\text{e}}$ and $\sigma_{\text{e}}$ in the literature 
(HCG 42A and HCG 59B), we calculate the predicted numbers of GCs and 
compare them to our estimates (Table~\ref{tab:GCC}) based on the observed 
GC luminosity function in each galaxy.

\section{Results and Discussion}\label{sec:Discuss}

Below, we present a short overview of the star cluster populations in
our sample of Hickson Compact Groups. Analyses of the data presented
in this catalogue for individual groups have been published in a 
number of publications \citep[e.g.,][]{Gallagher2010,Iraklis2013}. 
However, most of these analyses were on a case-by-case basis. In this 
publication, we aim for a systematic approach by applying the same 
criteria to the catalogue selection as whole.  Because of this, there 
will be some differences between already published results and the 
numbers obtained in this paper.  For example, the total number of GCCs 
may differ because we apply a different magnitude cut off or use a 
slightly modified distance modulus.  Throughout the paper, we carefully 
outline all of our steps so the reader can follow them and, if desired, 
modify them to apply their own criteria.

Furthermore, we examine the star cluster populations of compact groups
through the prism of the formation history and evolution of those
groups.  As mentioned previously, \citet{Iraklis2010} outlined a
proposed evolutionary sequence of CGs with respect to the amount and
spatial distribution of cold gas in these groups.  In brief, using the 
ratio of gas mass to the dynamical mass, the groups are divided into three 
types: I, II, and III for gas rich, intermediate, and gas poor groups.  
Moreover, these groups are further split into two parallel sequences, 
depending on the location of gas inside a group.  Sequence~A is 
for groups with gas contained in galaxies, and Sequence~B is for groups 
with gas being dispersed throughout the intra-group medium.  Our sample
represents all three types of groups in terms of gas content, and so
we can trace differences between the group types through the lens of
their star cluster populations.

\begin{figure}
\begin{center}
\includegraphics[width=1\columnwidth]{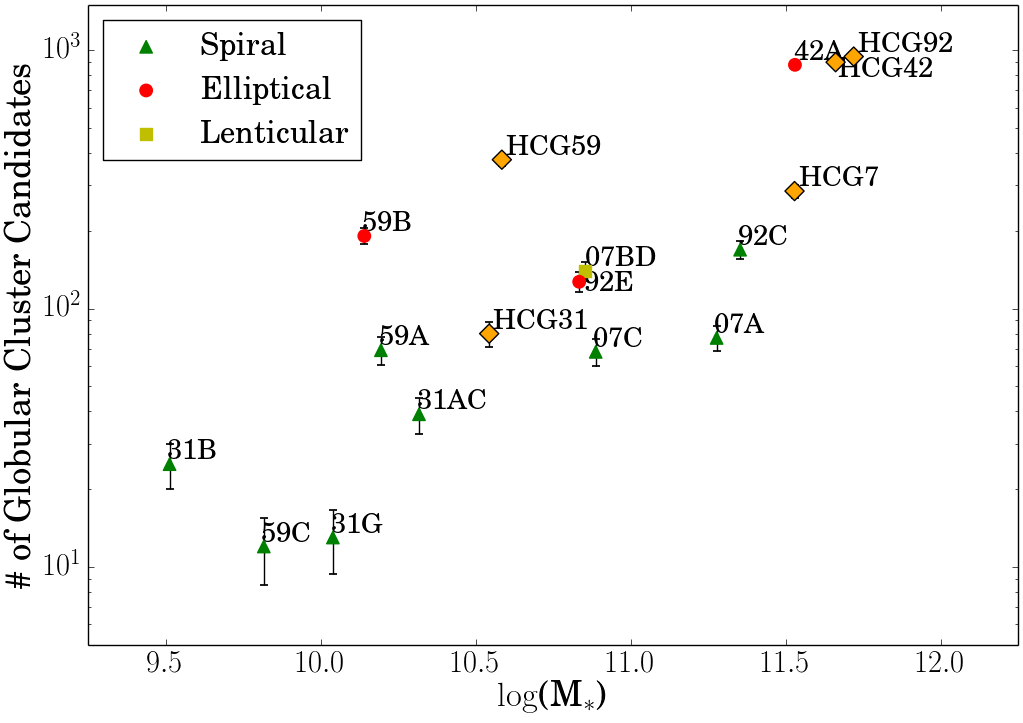}
\caption[Number of GCCs in a galaxy as a function of its stellar mass]
{Plot of the number of GCCs ($M_{V_{606}}<-7.8$) in a galaxy vs. its 
stellar mass. Large diamonds represent HCGs, where the total number of 
globular clusters in each group is the summation of globular clusters 
of its individual galaxies.  Similarly, the mass of a group is the 
summation of the stellar masses of all galaxies in that group from 
\citet{Tyler2014}.  The masses were determined by SED fitting and the 
reported errors (T. Desjardins, private communication, Apr. 30, 2014) 
are not derived from the photometric uncertainties but rather from the 
fitting, and are very small (not visible behind the symbols).  From this 
plot we can see that 59B has a very large globular cluster population, 
on a par with the total numbers for some individual groups. The region 
07BD is presented here as a lenticular galaxy because the dominant galaxy 
B -- expected to have the most GCCs of the two --  is lenticular.  (A 
colour version of this figure is available in the online journal.)}
\label{fig:GCs_vs_M*}
\end{center}
\end{figure}

To check the general properties of galaxies in our sample we plot two 
figures. First, we plot the number of detected GCCs in galaxies as 
a function of stellar masses of those galaxies, Fig.~\ref{fig:GCs_vs_M*}.
On average, the numbers of GCCs in a host galaxy are proportional to the 
stellar mass of that galaxy.  However, galaxy 59B, and as the result the 
whole HCG 59, appears to have an excess of GCs given its stellar mass.  
This is discussed in \S\ref{sec:HCG59} in more detail.

\begin{figure}
\begin{center}
\includegraphics[width=1\columnwidth]{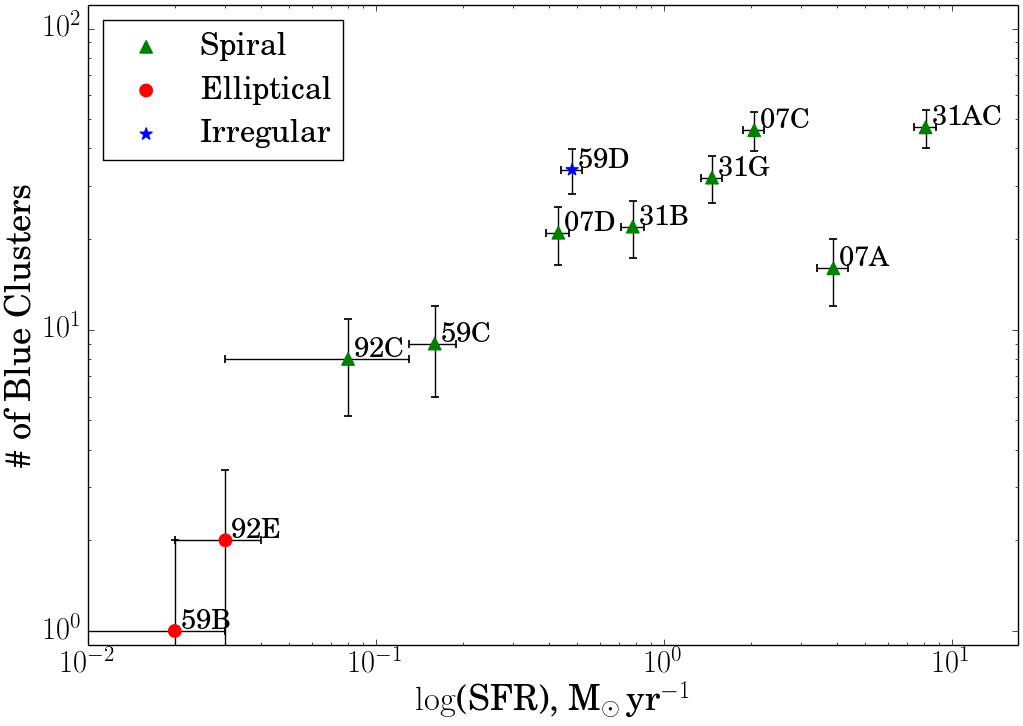}
\caption[Number of ``blue'' clusters in a galaxy as a function of galactic SFR]
{Plot of the number of ``blue'' clusters (with colour 
$V-I < 0.1$) in a galaxy versus its star formation rate. On 
average, the elliptical galaxies have low SFR with 
correspondingly low number of ``blue'' clusters; spirals have 
higher number of young clusters and higher SFRs.  The relatively 
high number of young clusters and relatively low SFRs of galaxies 
92C and 59C are consistent with a recent end to a star formation 
phase in these galaxies.  However, the young clusters are 
still numerous and bright enough to be detected.  Also, as was 
noted earlier, galaxies in HCG 07 exhibit elevated SFR, and 
correspondingly large numbers of young blue clusters, without 
visible signs of interactions. 59B has only one, 92E has two, 
and 07B has none of the blue clusters.  This is expected given 
that these are elliptical/lenticular galaxies.  The region 31AC 
is marked as a spiral because the largest of the two 
interacting galaxies, galaxy A, is classified as spiral. (A colour 
version of this figure is available in the online journal.)}
\label{fig:BCs_v_SFR}
\end{center}
\end{figure}

The second plot represents the number of ``blue'' star clusters (young
SCs with ages $\lesssim 10$~Myr) in each galaxy as a function of star
formation rate of the host galaxies, Fig.~\ref{fig:BCs_v_SFR}.  The
young SCs were selected by applying the colour cut of $V-I < 0.1$. 
This generous colour cut enables us to keep the maximum number of 
young clusters, even those that may have significant reddening, 
avoiding both the evolutionary track loop around ages of 10 and 100 
Myr and the old globular cluster region (see Fig.~\ref{fig:CC07} for 
reference).  The numbers of young SCs behave in a predictable manner 
as well: the galaxies with higher SFR have a larger number of young 
SCs.  Notably, the large irregular 59d has a very high number of young 
SCs given its mass and SFR.

In the subsections below, we discuss the star and globular cluster
populations of each group individually.

\subsection{HCG 07}\label{sec:HCG07}
\subsubsection{Star Cluster Candidates}

This group is classified as a Type II group in the evolutionary
sequence, with an intermediate amount of gas contained within the
individual galaxies. As first presented in \citet{Iraklis2010},
although we do not observe strong signs of interactions between the
galaxies in this group, the large number of young star clusters
indicates that star formation rates are at an elevated level (see
Fig.~\ref{fig:BCs_v_SFR}).  From the distribution of clusters within
the colour-colour diagrams compared to the simple stellar population
models (SSP) of \citet{Marigo2008}, galaxies A, C, and D appear to have the
youngest SCCs, while B has a more mature population. Most of the
youngest SCCs are located down and to the right of the dashed
evolutionary track of nebular emission for $<10$~Myr clusters along
the direction of the reddening vector, consistent with the hypothesis
that these clusters have $A_{V}=1$ to 3 mag. Similarly, 
\citet{Whitmore2002} found that the median extinction value for 
optically selected very young clusters ($\lesssim 4$ Myr) in the 
Antennae galaxies is 2.6 mag.  Moreover, from the distribution of 
SCCs it would appear that the star formation in galaxy C has a more 
extended history, whereas galaxies A and D exhibit an onset of more 
recent star formation (Fig.~\ref{fig:CC07}), as shown by lower cluster 
densities between the ages of 100~Myr and 1~Gyr.

\subsubsection{Globular Cluster Candidates}
The colour-colour plots of the GCC population for galaxies in HCG 07 are
presented in Fig.~\ref{fig:GC07} and their properties are presented
in Table~\ref{tab:GCC}.  Because galaxies A, C, and D are spiral galaxies 
(with C and D being face-on galaxies), the number of GCCs in them should 
be taken with caution. The GCCs located in the central regions and spiral 
arms may be contaminated by reddened young star clusters. For the GCCs of 
galaxies B and D, we considered the pair as a single object because of the
difficulty of distinguishing cluster ownership given the overlap of
the expected extent of the GC systems in these galaxies. As derived
from the $B-I$ colour, the average metallicities of the globular clusters 
in galaxy A (spiral) and the BD (B is lenticular, D is spiral) system are
below Solar metallicity and comparable to metallicities of galaxies of
comparable luminosity \citep[e.g.,][]{Barmby2000, Goudfrooij2003, 
Chandar2004, Kundu2001, Peng2006}.  Galaxy A has $43 \pm 7$ observed 
GCCs with $V < 25.5$~mag. Compensating for the missing portion of 
GC system extent, and assuming a circular symmetry, we estimate the 
total number of observed GCCs as $62 \pm 14$ (taking in consideration 
the foreground sources, Table~\ref{tab:GCC}).  At the given $V=25.5$~mag 
cutoff, assuming the GCLF turnover at $-7.4 \pm 0.2$~mag and width 
$\sigma=1.2\pm0.2$~mag, we sample $28\% \pm 6\%$ of the GCLF.  Taking 
our completeness fraction as $0.95\pm0.05$, we conclude that for 
galaxy A ($M_V=-21.31$~mag) the total number of GCCs in the system is 
$N_{\text{total}}=226\pm73$ and specific frequency is $S_N=0.7\pm0.2$.  
Applying the same  approach to other galaxies in the group we find 
that for the BD region ($M_V=-20.9$~mag for B, $M_V=-19.7$~mag for D) 
$N_{\text{total}}=580\pm150$ and $S_N=2.5\pm0.8$, for C 
($M_V=-20.85$) $N_{\text{total}}=140\pm39$ and $S_N=0.6\pm0.2$.  
We note that because galaxies A and C are spiral (A is highly inclined 
and C is a face-on), a significant number of objects with GC-like 
colours could, potentially, be reddened young star clusters.

\subsection{HCG 31}\label{sec:HCG31}
\subsubsection{Star Cluster Candidates}

This group -- classified as Type I with a cold gas-rich intra-group
medium -- consists of a number of small galaxies apparently coming
together for the first time.  The colour-colour plot of all detected
SCCs, including those in the intra-group medium, paints a picture of a
group that is actively forming stars for the last tens of Myrs (panel
(a) of Fig.~\ref{fig:CC31}).  Simultaneous interactions (e.g.,
galaxies AC and B) have triggered a very high star formation rate
of 8.11 M$_{\odot}\,$yr$^{-1}$ \citep{Tzanavaris2010}, as shown by the
large number of very young SCCs on the colour-colour plots. Virtually
all of the SCCs in regions E and F (24 SCCs combined) are younger than
10 Myr.  These regions are tidal features connecting spiral galaxy G,
interacting pair AC and B, placing a time constraint on the
interaction between those galaxies. There is a high concentration of
SCCs around the region where the evolutionary track makes a backward
loop, essentially making it impossible in this colour-space to
distinguish between SCs of 10 to 100 Myr old (panels (c) and (d) of 
Fig.~\ref{fig:CC31}). However, given the low density of SCs older 
than 300 Myr ($\log(t) \sim 8.5$~yr) and the large number of young SCs 
($\lesssim 10$ Myr), we consider it likely that most of the SCs in the 
vicinity of evolutionary track loop are closer to being a few tens of 
Myr old rather than 100 Myr.

\subsubsection{Globular Cluster Candidates}
The globular cluster population of HCG 31 as expected is rather small,
with only 77 sources with colours similar to the Milky Way's GCs. Most
of the clusters are situated within the boundaries of the major
galaxies AC, B, and G, and, in all likelihood, are reddened young SCs
(especially for galaxies B and G). Additionally, we do not expect a
large GC population in this group as the galaxies are not massive
enough to host a significant number of GCs with masses (and hence
luminosities) large enough to be detectable at such
distances. Furthermore, we observe a lack of old ($>1$~Gyr) star
clusters in the colour-colour plot for SCCs of HCG 31 (panel (a) of
Fig.~\ref{fig:CC31}).  

Determining the total number of GCCs and GC specific frequencies 
in the galaxies of HCG 31 is more straightforward -- there is no 
reason to extrapolate to areas outside the ACS field of view.  
At the $V < 25.5$~mag cutoff we observe $30\% \pm 7\%$ of the GCLF, 
assuming the GCLF turnover is at $-7.4 \pm 0.2$~mag and 
$\sigma=1.2\pm0.2$~mag.  Taking the completeness fraction as 
$0.95\pm0.05$, we obtain $N_{\text{total}}=67\pm23$ and 
$S_N=0.6\pm0.2$ for 31AC ($M_V=-20.5$), $N_{\text{total}}=35\pm14$ 
and $S_N=1.0\pm0.4$ for 31B ($M_V=-18.9$), and 
$N_{\text{total}}=28\pm13$ and $S_N=0.7\pm0.3$ for 31G 
($M_V=-19.0$).

\subsection{HCG 42}\label{sec:HCG42}
\subsubsection{Globular Cluster Candidates}

The HCG 42 group consists of four large galaxies, three of which are
elliptical and one lenticular, with low overall H\one\ content.  In the 
evolutionary sequence scheme this group qualifies as Type III.  More
details on this group configuration including the dwarf galaxy
population can be found in \citet{Iraklis2013}.  Because of the 
predominance of elliptical galaxies in the group and the low content of 
the cold gas we do not expect to find young and intermediate-age star 
clusters, as illustrated by Fig.~\ref{fig:CC42} panel (c).  Thus, we 
start our star cluster population analysis by looking at the GCCs.  The 
ACS observations of this group consist of only one pointing that covers 
the two elliptical galaxies, 42A and C (panel (b) of 
Fig.~\ref{fig:CC42}).  The brightest of these two galaxies, 42A, has an 
optical luminosity comparable to M87 \citep[$M_{\rm V,42a}=-22.8$ and 
$M_{\rm V,M87}=-22.77$;][]{Misgeld2011}.  The extent of the globular 
cluster system for this galaxy covers most of the field of view and 
overlaps the extent of the globular cluster system of the neighbouring 
galaxy 42C.  From the magnitude difference of these two galaxies 
($M_{\rm V,42c}-M_{\rm V,42a} = 2.11$ mag), most of the detected 
globular cluster candidates in this image are expected to belong to 
42A.  Additionally, northward of 42A lies a dwarf galaxy [VC94] 
095753$-$1922.2, which has been identified as a member of HCG 42
\citep{Carrasco2006,Iraklis2013}.  We observe a slight overdensity of
globular clusters in that region (panel (c) of Fig.~\ref{fig:GC42}).  
However, following the same line of argument as above, we consider 
those clusters to be part of the 42A globular cluster system.  Altogether, 
there are 878 detected GCCs that located inside the GC system extent. 
The photometric metallicity distribution as probed by $B_{435}-I_{814}$ 
was determined for 393 of them (for reasons mentioned in 
\S\ref{sec:Completeness}). The distribution has a very well defined 
bimodality with peaks at $\text{[Fe/H]} = -1.04 \pm 0.07$ and 
$\text{[Fe/H]} = 0.16 \pm 0.15$, for the `blue' and `red' peaks, 
respectively.  Both peaks appear to be more metal rich by $\sim 0.4$ 
as compared to the average peaks of GC metallicity distributions of 
approximately $-1.5$ and $-0.5$ measured for different types of 
galaxies \citep[][and references within]{VanDalfsen2004}.

From RC3, we find that the effective radius is $32 \farcs 9$ (estimated
from a Johnson $B$ image and corresponding to $\sim 9.4$ kpc); the
value for the velocity dispersion of $321.4 \pm 9.3$ was obtained from
HyperLeda\footnote{\url{http://leda.univ-lyon1.fr}} \citep{Makarov2014}.  Using the above
numbers and equation~\ref{eq:Harris}, we find that the predicted 
population of the GC system for 42A is $2498 \pm 146$ GCs.  At the cutoff 
magnitude of $M_V=25.0$ and with the assumption for GCLF turnover at 
$-7.4 \pm 0.2$~mag and $\sigma=1.2\pm0.2$~mag, we probe $14\% \pm 7\%$ of 
the GCLF. Taking a completeness fraction as $0.95 \pm 0.05$ we estimate 
$N_{\text{total}} = 3420 \pm 1710$ and $S_N=2.6\pm1.3$.  Although the 
$N_{\text{total}}$ number is $\sim 50\%$ larger that one predicted by 
equation~\ref{eq:Harris}, it is still within reasonable uncertainties.  
Given the large number of GCCs in a luminous, central dominant group 
elliptical, it seems likely that 42A is the product of a gas-rich merger 
from several Gyr ago.

\subsection{HCG 59}\label{sec:HCG59}
\subsubsection{Star Cluster Candidates}
According to \citet{Iraklis2012}, HCG 59 belongs to the Type III
groups, with low H\one\ content relative to its apparent dynamical
mass.  However, a number of young SCCs are found in this group,
located in the smaller galaxies 59C and D (panels (f) and (g) of 
Fig.~\ref{fig:CC59}).  The SCC population of the spiral 59C is somewhat 
small, with only 16 SCCs detected in the disc of the galaxy.  These 
clusters span a range of ages between a few Myr and 1 Gyr, similar to 
clusters in the large irregular 59D. The difference between the SCC 
populations of those galaxies is that 59D has a larger population of 
clusters detected and there is also a number of very young clusters 
present ($\sim 1 Myr$).  Given that these two galaxies have approximately 
the same stellar mass (Table~\ref{tab:HCG_info}) we can compare their 
sSFRs and see that 59D is forming stars over 8 times more efficiently 
than 59C (sSFRs are 0.024 Gyr$^{-1}$ and 0.200 Gyr$^{-1}$ for 59C and 
59D, respectively), which can be seen clearly in the colour-colour plots 
for each galaxy.  The irregular galaxy 59D has a higher sSFR likely
because of the larger amount of available cold gas.  Star formation in
59D may also be enhanced dynamically because of its proximity to 59A.

\subsubsection{Globular Cluster Candidates} 
The majority of GCCs in HCG 59 are part of the globular cluster 
system of the elliptical galaxy 59B (IC 0736) (panels (d) and (g) 
of Fig.~\ref{fig:GC59}).  Intriguingly, it appears that the GCC 
population is much richer than would be expected of a galaxy of its 
luminosity \citep[$M_{B}=-18.5$;][]{Sabater2012}.  We use the SDSS 
\citep{York2005} values for the velocity dispersion, $\sigma=99.9$ 
km$\,$s$^{-1}$ and the value of the effective radius $R_{\text{e}}= 
3 \farcs 18$ (as determined from a deVaucouleurs profile) which 
corresponds to 0.96 kpc.  Substituting these numbers into equation~
\ref{eq:Harris}, we find that the predicted number of GCs for this 
galaxy $N_{GC, \text{pred}} = 29 \pm 2$ is significantly smaller 
than the number of GCs estimated from the observed bright end of 
the GCLF ($N_{\text{total}} = 507 \pm 150$, details to follow), 
and is in fact even smaller than the number of detected GCCs 
($N_{\text{All GCCs}} = 191 \pm 14$ or $N_{\text{95\%}} = 112 \pm 11$ 
for sources with 95\% completeness level). However, we note that 
the velocity dispersion value taken from SDSS appears to be rather 
low for a galaxy with this mass (see Table~\ref{tab:HCG_info}).  
In addition, we point out the unusually dense population of 
extragalactic star clusters located to the south-west of 59B (panel 
(d) of Fig.~\ref{fig:GC59}), away from the visual center of the group 
and along the stellar stream that appears to connect galaxies A and 
B \citet{Iraklis2012}.  One of the possible explanations for this 
population of clusters is that they are possibility a remnant of a 
prior interaction between the A and B galaxies, approximately 1 Gyr 
ago (see \citet{Iraklis2012} for further discussion).

At the cutoff magnitude of $M_V=25.5$ and with the assumption of a 
GCLF turnover at $-7.4 \pm 0.2$~mag and width $\sigma=1.2\pm0.2$~mag, we 
probe $22\% \pm 6\%$ of the GCLF of galaxies in this group. Taking the 
completeness fraction as $0.95 \pm 0.05$ we estimate $N_{\text{total}} 
= 86\pm33$ and $S_N=0.7\pm0.3$ for 59A ($M_V=-20.1$; after removing 
clusters around the irregular 59D that are within the expected 59A 
GC system extent, but are most likely reddened young star clusters), 
and $N_{\text{total}} = 507\pm105$ and $S_N=8.7\pm2.6$ for 59B 
($M_V=-19.4$).  For galaxy 59C, the number of detected GCCs from the 
95\% completeness sub-catalogue, is on a par with the number of 
contaminating sources, $N_{obs}\leq2$.  That gives us the upper limits 
for $N_{\text{total}} < 10$ and $S_N < 0.1$.  We did not estimated 
$N_{\text{total}}$ and $S_N$ for 59D because the detected objects 
may well be reddened bluer objects, rather than GCCs.

Being the only elliptical with sufficient number of GCCs in this group, 
59B was checked for bimodality in its metallicity distribution.  The 
GMM statistical results do not support the idea of bimodality, rather, 
it would appear that the distribution is unimodal with a peak at 
$[\text{Fe/H}] = -1.04 \pm 0.05$. 

In all, the population of old clusters in HCG 59 is intriguing enough 
to warrant further study.\\

\begin{table}
\caption{General properties of GCC systems in galaxies in our sample.}
\label{tab:GCC}
\begin{minipage}{\columnwidth}
\begin{center}
\begin{tabu}{@{} CCCCC @{}}
\toprule
Galaxy  & $N_{\text{GCC}}$  & $N_{\text{contam}}$ & $N_{\text{total}}$ & $S_{N}$ \\
(1) & (2) & (3) & (4) & (5) \\
\midrule
\gr HCG 07A & $62 \pm 14$ & $2 \pm 1$ & $226 \pm 73$ & $0.7 \pm 0.2$ \\
HCG 07BD & $155 \pm 22$ & $2 \pm 1$ & $580 \pm 150$ & $2.5 \pm 0.8$ \\
\gr HCG 07C & $38 \pm 6$ & $2 \pm 1$ & $140 \pm 39$ & $0.6 \pm 0.2$ \\
HCG 31AC & $19 \pm 5$ & $1 \pm 1$ & $67 \pm 23$ & $0.6 \pm 0.2$ \\
\gr HCG 31B & $10 \pm 3$ & $1 \pm 1$ & $35 \pm 14$ & $1.0 \pm 0.4$ \\
HCG 31G & $8 \pm 3$ & $1 \pm 1$ & $28 \pm 13$ & $0.7 \pm 0.3$ \\
\gr HCG 42A & $465 \pm 26$ & $4 \pm 2$ & $3420 \pm 1710$ & $2.6 \pm 1.3$ \\
HCG 59A & $18 \pm 5\ ^* $ & $1 \pm 1\ ^{**}$ & $86 \pm 33$ & $0.7 \pm 0.3$ \\
\gr HCG 59B & $106 \pm 11$ & $1 \pm 1\ ^{**}$ & $507 \pm 150$ & $8.7 \pm 2.6$ \\
HCG 59C & $ \leq 2$ & $1 \pm 1\ ^{**}$ & $< 10$ & $< 0.1$ \\
\gr HCG 92BD & $10 \pm 4$ & $5 \pm 2$ & $\cdots$ & $\cdots$ \\
HCG 92C & $1 \pm 3$ & $4 \pm 2$ & $\cdots$ & $\cdots$ \\
\gr HCG 92E & $5 \pm 3$ & $2 \pm 1$ & $\cdots$ & $\cdots$ \\
\bottomrule
\end{tabu}\vspace{0.1cm}
\end{center}
\textbf{Notes.} Columns list: (1) galaxy id; (2) number of detected GCCs in the 
system's extent.  The GCCs presented here are all at or above the 90\% completeness level 
in all three filters.  Additionally, $V$-filter magnitude cutoffs were applied at 24.5~mag 
for HCG 92, 25.0~mag for HCG 42, and 25.5~mag for HCG 07, 31, and 59. The reasons for doing 
so are explained in \S\ref{sec:Completeness}.  If the full system extent is not visible, 
detected numbers of GCCs are scaled to estimate the numbers of the full system extent.  
Foreground contamination is taken into account, i.e. subtracted from the number of detected 
GCCs; (3) estimated number of contamination sources from Besan\c{c}on Milky Way stellar 
population model \citep{Robin2003} in the direction of the group up to the distance of 100 
kpc, with colours similar to MW GC colours and in the visible area of GC system extent; 
(4) estimated total number of GC population based on GC luminosity function; (5) specific 
frequency.  
$^*$~Because GCC extent of 59A overlap sources from 59D (which are, 
most likely, reddened young star clusters), the predicted 
number of GCCs for 59A galaxy was obtained by subtraction of 59D 
sources from all detected sources in that region.  $^{**}$~For HCG 59, 
the background contamination is much higher ($4 \pm 2$), due to the 
close proximity of the Sagittarius dwarf galaxy \citep{Iraklis2012}.
\end{minipage}
\end{table}

\subsection{HCG 92}\label{sec:HCG92}
\subsubsection{Star Cluster Candidates}
HGC 92, which also known as Stephan's Quintet and which classified as Type 
II in the proposed CG evolutionary sequence \citep{Iraklis2010}, is a 
group of five galaxies (including the foreground interloper NGC 7320) with 
numerous signs of past and ongoing interactions. Another galaxy associated 
with the group, NGC~7320C, is not in the HST field-of-view. 
As a result, this group exhibits the largest number of detected star 
clusters in our sample.  There are a number of interesting features 
singular to this group which are explored in depth in \citet{Fedotov2011}.  
For example, we were able to detect star clusters in two tidal tails, the 
Old Tail (OT) and Young Tail (YT) (Fig.~\ref{fig:CC92}). Because tidal 
tails typically have low gas and dust content \citep[$A_V \lesssim 0.5$ 
mag; e.g.,][]{Temporin2005}, star clusters in tails usually suffer minimal 
reddening and their ages estimated from $BVI$ colour-colour plots are more 
accurate than in galaxy discs. In our case, star clusters detected in 
these tails have compact distributions in the colour-colour plane, 
supporting the idea that SCCs within each tidal tail were formed coevally 
\citep{Trancho2012}, presumably during the interactions that caused the 
formation of those features. Thus, from overdensities of star clusters in 
colour-colour plots and supplemental information from the literature, we 
were able to estimate the ages for the young and old tidal tails to be 
150--200 Myr and 400--500 Myr, respectively \citep{Fedotov2011}.

HCG~92 is an ideal system to study populations of star clusters forming
outside of galaxies.  In particular, there are two areas of
extragalactic clusters labelled as the Northern Star Burst Region
(NSBR) and the Southern Debris Region (SDR).  The NSBR has a SCC
population that spans a wide range of ages, from young SCCs of a few
Myr to very old globular clusters of over 10 Gyr old.  This region
includes two intersecting tidal arcs, a byproduct of the interaction
between 92B (NGC 7318B) and 92D (NGC 7318A).  The young SCs detected
in the region were likely formed during that interaction. The presence
of a significant number of intermediate-age SCCs (ranging from 100 to
500 Myr old) could be indicative of earlier interactions involving 92C
(NGC 7319), NGC 7320C, and perhaps 92D \citep{Moles1997, Xu1999}. And
finally, there are a few old star clusters likely deposited into that
region through gravitational interactions between the galaxies.

Consideration of the SCC population of the Southern Debris Region, on
the other hand, paints a bit different picture.  To begin with, there
are not as many very young (1 to 8 Myr) SCs in that region. Also,
there is a well-defined separation between two groups of SCCs in the
colour-colour plane, one group consists of clusters that are
approximately 10 to 100 Myr old (unfortunately, it is impossible to
establish more precise ages based only on $BVI$ photometry) and the
second group, a collection of older star clusters ranging in age from
1 to 10 Gyr.  Interestingly, there appears to be a concentration of
clusters with ages between 6 and 8 Gyr.  Since it is highly unlikely
that a galaxy interaction would specifically launch into this
extragalactic region star clusters of such a limited age range, we
speculate that these clusters (with ages 6 to 8 Gyr) were formed
together at some location, and the population of the younger SCCs is
either the latest addition or a chance projection.  We consider the
former as more likely explanation, and as such, SDR could potentially
be the remnants of a dwarf galaxy that used up its last reservoir of
gas to form these younger population of clusters.  However, there are
a few reasons why this might not be true.  For example, the SDR region
appears to extend over a large area for a dwarf galaxy, and there is 
no detection of extended, diffuse light consistent with a dwarf galaxy.

The spiral galaxy 92C is the largest galaxy in the group. At the same
time, it has no detectable H\one\ \citep{Sulentic2001}, which most
likely was stripped during previous interactions among the group
members \citep{Moles1998}. Unless the galaxy manages to acquire more
cold gas, the intermediate age (100 to 500 Myr) population of SCCs is
likely the trace of the last epoch of star formation in that galaxy.

\subsubsection{Globular Cluster Candidates}

For the GCC population in SQ we focus on three regions (panel (c) of 
Fig.~\ref{fig:GC92}).  One region is associated with the elliptical 
galaxy 92E (NGC 7317), another one with the spiral galaxy 92C (NGC 7319), 
and the last one with the interacting pair 92B and D (NGC 7318B and A).  
For the last system, we expect that a large fraction of GCCs belong to 
92D: first, because it is an elliptical galaxy, and, secondly, it is more 
massive than the spiral 92B (Table~\ref{tab:HCG_info}).  According to our 
estimates of the GCC system extent, the GCC systems of 92C and 92BD 
overlap.  Moreover, the overlap adds virtually no GCCs to 92BD region, 
whereas the expected system extent of 92C adds an appreciable part of the 
eastern tidal feature of 92BD.

According to GMM statistics, the distribution of all GCCs in SQ has 
detectable bimodality (panel (b) of Fig.~\ref{fig:GC92}).  However, the 
individual distributions for each galaxy/region do not have well defined 
bimodality, with the possible exception in the case of 92C.  However, 
this galaxy does not have a large enough number of GCCs for statistically 
significant analysis.  If we look at the unimodal peaks of $-0.55 \pm 0.12, 
-0.38 \pm 0.08$, and $-0.63 \pm 0.10$, for the GCC systems in 92C, 92BD, 
and 92E, respectively, we notice that these values are a bit larger (less 
negative) compare to the average value of Galactic GCs, the bulk of which 
have metallicities around [Fe/H] $= -1.3$ \citep{Murdin2001}.

\subsection{Bimodality of GCC Population of Elliptical Galaxies}

Above we discussed bimodality in the galaxies from our sample and the 
results are outlined in Table~\ref{tab:GCC}.  We compare our results with  
those in the literature, in particular, the bimodality analysis in the 
sample of 92 elliptical galaxies in the Virgo cluster \citep{Peng2006}.  
The authors found a relationship between the GCs colour distribution 
modality and absolute magnitude of the host galaxy.  According to their 
fig. 5, which plots the colour distributions of GCs in seven bins of 
host galaxy magnitude, for the luminous HCG 42A we expect a well defined  
bimodality, whereas with the much lower luminosity of HCG 59B, only a weak 
bimodality is expected.  Given that we observe only $22\% \pm 6\%$ of the 
total GC population in that galaxy, it is not surprising that we can not 
confirm bimodality of GC distribution.  Thus, our findings compare well 
with the conclusions of \citet{Peng2006}: HCG 42A has a well defined 
bimodal distribution and the colour distribution of HCG 59B GCCs is 
unimodal.  The question of the nature of colour distribution for HCG 92E 
is still open.  The luminosity of HCG 92E, according to \citet{Peng2006}, 
corresponds with a relatively well defined bimodality.  However, we 
currently do not posses the required number of GCCs to definitively answer 
that question.  As it is, with the currently observed number of GCCs, HCG 
92E has a unimodal distribution.

\subsection{Star Cluster Populations in Compact Groups}

\begin{figure*}
\begin{center}
\includegraphics[width=0.9\textwidth]{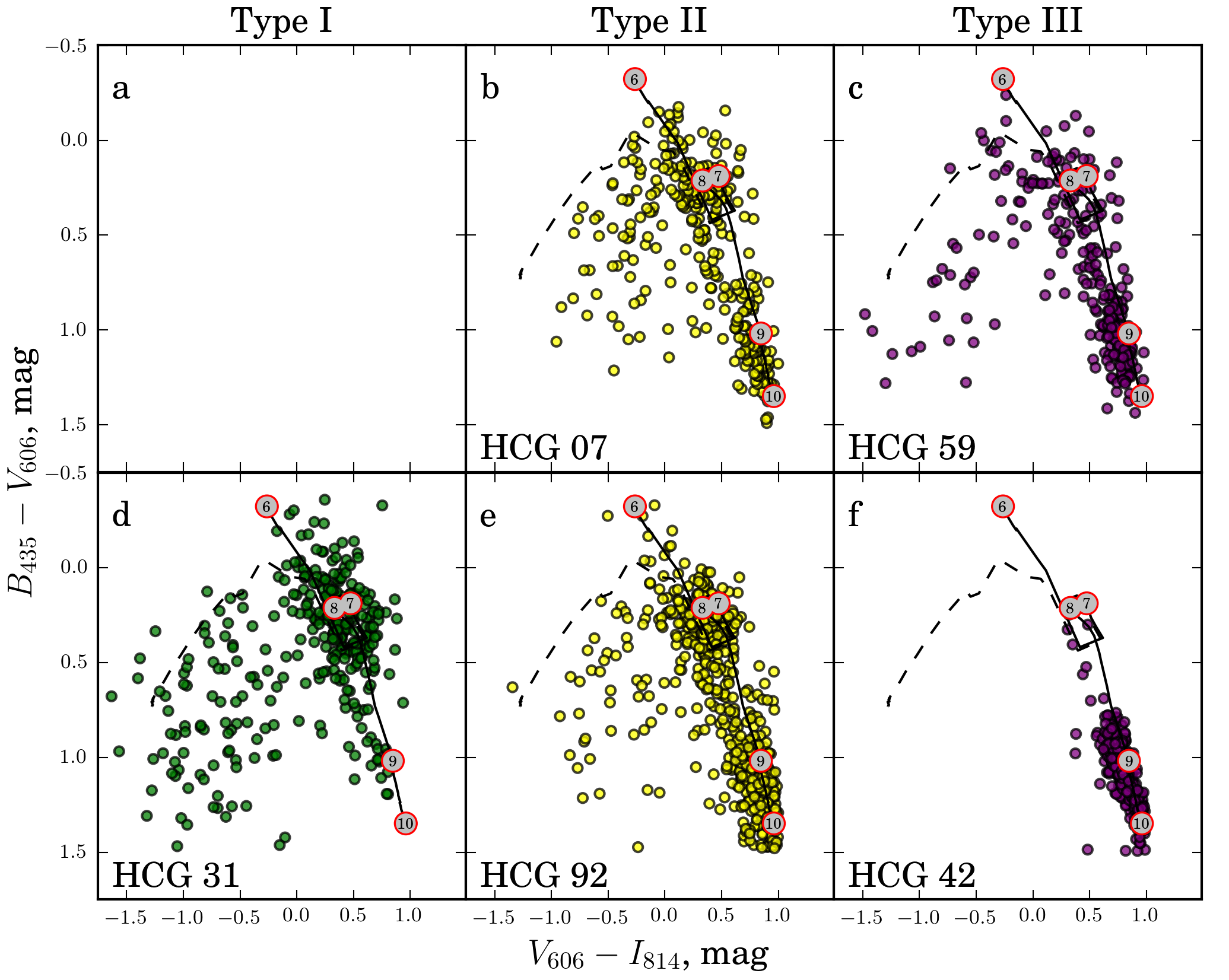}
\caption[Colour-colour plot of the SCC populations of the 
compact groups in the sample]
{Colour-colour plots of the SCC populations of the 
compact groups in our sample, arranged similarly to fig.~1 in 
\citet{Iraklis2010}.  The solid black line trace the evolution of SSP 
models of 1.0 Z$\odot$ \citep{Marigo2008}. The dashed line to the left of 
the main evolutionary track represents a track that incorporates a model 
of nebular emission \citep[Starburst99;][]{Leitherer1999}, common for 
young star clusters \citep[e.g.][]{Vacca1992, Conti1996}.  The numbers 
on the track denote age represented in $\log(\text{age}/\text{yr})$.  
The upper panels represent groups with the H\one\ gas contained within 
the member galaxies, whereas the lower panels represent groups with the 
H\one\ gas stripped from the galaxies.  These groups tend to have a rich 
intra-group medium.  Although \citet{Iraklis2013} have classified HCG 
59 (panel c) as a Type III group based on its estimated dynamical mass, 
the SC population is more consistent with a Type II group. (A colour 
version of this figure is available in the online journal.)}
\label{fig:SC_evol}
\end{center}
\end{figure*}

Here, we relate the star cluster populations to the CG evolutionary sequence
proposed by \citet[][see their fig. 1]{Iraklis2010}.  Upon inspection,
the SCC populations of different group types have qualitative differences.  
Specifically, the population of HCG 31 (panel (d) of Fig.~\ref{fig:SC_evol}), 
classified as a Type I group (gas-rich and dynamically young), is 
distinguished by a large number of young and intermediate-aged SCs. 
Type II groups, represented in our sample by HCG 07 and 92 (panels (b) and (e), 
respectively), still have a large number of young and intermediate SCs; 
however, a population of $> 500$\,Myr -- 10\,Gyr is also well defined. 
For compact groups of Type III, which predominantly consist of early type 
galaxies, we can observe the lack of young SCs, and a significant reduction 
in the numbers of intermediate-aged SCs (HCG 42, panel (f)). Initially, 
\citet{Johnson2007} classified HCG 59 (panel (c)) as Type II group. The recent 
work of \citet{Iraklis2013} has associated five additional dwarf galaxies with 
that group.  The updated information on velocity dispersion and H\one\ mass 
led to a change of the HCG 59 from Type II to Type III. However, based on the 
observations outlined above, the SC population of HGC 59 would appear to be 
more consistent with that of Type II groups.  


These discrepancies between type classification based on the SC
populations and on the ratio of H\one\ to dynamical mass point to the
issue of accuracy in measurements of the dynamical masses.  In
particular, it is quite difficult to accurately measure a dynamical
mass of a group given only a handful of radial velocities and the
projected separations between galaxies \citep{McConnachie2008}. It may
be more appropriate in this case to classify the groups based on the
ratio of H\one\ to total stellar mass, which can be much more reliably 
measured from near-to-mid infrared photometry \citep{Tyler2014}.

\subsection{Cluster Luminosity Function}

\begin{figure*}
\begin{center}
\includegraphics[width=0.3\textwidth]{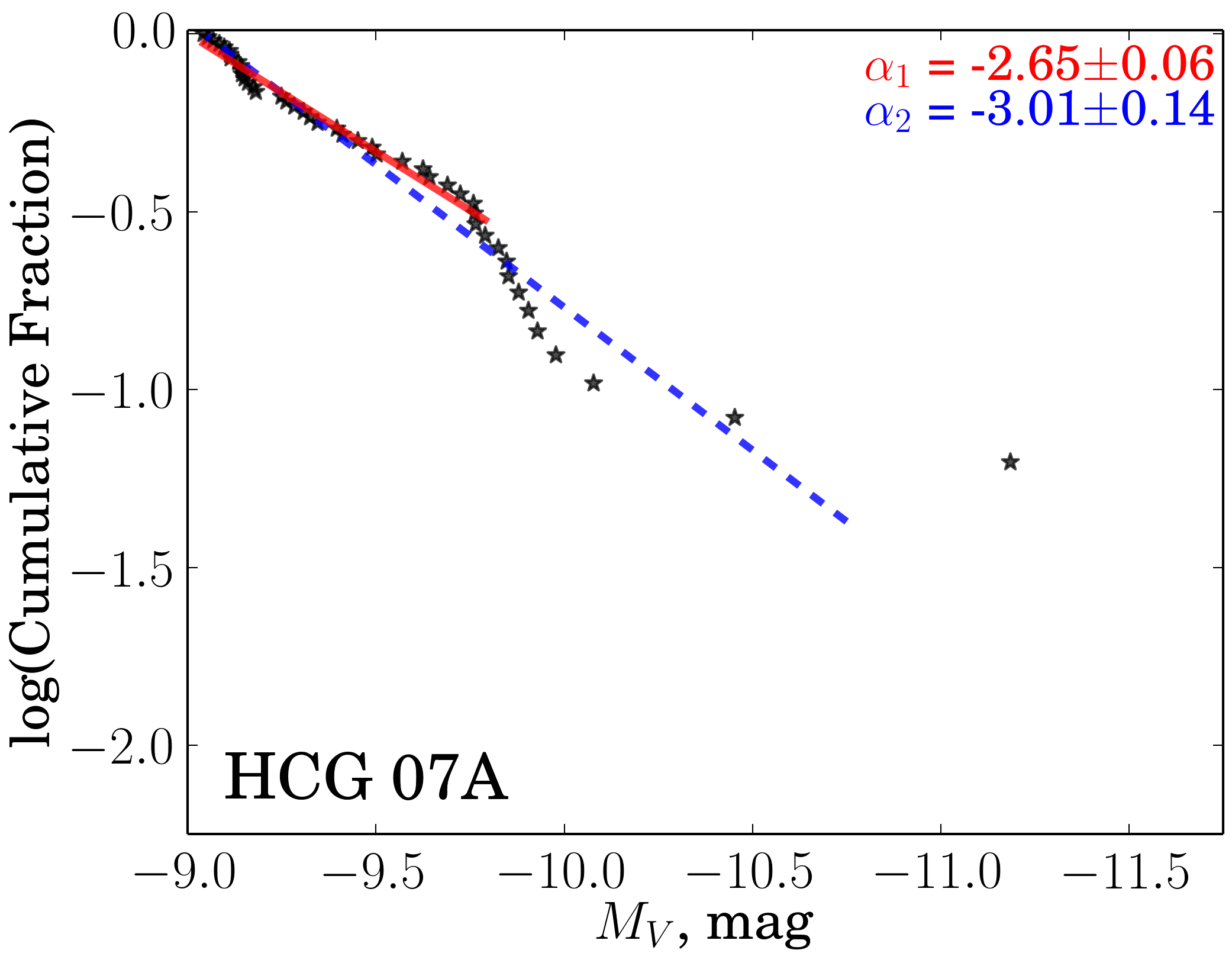}
\includegraphics[width=0.3\textwidth]{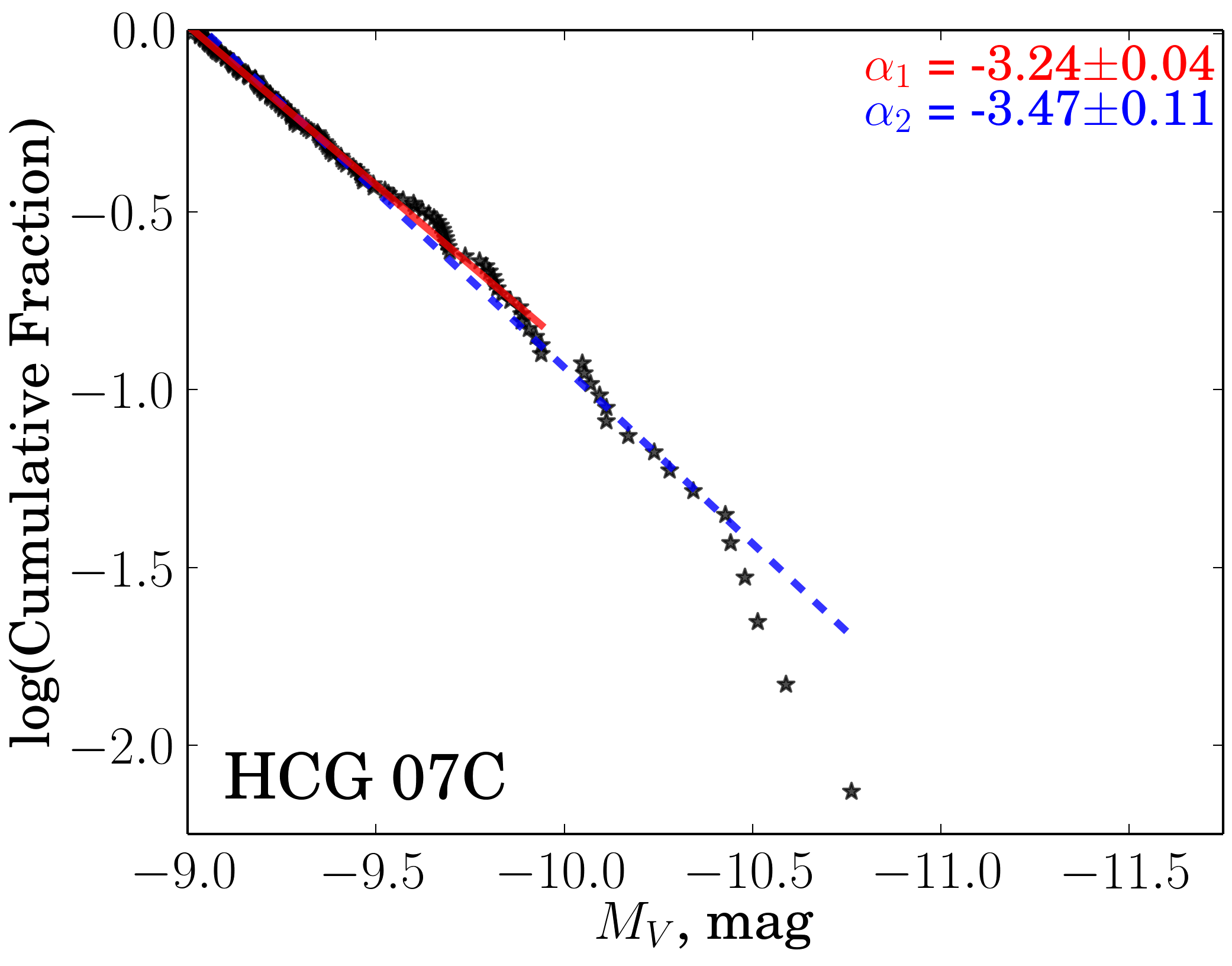}
\includegraphics[width=0.3\textwidth]{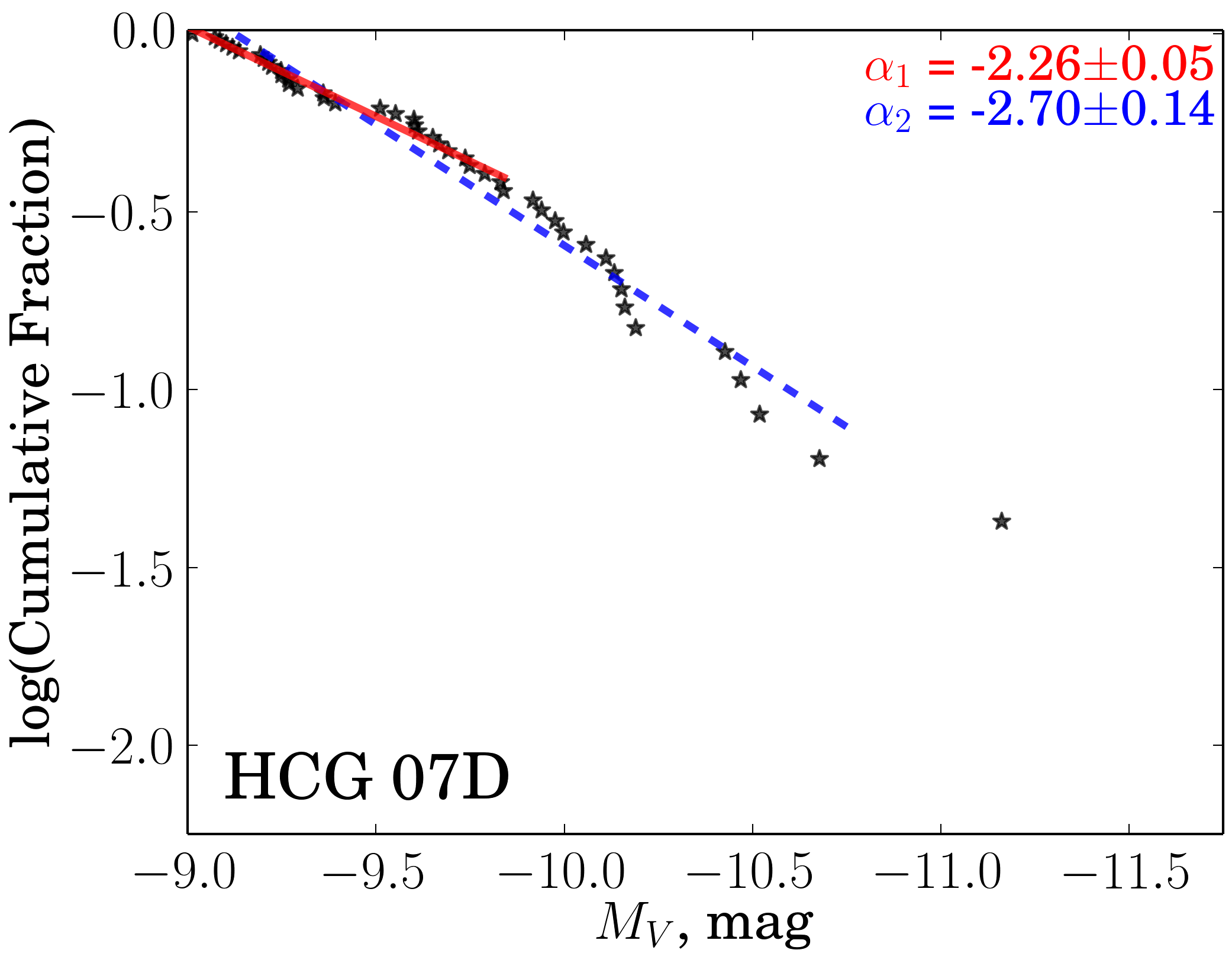}\\
\includegraphics[width=0.3\textwidth]{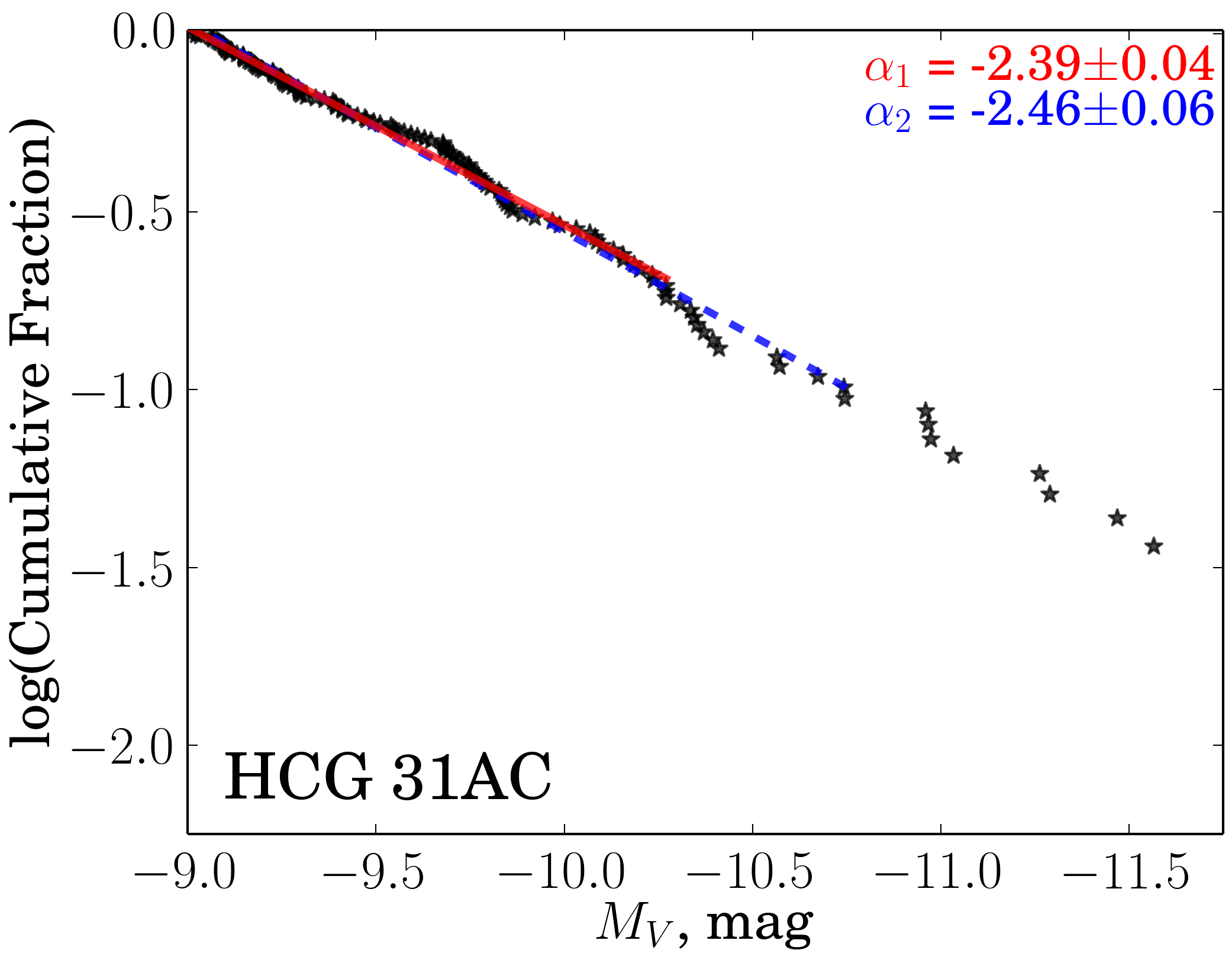}
\includegraphics[width=0.3\textwidth]{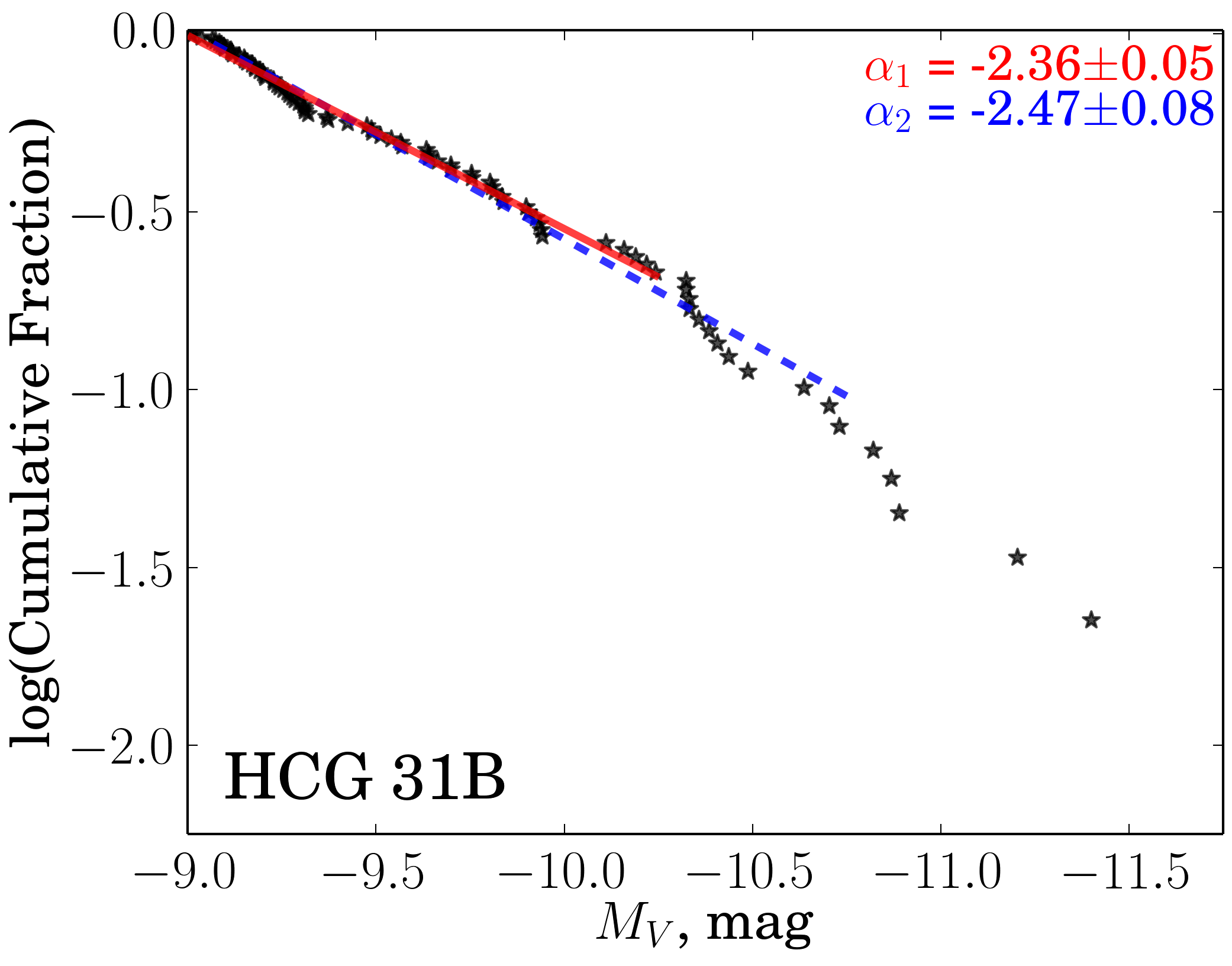}
\includegraphics[width=0.3\textwidth]{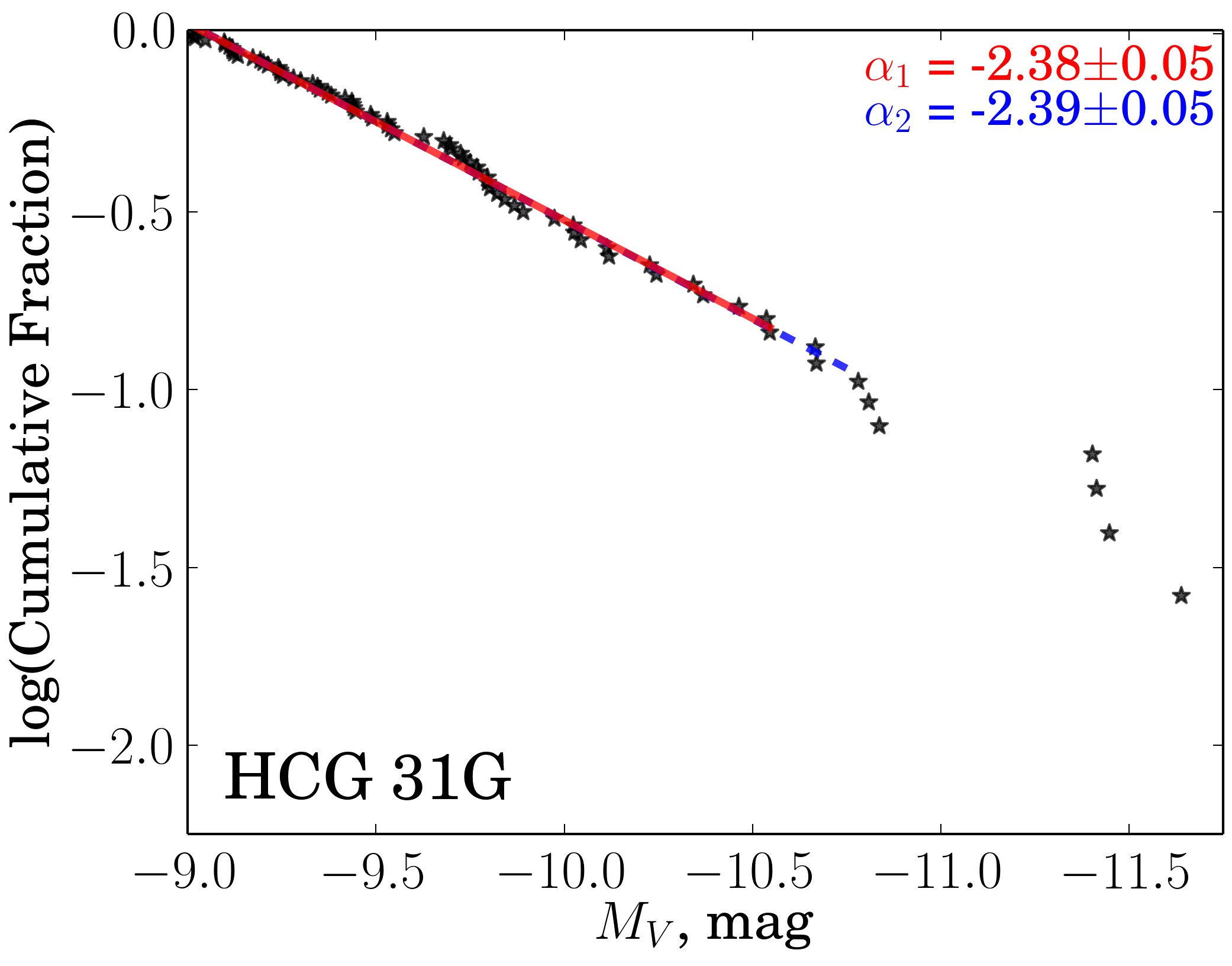}\\
\includegraphics[width=0.3\textwidth]{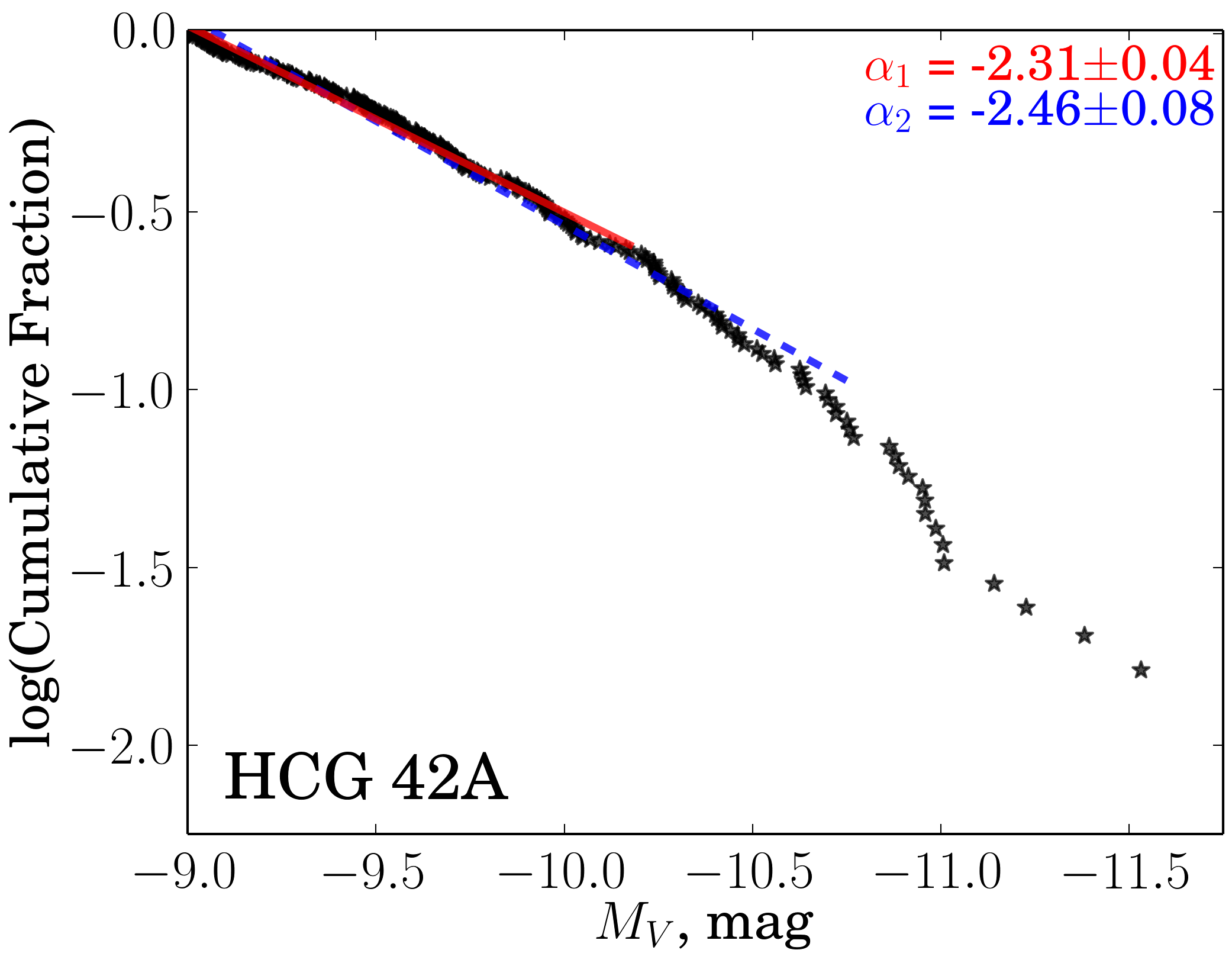}
\includegraphics[width=0.3\textwidth]{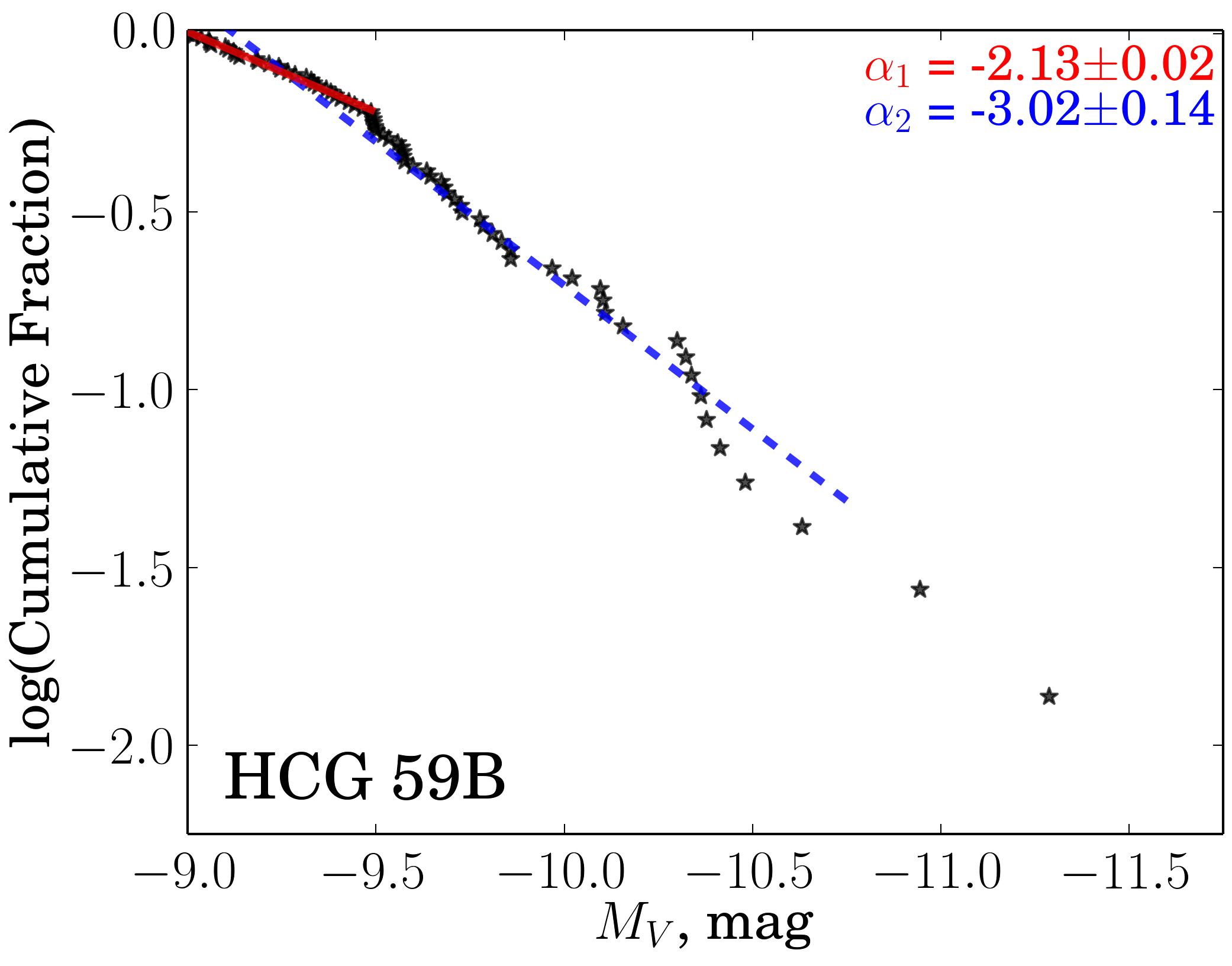}
\includegraphics[width=0.3\textwidth]{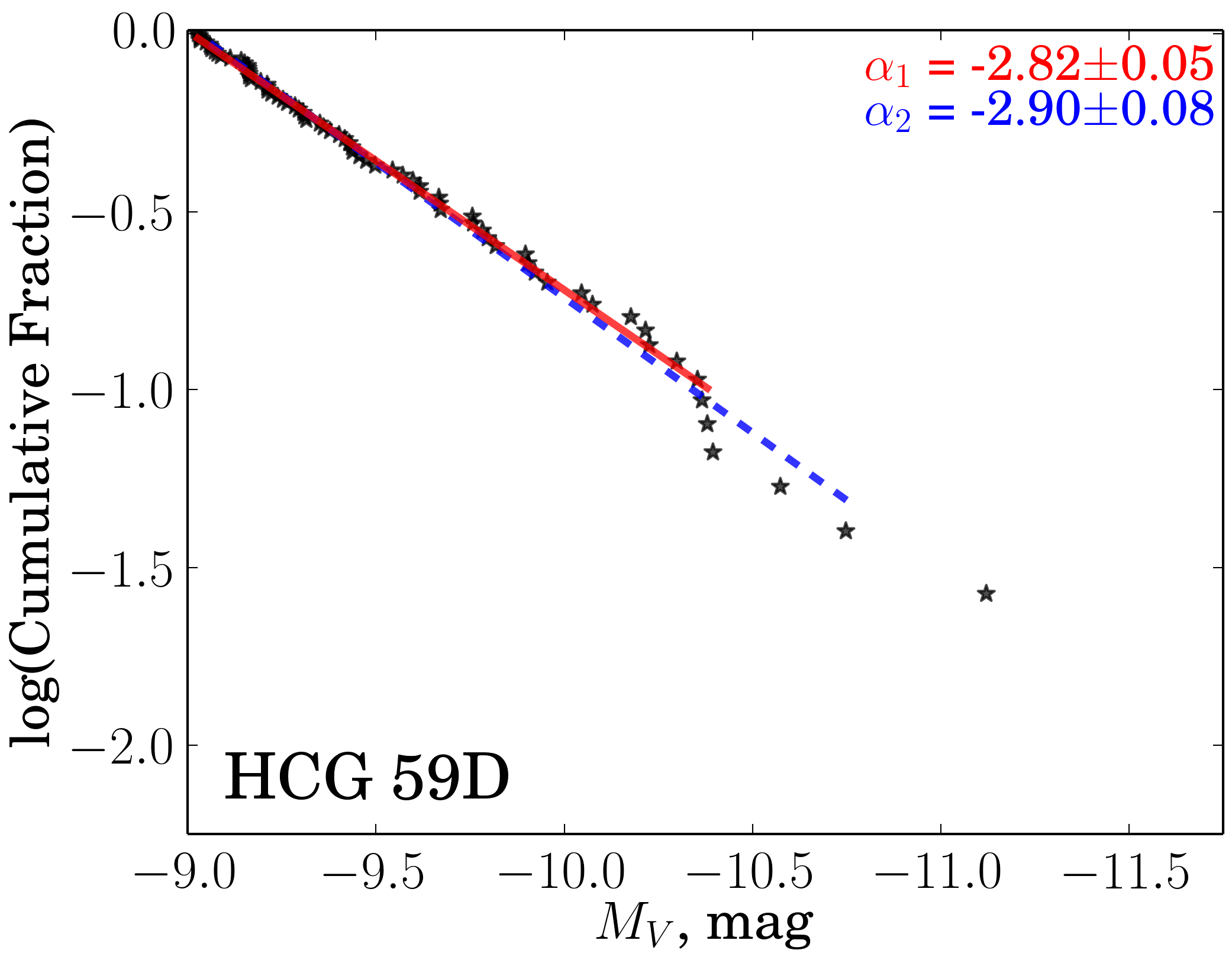}\\
\includegraphics[width=0.3\textwidth]{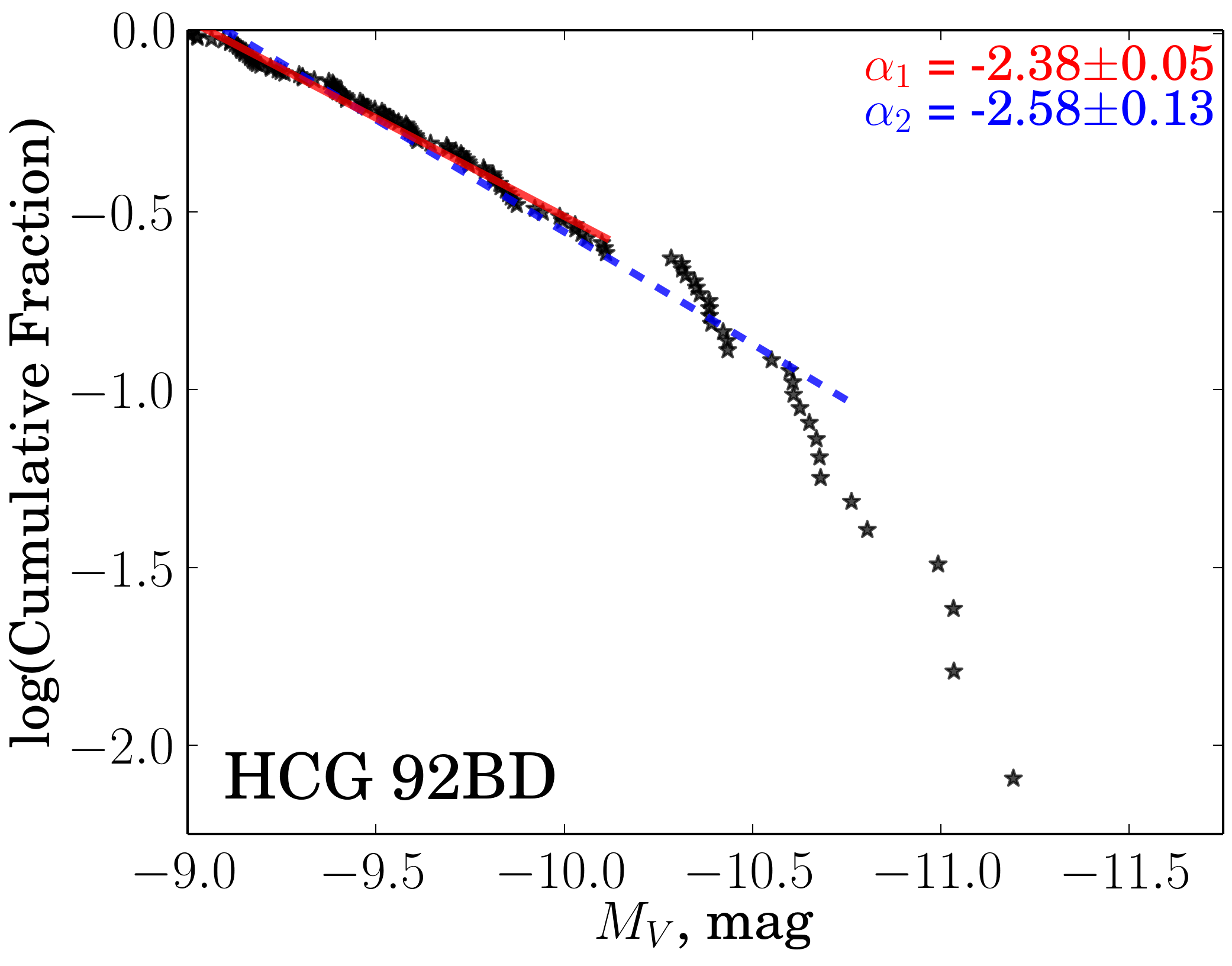}
\includegraphics[width=0.3\textwidth]{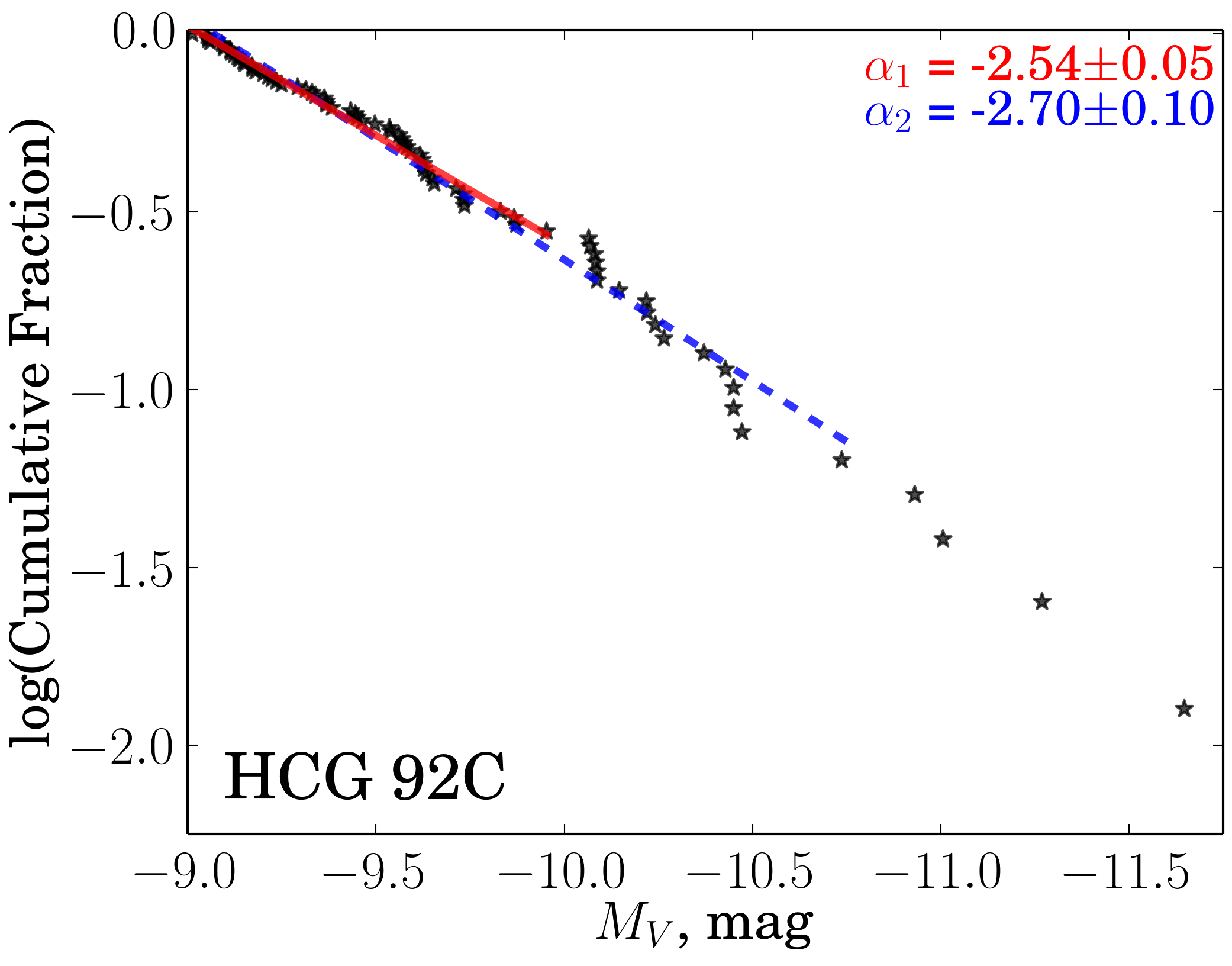}
\includegraphics[width=0.3\textwidth]{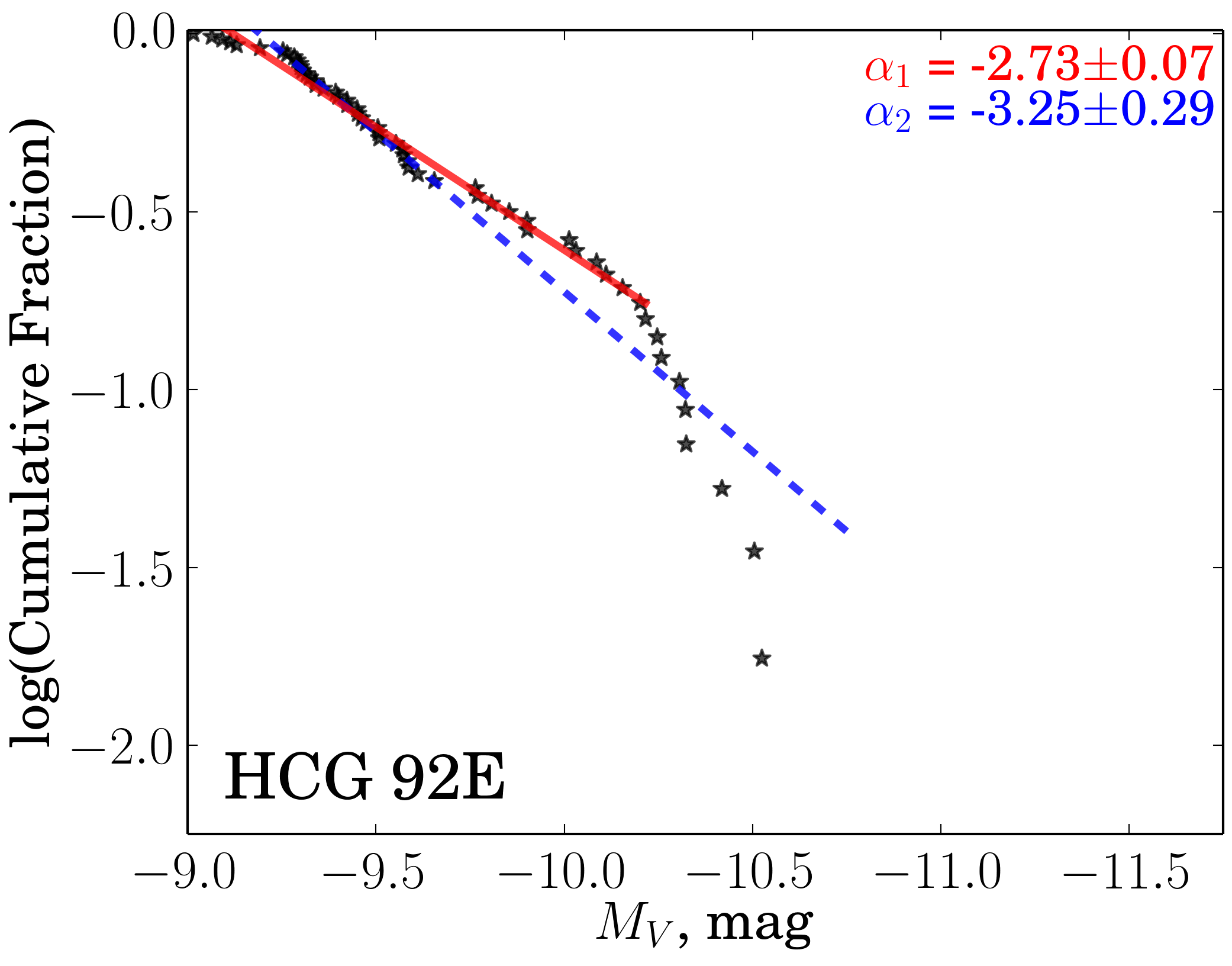}\\
\includegraphics[width=0.3\textwidth]{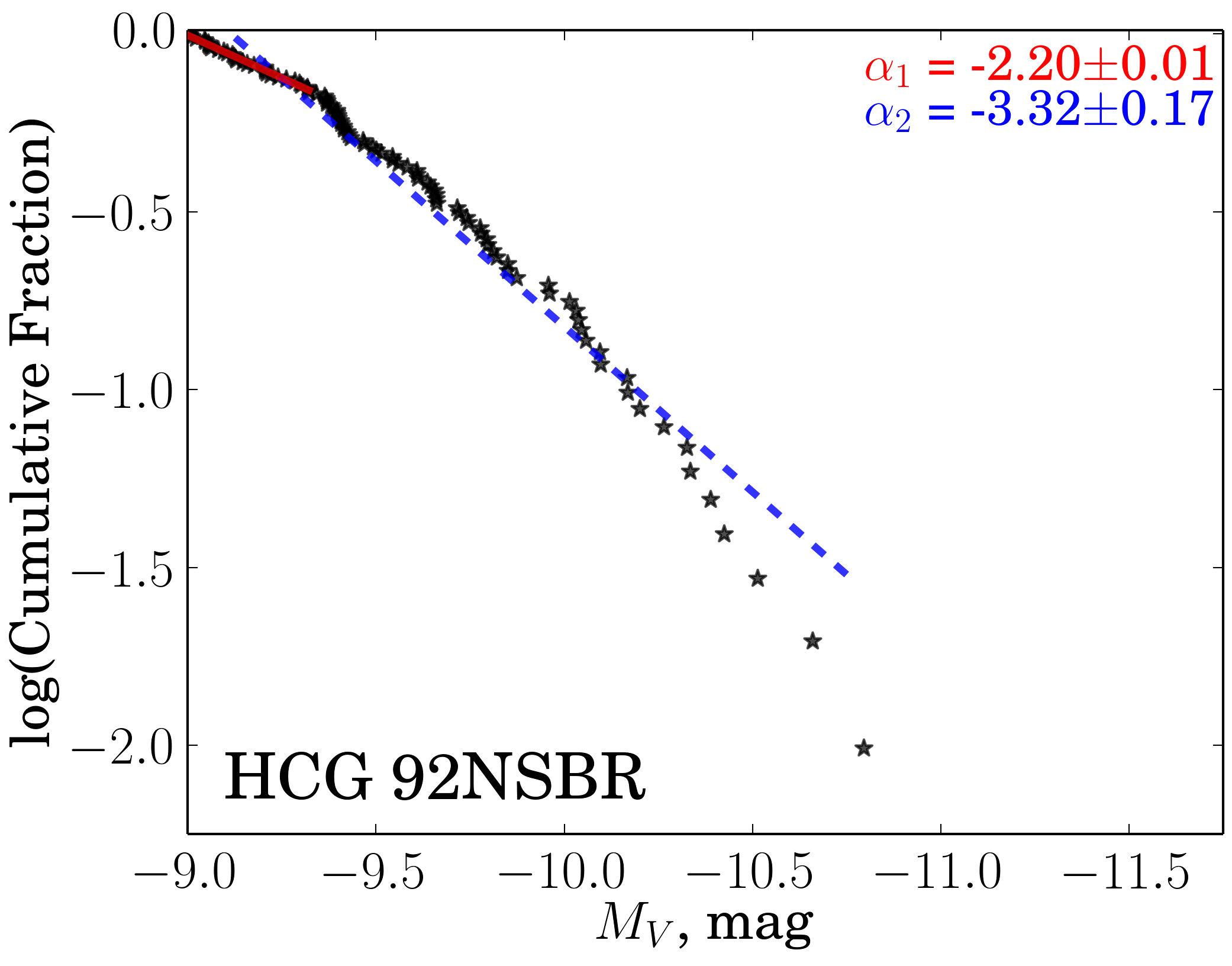}
\caption[Cumulative luminosity function for each galaxy in the sample]
{Cumulative luminosity functions of galaxies in our sample.  
For statistically significant results, only galaxies with more than 
40 SCCs were used.  The slopes were determined by a least squares 
fit over the range covered by the line.  The solid line represents 
the best-fitting slope over the range that was chosen manually, 
whereas the dashed line represents the best-fitting slope over 
a common range for all CLFs, from $-9$~mag to $-10.75$~mag.
The corresponding $\alpha$ values for each slope are displayed 
in the upper right corner of each panel. (A colour version of this 
figure is available in the online journal.)}
\label{fig:LF}
\end{center}
\end{figure*}

\begin{figure}
\centering
\includegraphics[width=1.0\columnwidth]{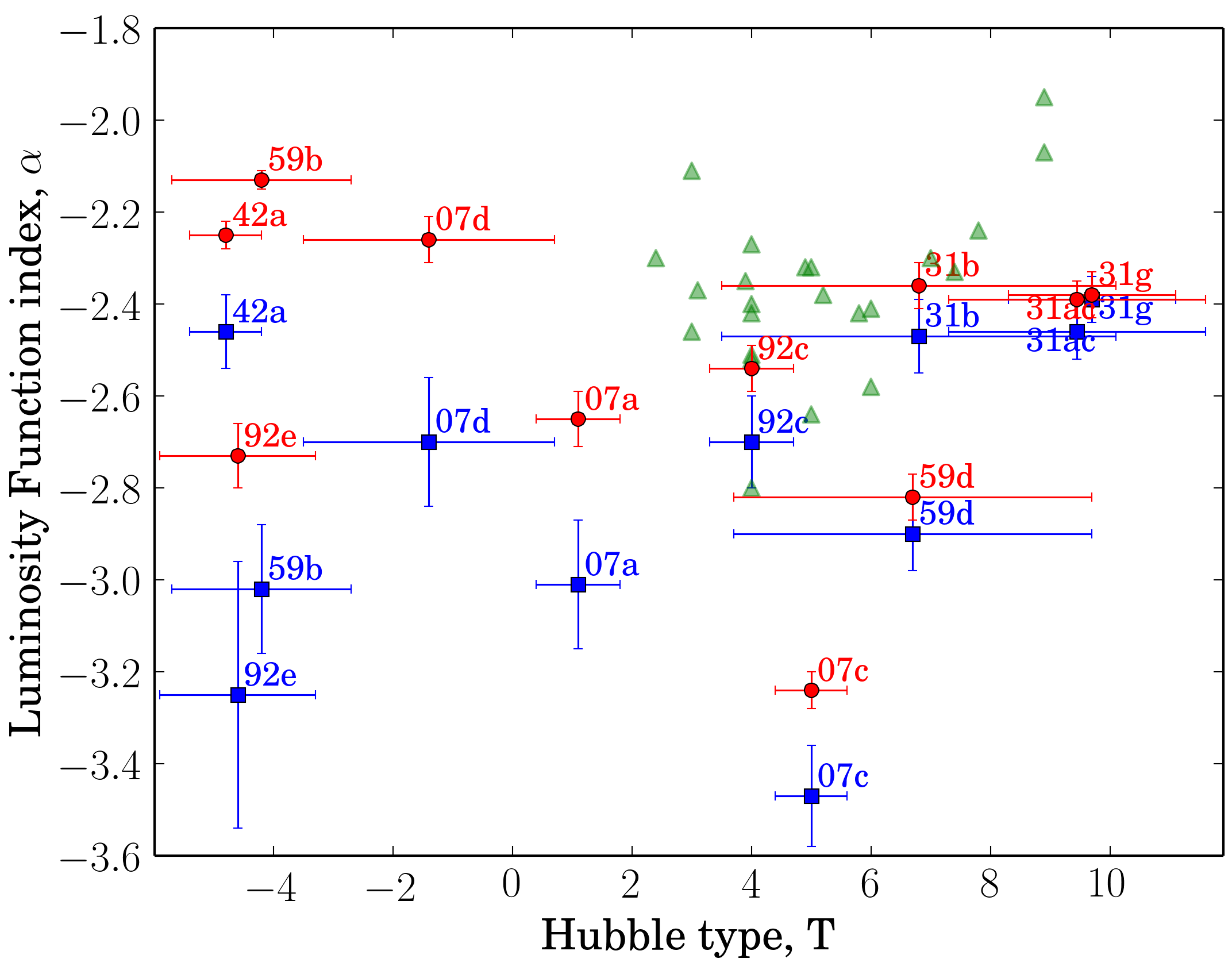}
\caption[Cluster luminosity function index $\alpha$ as a function of Hubble type]
{Plot of CLF index $\alpha$ as a function of Hubble type (T). 
Red circles represent indices based on fitting custom ranges of CLFs, 
and blue squares represent indices of common range fittings.  In all 
cases, extending the fitting range makes the indices more negative.  
Green triangles are the indices obtained from \citet{Whitmore2014}, 
who studied a sample of 20 nearby star-forming spiral galaxies and 
found a correlation between $\alpha$ and T, for T ranging from 2 to 9.  
Based on our sample, it appears that no significant correlation between 
$\alpha$ and T can be observed.  Out of 11 data points in this plot, 1 
point does not represent a galaxy itself but rather a pair of close 
interacting galaxies: 31AC -- a combination of spiral and irregular 
galaxies. Because their Hubble type is very close in value (for 31A 
${\rm T} = 8.9 \pm 0.9$, for 31C ${\rm T} = 10 \pm 2$), for the T value 
of 31AC we adopted the value of $9.45 \pm 2.15$. For another interacting 
pair, 92BD -- a close pair of a spiral and an elliptical, the 
morphological types are very different and the ``average'' value would 
not be meaningful.  For this reason, we do not include this data point 
in our plot. The morphological types are taken from HyperLeda. (A 
colour version of this figure is available in the online journal.)}
\label{fig:LF_v_Morph}
\end{figure}

Figure \ref{fig:LF} shows the cluster luminosity functions (CLFs) for
selected galaxies from our sample.  To obtain statistically significant 
results, only galaxies with more than 40 SCCs were used.  In each plot, 
the CLF is shown as a cumulative distribution function of the absolute 
magnitude, $M_{\rm V_{606}}$.  The solid line represents the best-
fitting slope of the distribution that was determined by the least 
squares fit over the range covered by the line.  This range was chosen 
manually, based on our assessment of the linear region of each CLF 
starting at the faint-end cutoff.  The dashed line represents the best-
fitting slope over a common range for all CLFs, from $-9$~mag to 
$-10.75$~mag.  The slope is for a power-law distribution index $\alpha$ 
from $NdL \propto L^{-\alpha}dL$ as $2.5 \times \text{slope} + 1$. The 
overall range of the indices (from here on we are using slopes fitted to 
the custom ranges) for CG galaxies in our sample is from $-2.13$ to 
$-3.24$ (see Table~\ref{tab:CLF} for all the numbers).  
Figure~\ref{fig:LF_v_Morph} represents a plot of the LF index as a 
function of Hubble T-type.  For comparison, on the same figure, we 
overplot the data for 20 nearby star-forming spiral galaxies, which span 
Hubble T-types from 2 to 9, obtained from \citet{Whitmore2014}.  As can 
be seen, our results do not reproduce the shape of the LF index 
distribution of \citet{Whitmore2014}.  Moreover, it appears that no 
significant correlation between $\alpha$ and T can be observed. 


Overall, the CLF indices of spirals in our sample of galaxies are in 
reasonable agreement with the values reported in other works 
\citep[e.g.,][]{Larsen2002, Whitmore2014, Ryon2014}, except for a 
noticeable outlier 07C, which is a bit more negative than in 
\citet{Gieles2006}.  In addition, the galaxies in HCG~31 have very 
similar $\alpha$-values ($-2.39 \pm 0.04$, $-2.36 \pm 0.05$, and $-2.38 
\pm 0.05$, for galaxies 31AC, 31B, and 31G, respectively); the 
irregular 59D has a rather high negative $\alpha$-value of $-2.82 \pm 
0.05$.  As in the case of 07A and 07C, recent (and on-going) star 
formation could be responsible for the steeper value of $\alpha$ in 59D.  
A sustained star-formation episode could cause a build-up of clusters 
near the low-luminosity end of the luminosity function as old clusters 
fade with age.  

Some CLFs for the galaxies in our sample exhibit a bend at the bright 
part of the distribution, with that part of the distribution being 
steeper, a trend that was also  noticed in the aforementioned studies.  
\citet{Gieles2006} argues that the bend in the CLF corresponds to the 
upper mass limit in the cluster initial mass function.  For example, 
their linear fit to the CLF of the Antennae galaxies suggests a bend 
at $M_V = -10.3$, which corresponds to a maximum mass of $\sim 2.5 
\times 10^6$~\msun (with the assumption that the oldest cluster in the 
CLF is 3 Gyr).  In our sample, the galaxies 42A, 92BD, 92E display a 
bend at magnitudes of approximately $-10.2, -10.4$, and $-10.3$, 
respectively.  Galaxy 59B exhibits two possible bends, one at $-9.5$ and
the other one at $\sim -10.3$.  Because there is the possibility that
the GC system for this galaxy consists of two populations, its own GCs
and clusters stripped from 59A (see \S~\ref{sec:HCG59}), we speculate
that each bend is imprinted on the overall CLF by the two constituent
cluster populations. HCG~31B and HCG~92C might also have multiple bends.

Also worth noting is that the galaxies with active star formation 
according to the cluster colour-colour plots (in particular galaxies C 
and D in HCG 07; AC, B, and G in HCG 31, and irregular galaxy D in HCG 
59) show the most linear CLFs over a broad range of absolute magnitudes.

\begin{table*}
\caption{CLF indicies for galaxies in our sample that have over 40 SCCs.}
\label{tab:CLF}
\begin{minipage}{\textwidth}
\begin{center}
\begin{tabu}{@{} CCCCCCC @{}}
\toprule
Galax/Region  & $N_{\text{SCC}}$  & Type & \multicolumn{2}{C}{$\alpha$-index} & \multicolumn{2}{C}{Magnitude range} \\
 & & & $\alpha_1$ & $\alpha_2$ & custom & common\\
\midrule
\gr HCG 07A & 48 & Sb & $-2.65 \pm 0.06$ & $-3.01 \pm 0.14$ &  $-9.00 \ldots -9.81$ & $-9.00 \ldots -10.75$ \\
HCG 07C & 135 & SBc & $-3.24 \pm 0.04$ & $-3.47 \pm 0.11$ & $-9.00 \ldots -9.96$ & \\
\gr HCG 07D & 47 & SBc & $-2.26 \pm 0.05$ & $-2.70 \pm 0.14$ & $-9.00 \ldots -9.91$ & \\
HCG 31AC & 138 & Sdm + Im & $-2.39 \pm 0.04$ & $-2.46 \pm 0.06$ & $-9.00 \ldots -10.28$ & \\
\gr HCG 31B & 89 & Sm & $-2.36 \pm 0.05$ & $-2.47 \pm 0.08$ & $-9.00 \ldots -10.28$ & \\
HCG 31G & 76 & Sbc & $-2.38 \pm 0.05$ & $-2.39 \pm 0.05$ & $-9.00 \ldots -10.60$ & \\
\gr HCG 42A & 246 & E3 & $-2.31 \pm 0.04$ & $-2.46 \pm 0.08$ & $-9.00 \ldots -10.20$ & \\
HCG 59B & 73 & E0 & $-2.13 \pm 0.02$ & $-3.02 \pm 0.14$ & $-9.00 \ldots -9.49$ & \\
\gr HCG 59D & 75 & Im & $-2.82 \pm 0.05$ & $-2.90 \pm 0.08$ & $-9.00 \ldots -10.39$ & \\
HCG 92BD & 124 & Sbc + E2 & $-2.38 \pm 0.05$ & $-2.58 \pm 0.13$ & $-9.00 \ldots -10.14$ & \\
\gr HCG 92C & 79 & SBc & $-2.54 \pm 0.05$ & $-2.70 \pm 0.10$ & $-9.00 \ldots -10.24$ & \\
HCG 92E & 57 & E4 & $-2.73 \pm 0.07$ & $-3.25 \pm 0.29$ & $-9.00 \ldots -9.91$ & \\
\gr HCG 92NSBR & 102 & -- & $-2.20 \pm 0.01$ & $-3.32 \pm 0.17$ & $-9.00 \ldots -9.34$ & \\
\bottomrule
\end{tabu}\vspace{0.1cm}
\end{center}
\textbf{Notes.} Types of galaxies are taken from \citet{Hickson1989}.  
Region 92NSBR represent a collection of intergroup clusters and as such 
does not have a morphological type.  Magnitude range column specifies 
the range over which the slope was fitted.
\end{minipage}
\end{table*}

\subsection{Spatial Distribution of Globular Clusters in Elliptical Galaxies}

For the three elliptical galaxies in our sample with significant GC 
systems (HCG 42A, 59B, and 92E), we examined the physical distribution of 
the globular clusters in those galaxies as a function of metallicity 
(Fig.~\ref{fig:MD}).  We divided the clusters in each system into three 
groups, based on their metallicity distribution plots (e.g., panel (e) 
of \ref{fig:GC42}).  Specifically, all cluster to the left of the 
``blue'' peak are considered to be relatively metal-poor.  Similarly, 
all clusters rightward of the ``red'' peak are considered relatively 
metal-rich.  The clusters between the peaks are tagged as having an 
intermediate metallicity content.  Thus, the metal-poor cluster have 
metallicities below $-1.04$, $-1.05$, and $-0.93$ for galaxies HCG 42A, 
59B, and 92E, respectively. The metal-rich clusters have metallicities 
above $0.16$, $-0.43$, and $0.09$, for the same galaxies. Accordingly, 
the intermediate metallicity clusters in these galaxies have metallicities 
ranging between the values cited above. Then we plot the cumulative 
distribution of the clusters as a function of projected distance from 
the center of the galaxy.  We find that 42A and 59B have more 
metal-rich clusters concentrated closer to their galaxy centres, 
clusters with intermediate metal content are distributed throughout 
the galaxies, and metal-poor clusters tend to have higher concentrations 
in the outer regions of the galaxies.

A Kolmogorov-Smirnov (KS) statistical test was used to determine if
the cumulative radial distributions of the different metallicity
clusters are consistent with each other.  The results showed that 
the $p$-value for comparing the low and high metallicity distributions 
in 42A was $\sim10^{-4}$, which allows us to reject the null 
hypothesis (that they are drawn from the same parent population) 
with $>99.9\%$ confidence.  Similarly, $p=4\times10^{-4}$ for the rich 
and poor clusters of 59B, and so their radial distributions are also 
significantly different.  For galaxy 92E the $p$-values for all 
combinations of distributions are not rejecting the null hypothesis.  
From the lower panel of Fig.~\ref{fig:MD} it appears that the GCs 
of different metallicities are mixed throughout the galaxy.  
However, we have to note that the statistics of small numbers might 
be at play here.  The way we split the GC population into three groups 
with the different metallicity content leaves the the metal-rich 
population of 92E with only 7 GCs (20 and 26 GCs in metal poor and 
medium metal content populations, respectively).  Thus, it is highly 
speculative to talk about the radial distribution of the metal-rich 
clusters.  To try to avoid this situation, we split GC sample into 
two roughly equal groups, metal rich and metal poor, at $[\text{Fe/H}] 
= -0.63$ (panel (i), fig.~\ref{fig:GC92}).  The KS test does not 
reject the null hypothesis, and so it would appear that GCs do not 
have a preferential distribution based on their metallicities.  One of 
the possible explanations for this apparently well-mixed distribution 
would be a dry merger between two galaxies of similar mass.  An 
examination of the unsharp-masked image of 92E to look for signs of 
recent interaction such as shells or streams did not reveal any such 
features.

\section{Conclusions}

We present a catalogue of star clusters detected in five compact
galaxy groups (HCG 07, 31, 42, 59, and 92), based on sensitive,
high-resolution multi-colour images from the {\it Hubble Space 
Telescope} Advanced Camera for Surveys and Wide Field Camera 3 (in
the case of HCG~92) with the goal of examining the properties of the 
star cluster systems of compact group galaxies overall and further 
assisting researchers in star cluster-related studies. Altogether, 
the catalogue consists of 18,292 objects. After applying a number 
of criteria, we left with 1963 star cluster candidates and 1505 
globular cluster candidates detected in 16 galaxies in the high 
confidence samples. A sample of the photometric data from this 
catalogue is presented in the electronic Table~\ref{tab:catalog}. 

\thispagestyle{empty}
\begin{table*}
\tabcolsep=0.125cm
\caption{A sample table of the $BVI$ catalogue for star clusters in Hickson compact groups.  The full catalogue is available online.}
\label{tab:catalog}
\begin{minipage}{\textwidth}
\begin{center}
\resizebox{\columnwidth}{!}{%
\begin{tabu}{CCCCCCCCCCCCCCCCCCCCCCCCCCCCCCCC}
\toprule
\rowfont{\scriptsize} RA & Dec & \multicolumn{3}{C}{B} & \multicolumn{3}{C}{V} & \multicolumn{3}{C}{I} & $\rchi_{\text{I}}$ & S1 & S2 & S3 & S4 & S5 & G1 & G2 & G3 & G4 & SCC & GCC & [Fe/H] & HCG \\
\rowfont{\scriptsize} deg & deg & mag & err & sharp & mag & err & sharp & mag & err & sharp & & & & & & & & & & & & & &\\
\cmidrule{3-5} \cmidrule{6-8} \cmidrule{9-11} 
\rowfont{\scriptsize} (1) & (2) & (3) & (4) & (5) & (6) & (7) & (8) & (9) & (10) & (11) & (12) & (13) & (14) & (15) & (16) & (17) & (18) & (19) & (20) & (21) & (22) & (23) & (24)& (25)\\
\midrule
\rowfont{\scriptsize} \gr 338.9945 & 33.936701 & 28.921 & 0.502 & -0.122 & 27.844 & 0.203 & -0.67 & 26.716 & 0.107 & 0.143 & 1.35 & 0 & 0 & 1 & 1 & 0 & 1 & 1 & 0 & 1 & - & - & 0.1232 & 92 \\
\rowfont{\scriptsize}339.04261 & 33.936755 & 25.844 & 0.057 & 0.112 & 25.462 & 0.048 & 0.224 & 24.997 & 0.055 & 0.218 & 1.784 & 1 & 1 & 1 & 1 & 1 & 1 & 1 & 1 & 0 & yes & - & -3.441 & 92 \\
\rowfont{\scriptsize} \gr 339.02544 & 33.936792 & 26.918 & 0.1 & 0.111 & 26.642 & 0.123 & -0.615 & 26.429 & 0.183 & -0.408 & 2.811 & 0 & 1 & 1 & 1 & 1 & 1 & 0 & 1 & 0 & - & - & -4.399 & 92 \\
\rowfont{\scriptsize}338.96391 & 33.936928 & 27.076 & 0.13 & -0.348 & 25.919 & 0.062 & -0.079 & 25.111 & 0.054 & 0.075 & 1.606 & 0 & 1 & 1 & 1 & 1 & 1 & 1 & 1 & 1 & - & yes & -0.499 & 92 \\
\rowfont{\scriptsize} \gr 338.97734 & 33.937033 & 26.903 & 0.116 & 0.033 & 26.369 & 0.075 & -0.046 & 25.716 & 0.158 & 0.122 & 3.819 & 0 & 1 & 1 & 0 & 1 & 1 & 1 & 1 & 0 & - & - & -2.539 & 92 \\
\rowfont{\scriptsize}339.0115 & 33.937167 & 28.51 & 0.365 & 0.652 & 27.192 & 0.105 & -0.22 & 27.058 & 0.208 & -0.034 & 2.137 & 0 & 0 & 1 & 1 & 1 & 1 & 1 & 0 & 0 & - & - & -1.840 & 92 \\
\rowfont{\scriptsize} \gr 339.03373 & 33.937201 & 28.678 & 0.388 & -1.17 & 26.171 & 0.075 & 0.168 & 26.308 & 0.128 & 0.439 & 2.171 & 0 & 0 & 1 & 1 & 0 & 1 & 1 & 1 & 0 & - & - & 0.5495 & 92 \\
\rowfont{\scriptsize}338.95818 & 33.937251 & 27.208 & 0.144 & -0.517 & 26.064 & 0.066 & -0.137 & 25.33 & 0.074 & -0.072 & 2.078 & 0 & 1 & 1 & 1 & 1 & 1 & 1 & 1 & 1 & - & yes & -0.725 & 92 \\
\rowfont{\scriptsize} \gr 339.00239 & 33.93731 & 26.475 & 0.063 & 0.313 & 25.395 & 0.057 & 0.369 & 24.446 & 0.063 & 0.386 & 2.433 & 1 & 1 & 1 & 1 & 1 & 1 & 1 & 1 & 1 & yes & yes & -0.333 & 92 \\
\rowfont{\scriptsize}338.96115 & 33.93742 & 25.62 & 0.046 & 0.182 & 24.629 & 0.026 & -0.035 & 23.736 & 0.042 & 0.19 & 1.955 & 1 & 1 & 1 & 1 & 1 & 1 & 1 & 1 & 1 & yes & yes & -0.710 & 92 \\
\rowfont{\scriptsize} \gr 339.02347 & 33.93746 & 28.012 & 0.228 & -1.068 & 26.94 & 0.093 & 0.051 & 25.747 & 0.102 & 0.029 & 2.271 & 0 & 1 & 1 & 1 & 0 & 1 & 1 & 1 & 1 & - & yes & 0.2784 & 92 \\
\rowfont{\scriptsize}338.99922 & 33.937483 & 27.763 & 0.155 & 0.289 & 25.791 & 0.051 & 0.236 & 24.743 & 0.064 & 0.337 & 2.284 & 0 & 1 & 1 & 1 & 0 & 1 & 1 & 1 & 0 & - & - & 2.2147 & 92 \\
\multicolumn{25}{C}{\ldots}\\
\bottomrule
\end{tabu}\vspace{0.1cm}
}
\end{center}
\textbf{Notes.} Columns list: (1) Right Ascension (J2000); (2) 
Declination (J2000); (3)---(5) Magnitude, error in magnitude, and 
sharpness values for B-band (F435W; F438W for HCG 92); (6)---(8) 
Magnitude, error in magnitude, and sharpness values for V-band (F606W); 
(9)---(11) Magnitude, error in magnitude, and sharpness values for I-
band (F814W); (12) Goodness of fit factor $\chi$ from PSF-fitting in I-
band; (13)---(17) Star cluster candidates selection criteria, see 
Section~\ref{sec:SCC_selection} for full description. 1 means that a 
given criterion is satisfied; (18)---(20) Globular cluster candidate
selection criteria, see Section~\ref{sec:GCC_selection} for full 
description. 1 means that given criterion is satisfied; (21) Star 
cluster candidate flag. If `yes', all SCC criteria are satisfied; (22) 
Globular cluster candidate flag. If `yes', all GCC criteria are 
satisfied; (23) Metallicity value derived from $B-I$ color, see 
Section~\ref{sec:Discuss} for more details; (24) Hickson Compact group 
number.
\end{minipage}
\end{table*}

In particular, a detailed examination of our catalogue revealed 
the following:

\begin{itemize} 
\item Star clusters are powerful tracers of episodes
of star formation activity.  Careful study of the distribution of
cluster colours can lead to a better understanding of the evolutionary
state of their hosts and can help to constrain (and in some cases to
reconstruct) the sequence of events in the host groups
\citep[e.g.][]{Gallagher2010, Iraklis2010, Fedotov2011, Iraklis2012,
Iraklis2013}.  Thus, the analysis of star cluster populations in CGs 
allowed us to propose a reclassification of HCG 59 from a Type III to a 
Type II group.  Most galaxies in Type III groups appear to be `red' and 
`dead' (e.g., HCG 42).  However, the galaxy morphologies of HCG 59 do 
not comply with that statement.  Moreover, its population of star 
clusters is more consistent with Type II groups.

\item In general, the cluster luminosity functions of the CG spiral
galaxies were consistent with spirals studied in the literature
\citep[e.g.][]{Larsen2002, Whitmore2014}.  In particular, their CLF
$\alpha$-values ranged from $-2.26\pm0.05$ to $-2.54\pm0.05$.  A
notable exception were the large negative $\alpha$-values for the
spirals HCG~07A and 07C with $\alpha=-2.65 \pm 0.06$ and
$\alpha=-3.24 \pm 0.04$, respectively.

\item  We have examined the metallicity distributions of GCCs in the five 
groups overall and individually in the elliptical galaxies (nominally 
elliptical 07BD and 92BD, 42A, 59B, 92C, and 92E) with sufficient numbers 
of GCCs.  Only in galaxy 42A do we detect a metallicity distribution 
with well-defined bimodality peaking at [Fe/H]$=-1.04 \pm 0.07$ and 
[Fe/H]$=-0.16 \pm 0.15$.  The galaxy 92C may also host a bimodal 
distribution, but the statistical results were not conclusive.

\item The number of GCs in each galaxy is proportional to
the total stellar masses of galaxies (Fig.~\ref{fig:GCs_vs_M*}).
Notably, we detect a rather large number of GCs in 59B (and its immediate
environs) and a small number of GCs in 59A.  It is possible that these
two galaxies have interacted before given their morphologies (59B is
elliptical, 59A is lenticular) and apparent proximity; the GC
population distribution may be the only record of that interaction.  A
low surface brightness stream of material between galaxies 59A and B,
reported in \citet{Iraklis2013}, as well as their line-of-sight
velocities (within 3\% of eachother) also support the idea of previous
interactions.

\item  The number of ``blue'' clusters (with colour $V-I < 0.1$, 
representing the population of young star clusters) is well
correlated with the star formation rate.  Two spiral galaxies 
(92C and 59C) have a large number of ``blue'' clusters given 
their relatively lower star formation rates.  That likely indicates 
that a recent bout of star formation triggered by interactions 
is coming to an end. However, the young star clusters have not 
been significantly age-dimmed yet, and so a large number of 
them are still detectable.

\item For the three elliptical galaxies in our sample (42A, 59B, and
92E) we looked at the radial distribution of GCs of different
metallicities.  According to the KS-test, metal rich and metal poor
populations of 42A and 59B are drawn from different distributions
(with confidence of $>99\%$).  The GCs of different metallicities in
92E appear to be well mixed throughout the galaxy.  Dry mergers of
galaxies with a similar mass could explain this last observation.
However, the characteristic features for such a merger (such as shells
and streams) are not detected.

\end{itemize}

The star cluster populations of Hickson Compact Group galaxies are
particularly interesting given their potential to reveal the history
of dynamical interactions that have clearly been so important in
shaping the evolution of individual groups as a whole and the current
state of their member galaxies.  The sensitive $BVI$ imaging and
photometry presented in this catalogue illustrate this point.
However, the advantage of adding $U$-band to photometric studies of
star clusters is considerable. It breaks the age-extinction degeneracy
of the $BVI$ photometry and allows a shift from a qualitative
description of cluster ages to a more quantitative analysis, along
with determinations of intrinsic reddening and masses.  This
additional information adds considerably to the descriptive power of
star cluster populations in specific environments.

Our future work on star cluster populations of Stephan's Quintet 
(HCG 92; Fedotov et al., in prep.) with $UBVI$ photometry will 
illustrate that.

\section*{Acknowledgements}
K.F. and S.C.G. thank the Natural Sciences and Engineering Research 
Council of Canada and the Ontario Early Researcher Award Program for 
support.  This research has made use of the NASA/IPAC Extra- galactic 
Database (NED) which is operated by the Jet Propulsion Laboratory, 
California Institute of Technology, under contract with the National 
Aeronautics and Space Administration.  Additional support for this work 
was provided by NASA through grant No. HST-GO-10787.15-A from the Space 
Telescope Science Institute which is operated by AURA, Inc., under 
NASA contract NAS 5-26555.  We thank Alan McConnachie, John Blakeslee, 
Pat C\^ote, and Ruben Sanchez-Janssen for illuminating discussions on 
the properties of GCs and Peter Stetson for sharing his insight in 
aperture and PSF photometries. We also thank Theodoros Bitsakis, the 
referee for this paper, for the supportive encouragement of our work and for 
suggestions that improved the manuscript.  We thank Robert Corless, 
Gretchen Harris, John Landstreet, and Aaron Sigut for a careful read 
of the manuscript and their valuable suggestions.

\textit{Facilities: HST}
\clearpage


\begin{figure*}
\centering
\subfloat[]{\includegraphics[width=0.95\textwidth]{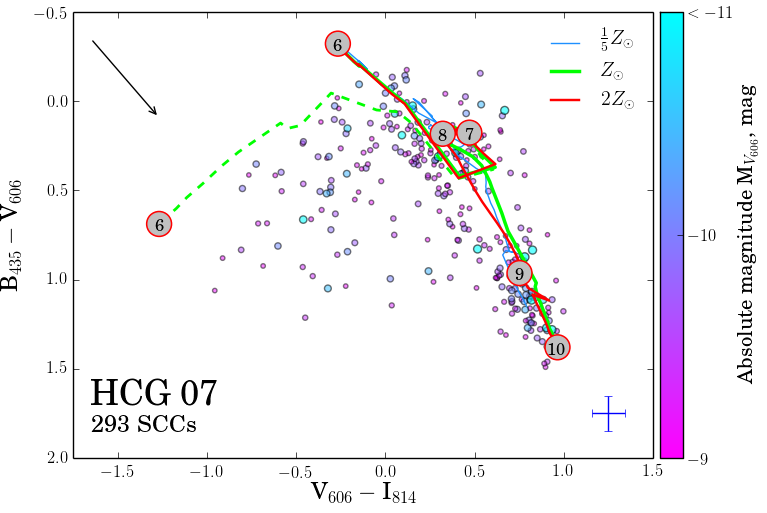}}
\caption[Colour-colour plot for all SCCs detected in HCG 07, and for SCCs 
in each galaxy. SCCs spatial distribution and system extent for each galaxy 
in HCG 07]
{Colour-colour plots of all the star cluster candidates in 
HCG 07 (a), including clusters located in the intra-group medium. The 
thin solid line, solid line, and dashed line trace the evolution of SSP 
models of [0.2, 1.0, 2.0] Z$\odot$ \citep{Marigo2008}. The thin dashed 
line to the left of the main evolutionary track represents a track 
that incorporates a model of nebular emission \citep[Starburst99;][]
{Leitherer1999}, common for young star clusters \citep[e.g.][]
{Vacca1992, Conti1996}.  The numbers on the track denote age represented 
in $\log(\text{age}/\text{yr})$.  In the upper left corner one can find a 
reddening vector with length equivalent to V$_{606}=1$ mag. The number 
of star cluster candidates detected in this group is marked in the lower 
left corner. A typical photometric error bar, based on the median errors, 
is located in the 
lower right corner. A colour bar that represents the absolute magnitude 
of SCCs is given on the right. For ease of reading the plot, the 
sizes of SCCs in the plot are linearly proportional to their magnitude: 
the larger the dot, the brighter the SC. All SCCs on this and following 
plots have absolute magnitude $\leq -9$ mag.  This figure is continued 
on the next page. (A colour version of this figure is available in the 
online journal.)}
\label{fig:CC07}
\end{figure*}
\clearpage

\begin{figure*}
\ContinuedFloat
\centering
\subfloat[]{\includegraphics[width=0.49\textwidth]{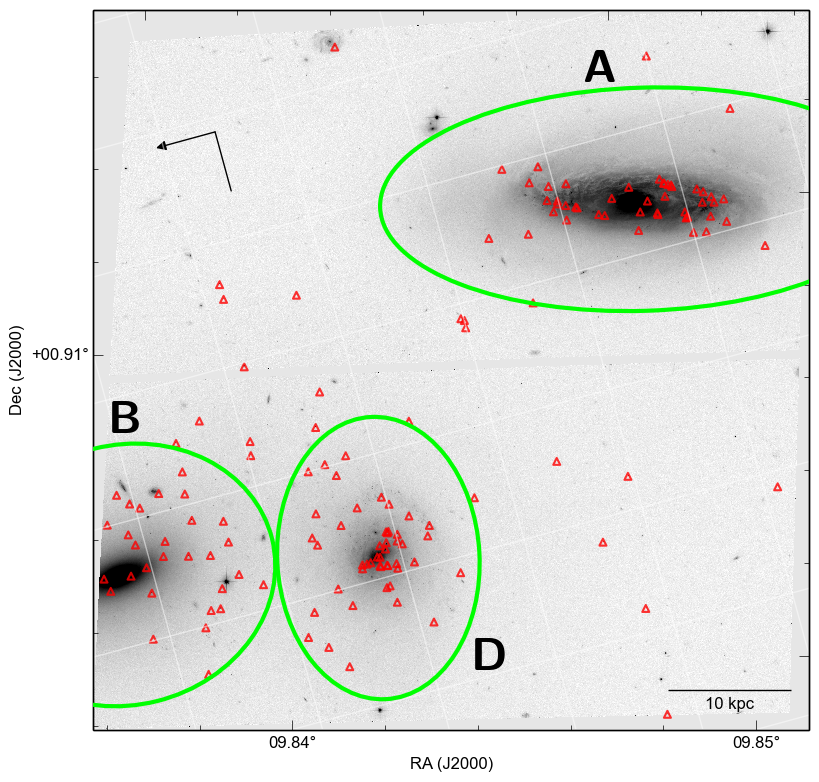}}
\hspace{0.075cm}
\subfloat[]{\includegraphics[width=0.49\textwidth]{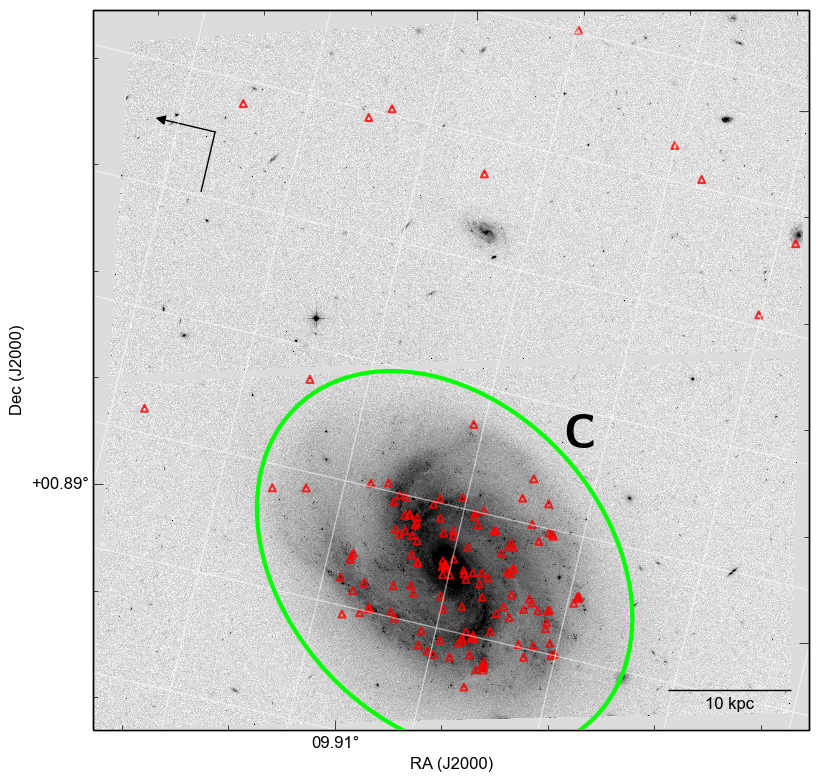}}\\
\subfloat[]{\includegraphics[width = 2.9in]{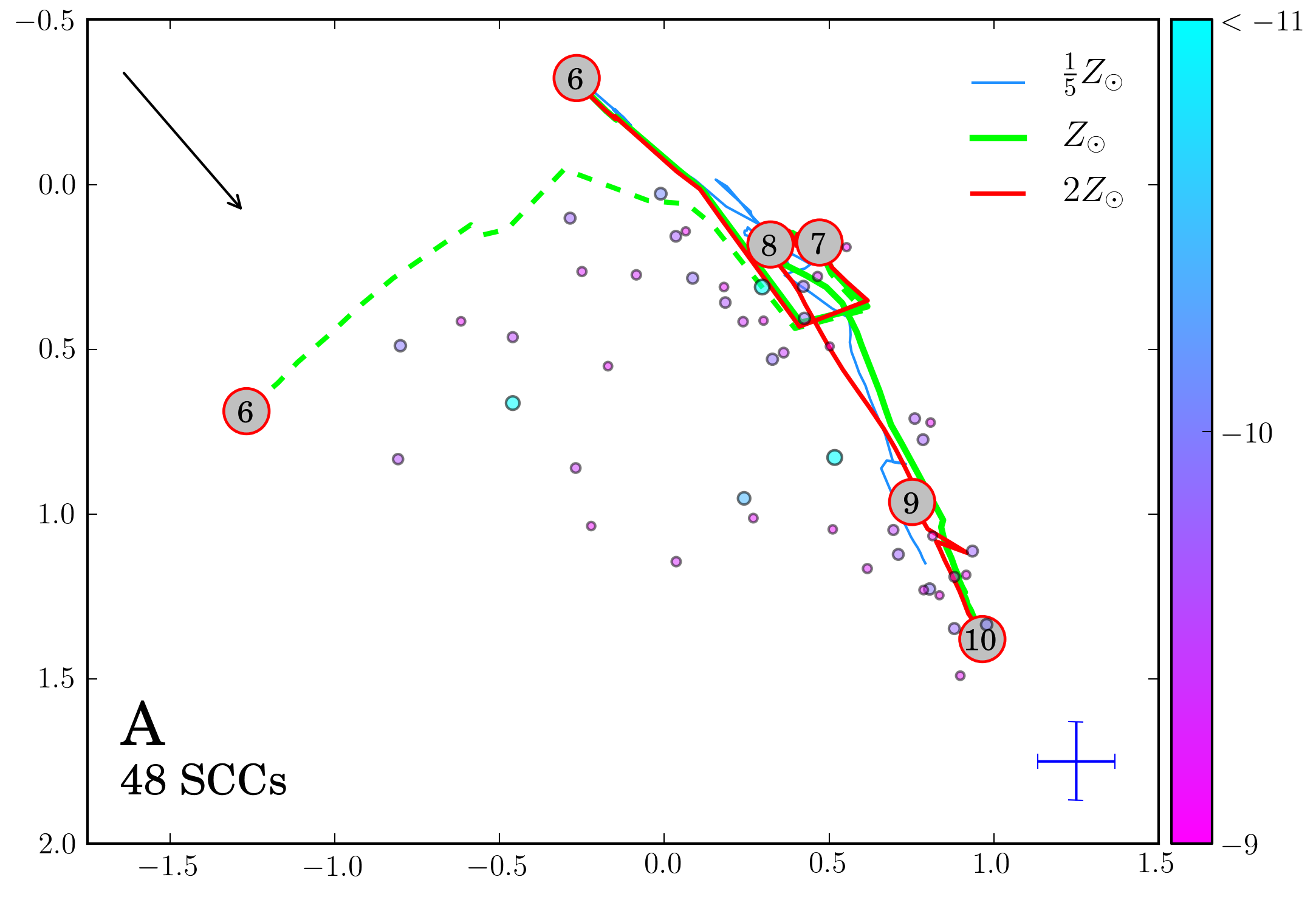}}\hspace{0.4cm}
\subfloat[]{\includegraphics[width = 2.9in]{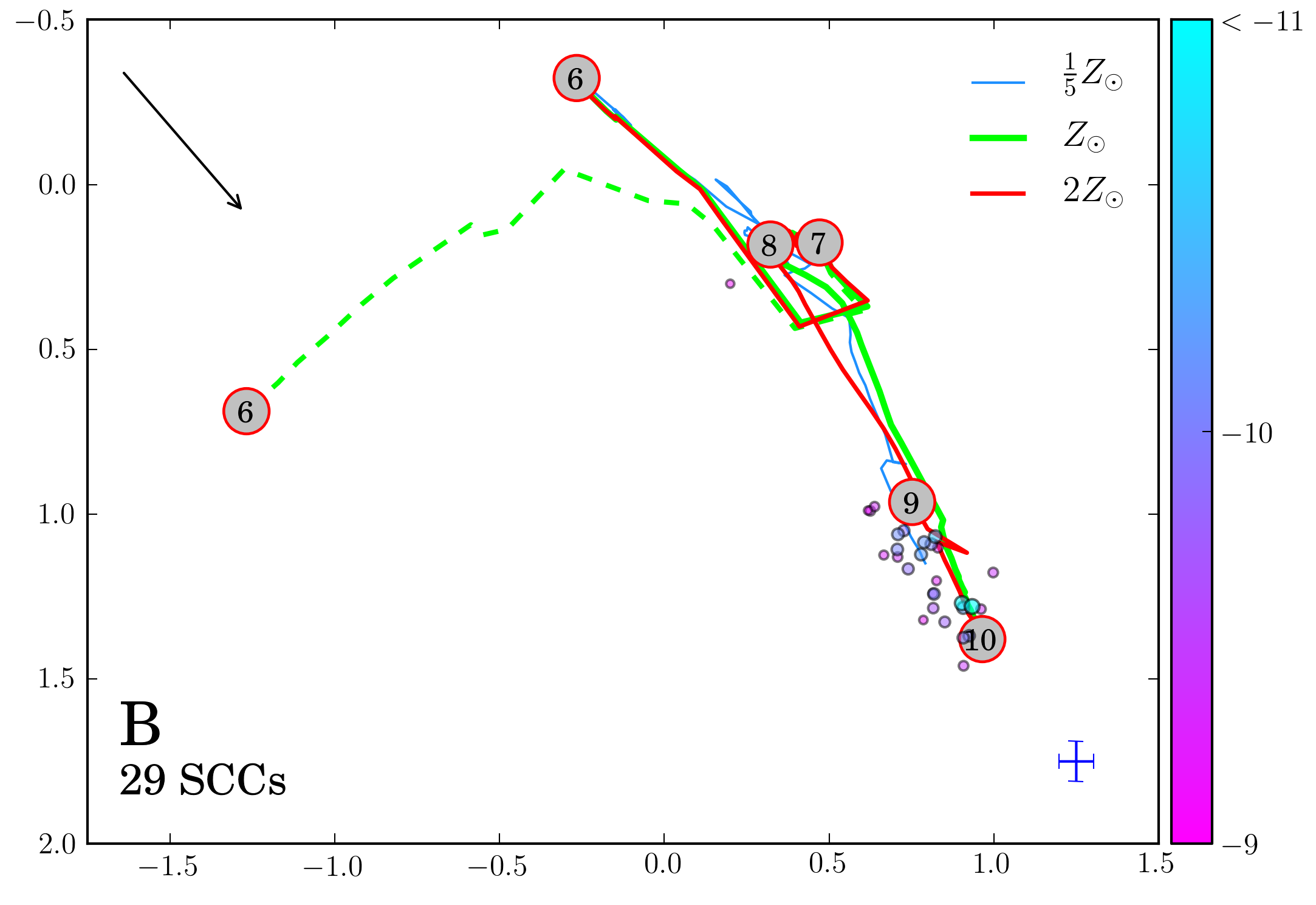}}\\
\subfloat[]{\includegraphics[width = 2.9in]{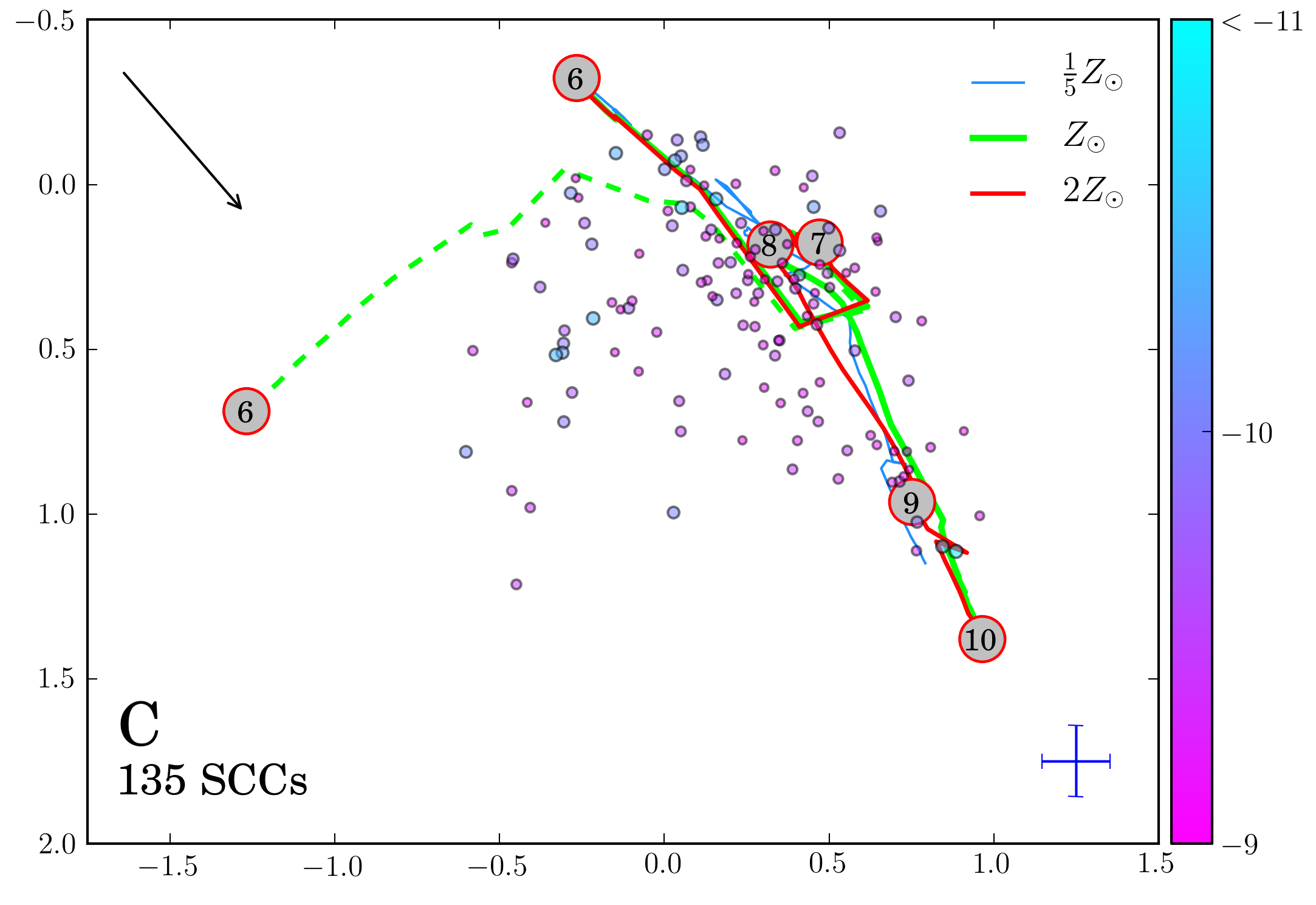}}\hspace{0.4cm}
\subfloat[]{\includegraphics[width = 2.9in]{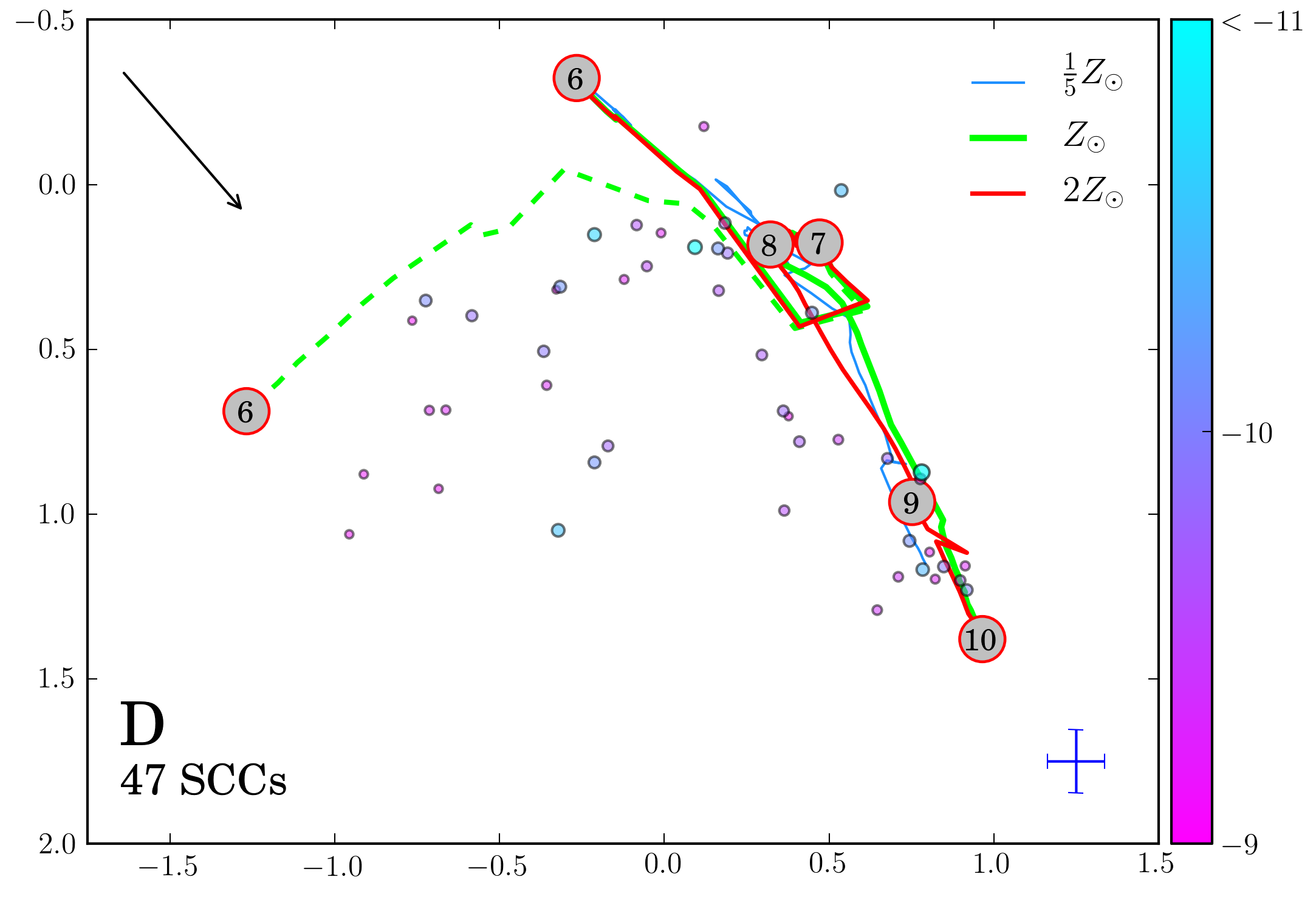}} 
\caption[continuation of the previous plot]
{Figure continued from the previous page.  Panels (b) and 
(c) are inverted $V_{606}$ images which show the star cluster system 
extent of each galaxy as defined by a brightness contour in $V_{606}$ 
of $\sim 1.25 \sigma$ above the background level.  Here and in 
subsequent plots, a compass indicates North (with an arrowhead) and 
East (without the arrowhead).  Panels (d)--(g) are $BVI$ 
colour-colour plots for individual galaxies and regions in the 
group. The large spiral HCG~7C hosts the greatest number of young 
clusters, while the quiescent elliptical HCG~7B has only globular 
clusters. (A colour version of this figure is available in the 
online journal.)}
\end{figure*}
\clearpage

\begin{figure*}
\centering
\subfloat[]{\includegraphics[width=0.95\textwidth]{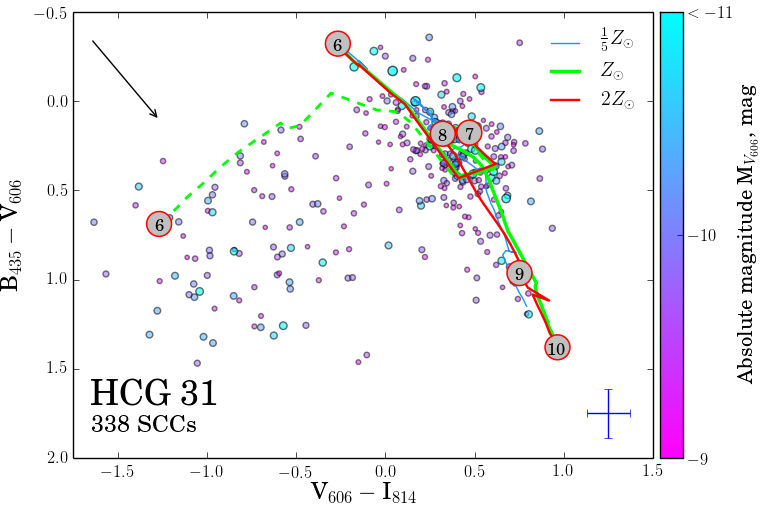}}\\
\caption[Colour-colour plot for all SCCs detected in HCG 31, and for SCCs 
in each galaxy. SCCs spatial distribution and system extent for each galaxy 
in HCG 31]{A colour-colour plot of all the star cluster candidates 
in HCG 31 (a), including clusters located in intra-group medium. 
Symbols are as in Fig.~\ref{fig:CC07}. (A colour version of this 
figure is available in the online journal.)}
\label{fig:CC31}
\end{figure*}
\clearpage

\begin{figure*}
\ContinuedFloat
\centering
\subfloat[]{\includegraphics[width=0.49\textwidth]{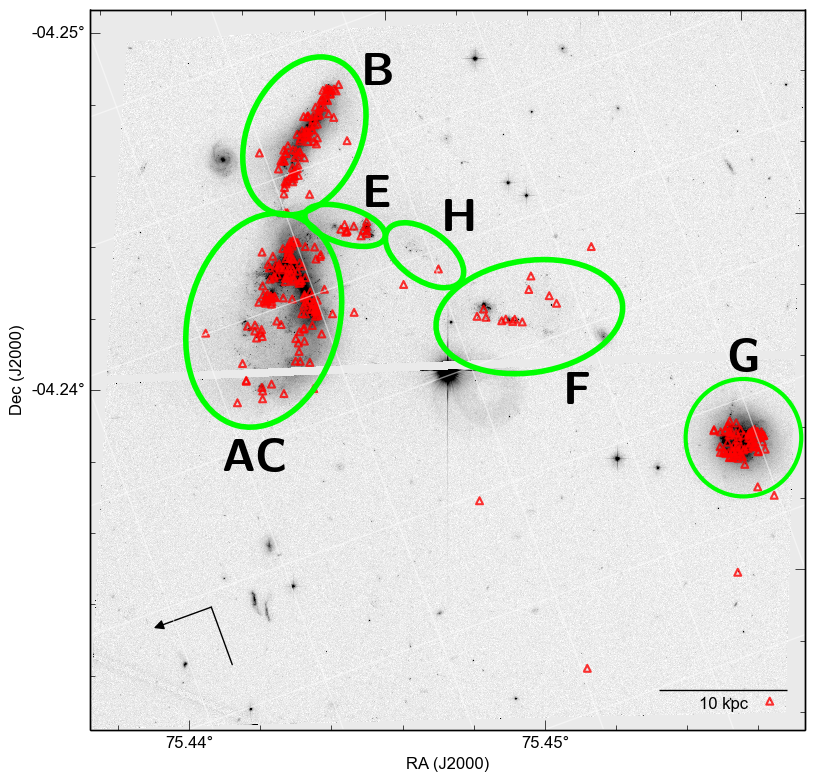}}\\
\subfloat[]{\includegraphics[width = 2.9in]{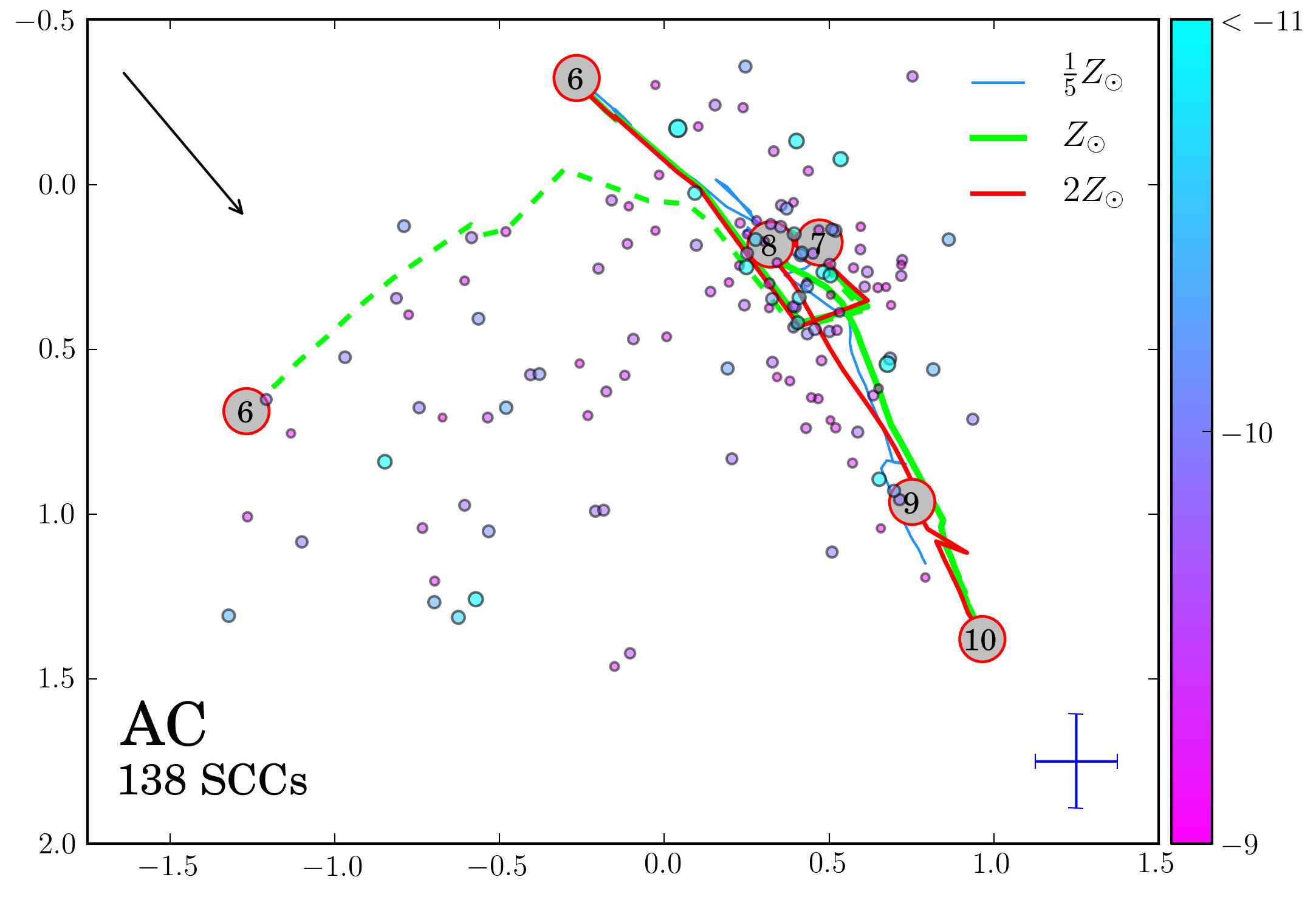}}\hspace{0.4cm}
\subfloat[]{\includegraphics[width = 2.9in]{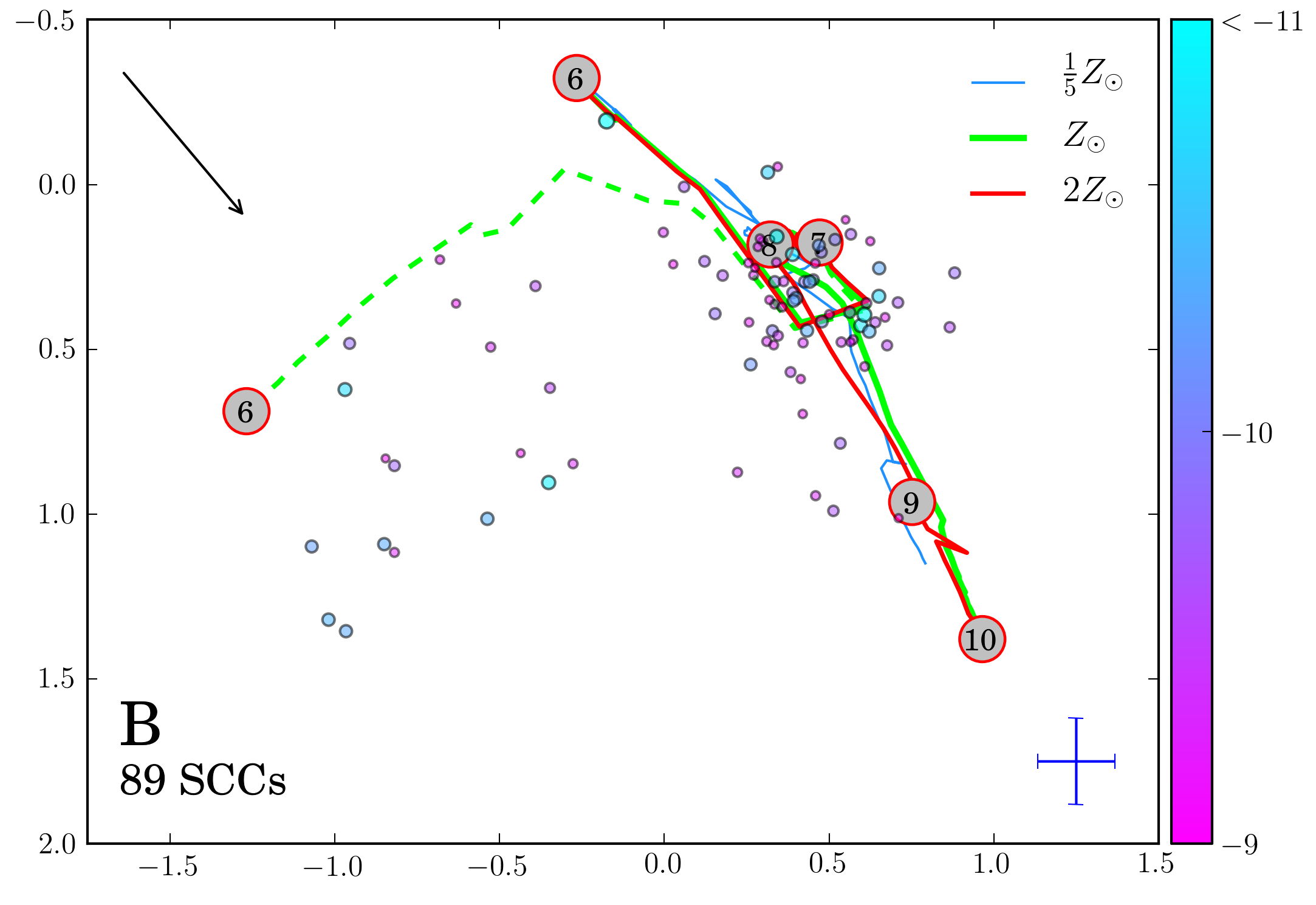}}\\
\subfloat[]{\includegraphics[width = 2.9in]{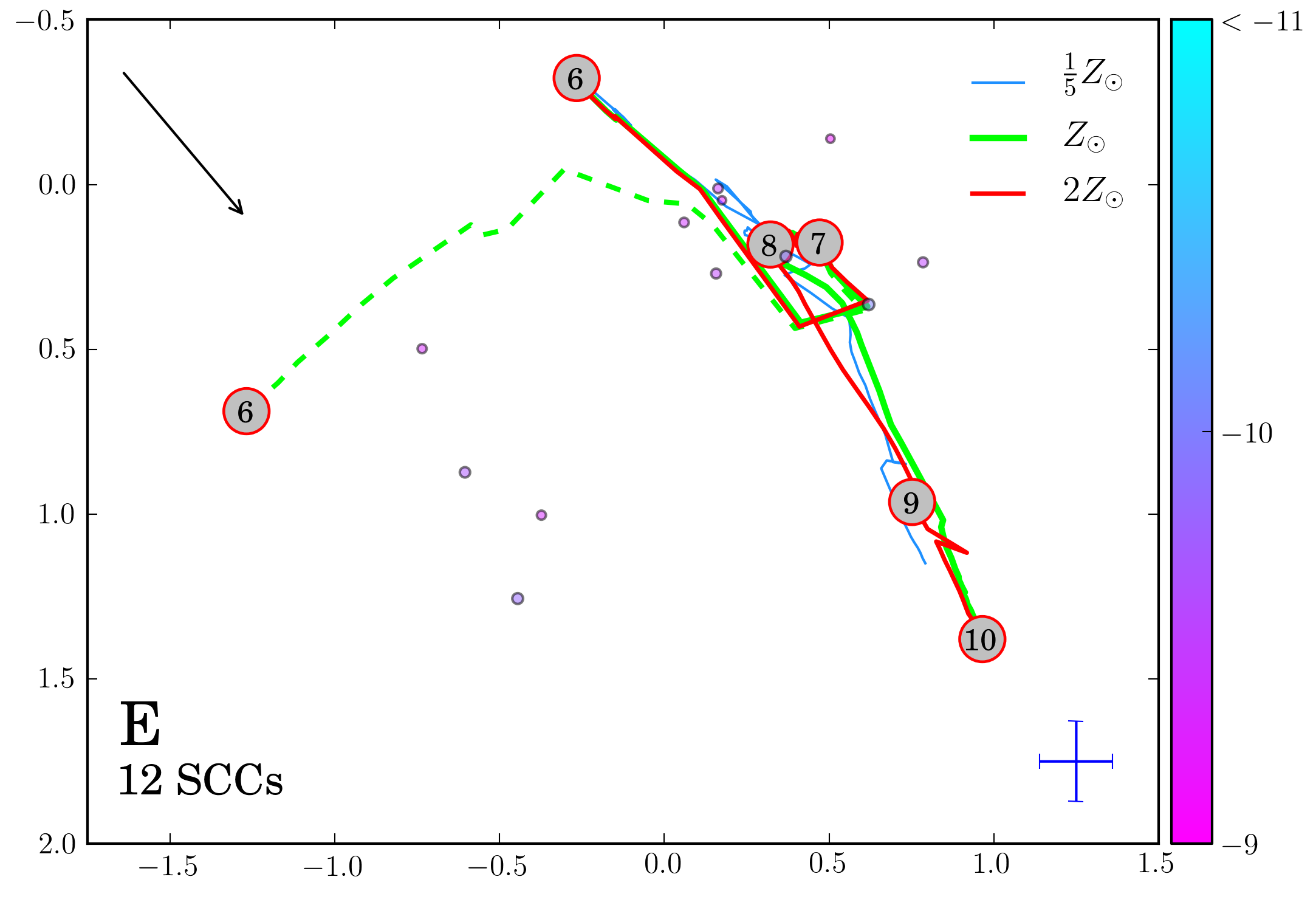}}\hspace{0.4cm}
\subfloat[]{\includegraphics[width = 2.9in]{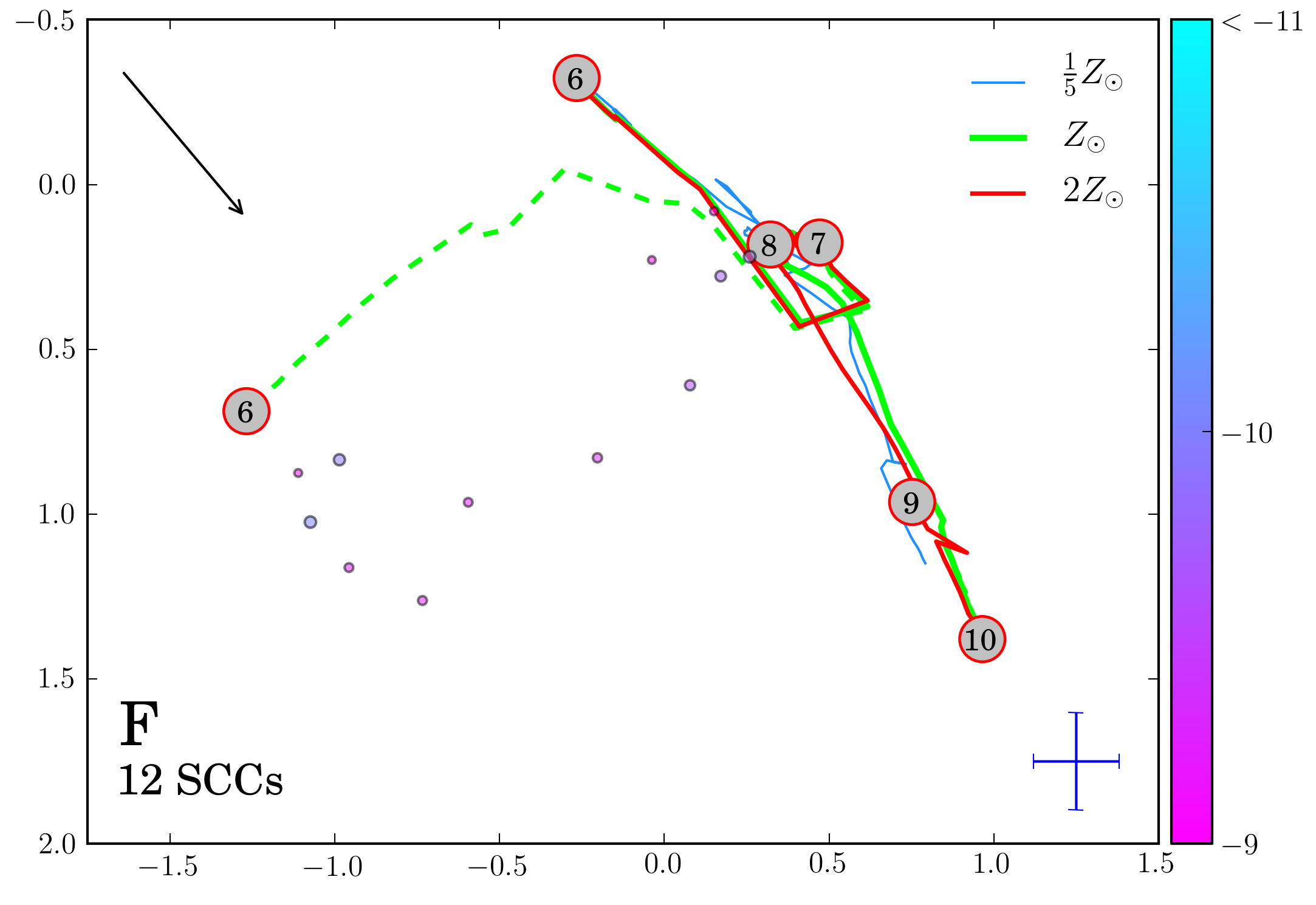}}\\
\caption[continuation of the previous plot]
{Figure continued from the previous page.  Image (b) is the 
inverted $V_{606}$ image which shows the SCC system extent as defined 
by a brightness contour of $\sim 1.25 \sigma$ in $V_{606}$ above the 
background level.  Panels (c)--(f) are colour-colour plots for 
particular galaxies and regions in the HCG 31 group.   All galaxies 
(31AC, B, and G) and tidal regions (31E and F) in this group host 
young clusters; the entire system is suffused with star formation 
triggered by strong, recent galaxy interactions. (A colour version 
of this figure is available in the online journal.)}
\end{figure*}
\clearpage

\begin{figure*}
\centering
\ContinuedFloat
\subfloat[]{\includegraphics[width = 2.9in]{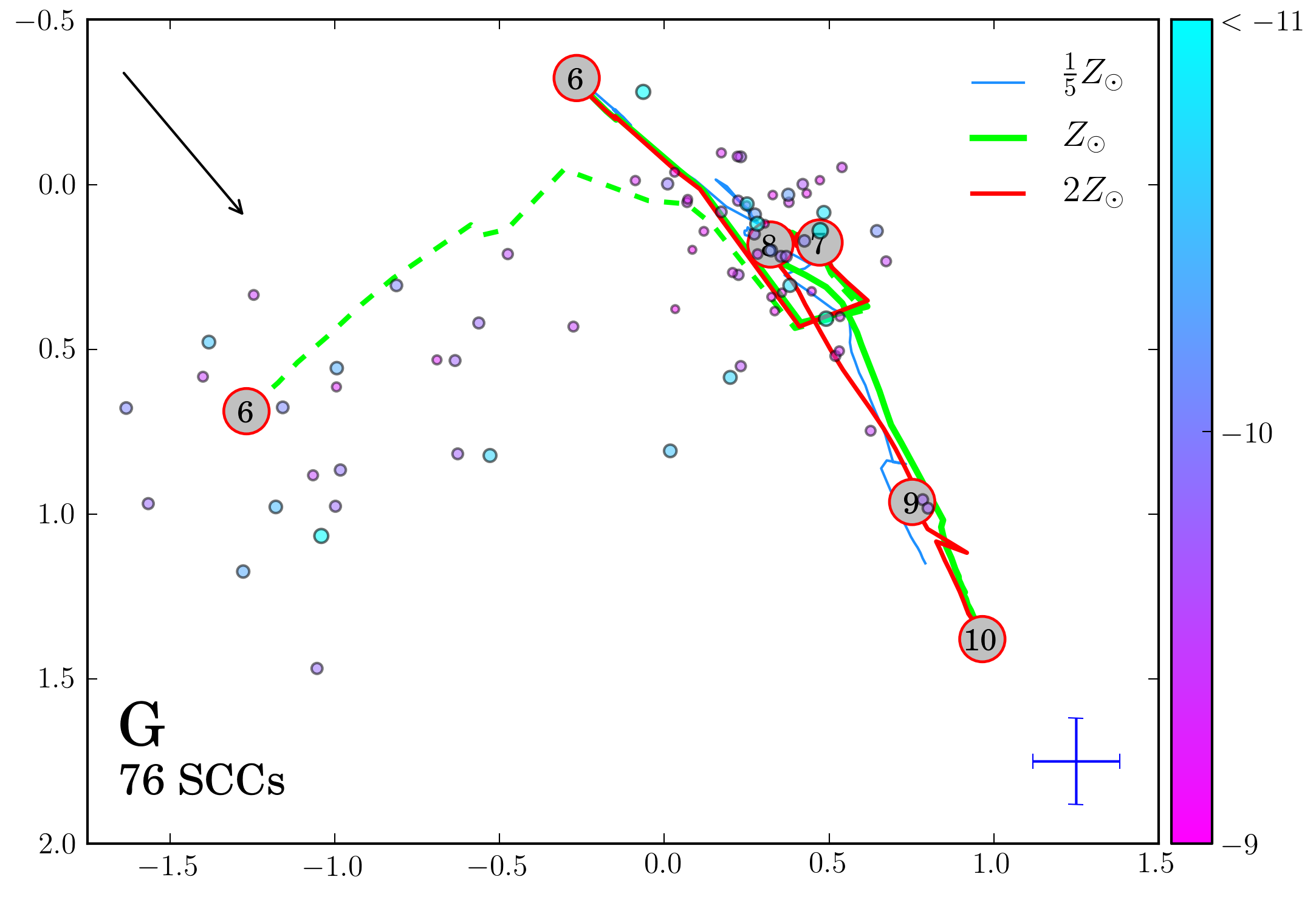}}
\caption[continuation of the previous plot]{$\ldots$ Figure continued from the previous page. A 
colour-colour plot for the galaxy G. (A colour version of this 
figure is available in the online journal.)}
\end{figure*}
\clearpage

\begin{figure*}
\centering
\subfloat[]{\includegraphics[width=0.95\textwidth]{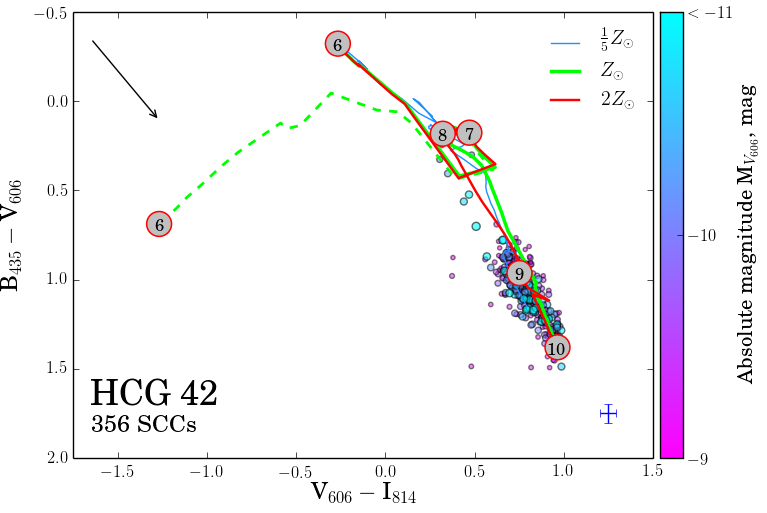}}
\caption[Colour-colour plot for all SCCs detected in HCG 42, and for SCCs 
in HCG 42A. SCCs spatial distribution and system extent for HCG 42A]
{A colour-colour plot of all star clusters candidates in 
the ACS image of HCG 42 (a), including clusters in intergroup medium, 
and subplot for a particular galaxy in that group (b), continued on 
the next page. For more details see caption for Fig.~\ref{fig:CC07}. 
(A colour version of this figure is available in the online journal.)}
\label{fig:CC42}
\end{figure*}
\clearpage

\begin{figure*}
\ContinuedFloat
\centering
\subfloat[]{\includegraphics[width=0.49\textwidth]{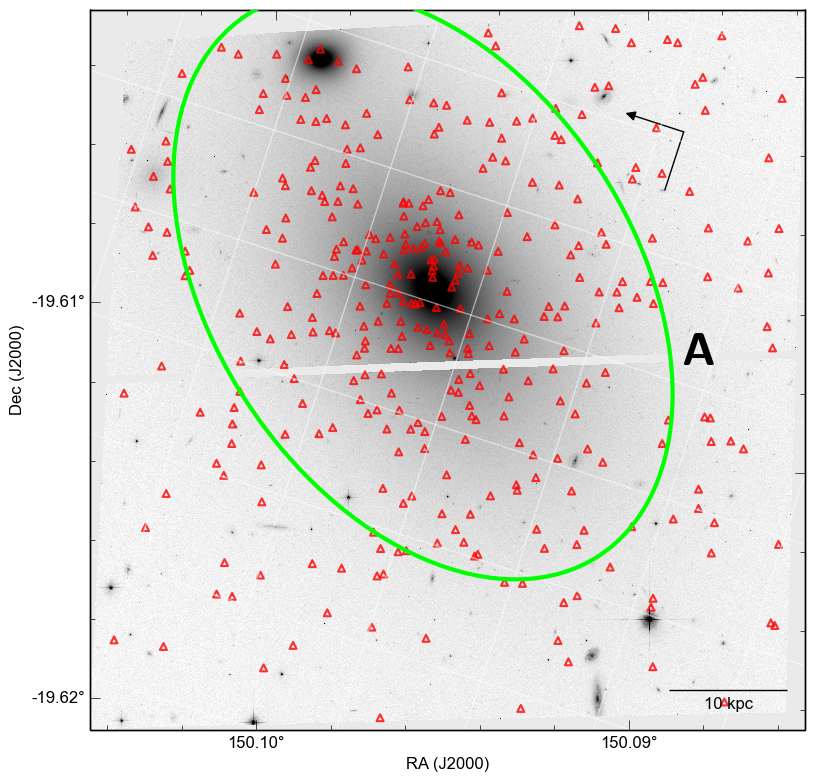}}\\
\subfloat[]{\includegraphics[width = 2.9in]{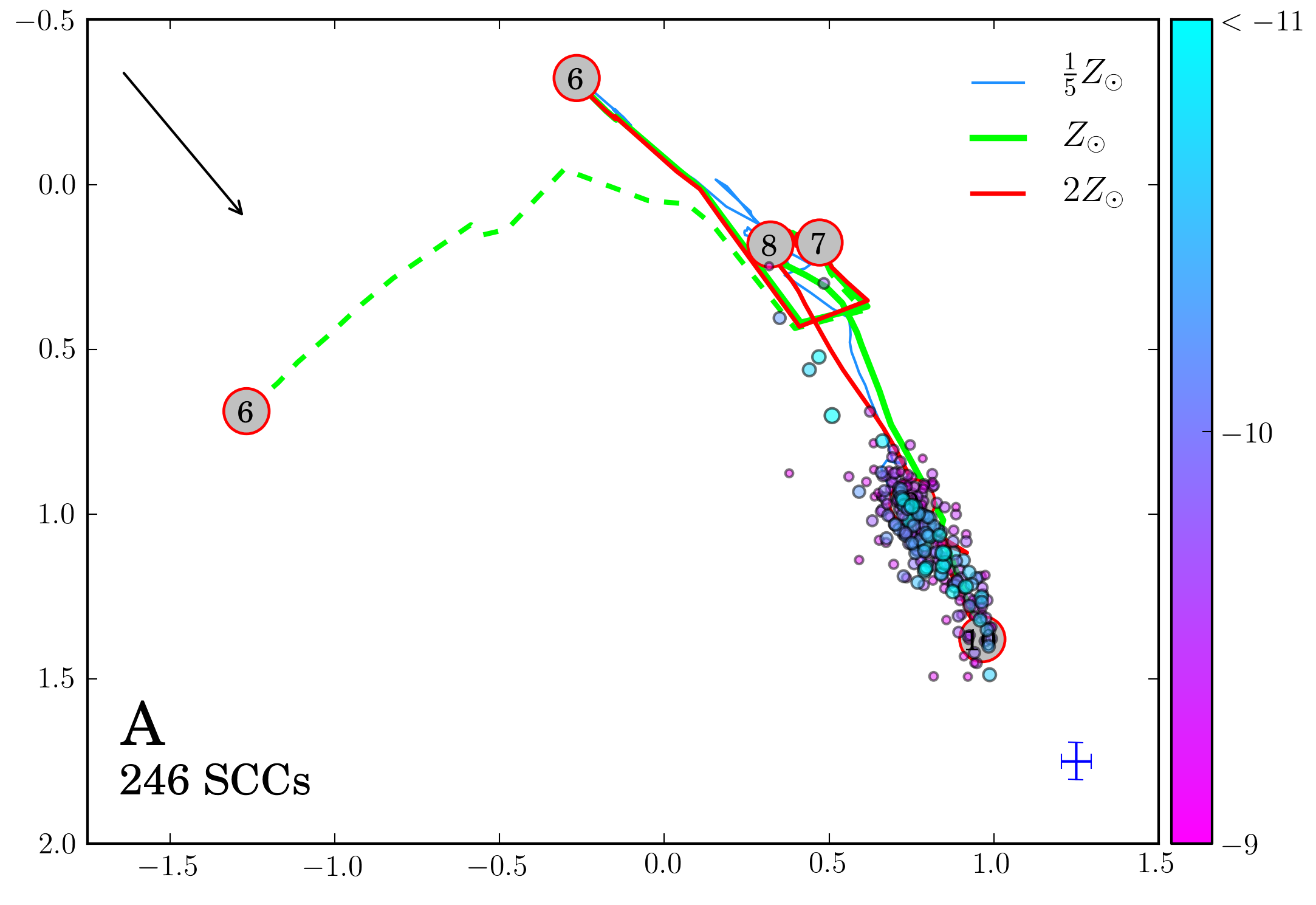}}
\caption[continuation of the previous plot]
{Figure continued from the previous page.  Panel (b) is the 
inverted $V_{606}$ image which shows the SCC system extent as defined 
by the $V_{606}$ brightness contour of $\sim 1.25 \sigma$ above the 
background level.  Panel (c) is a colour-colour plot for the 
luminous elliptical 42A.  In this colour-space, all of the clusters 
are consistent with being old globular clusters. (A colour version 
of this figure is available in the online journal.)}
\end{figure*}
\clearpage

\begin{figure*}
\centering
\subfloat[]{\includegraphics[width=0.95\textwidth]{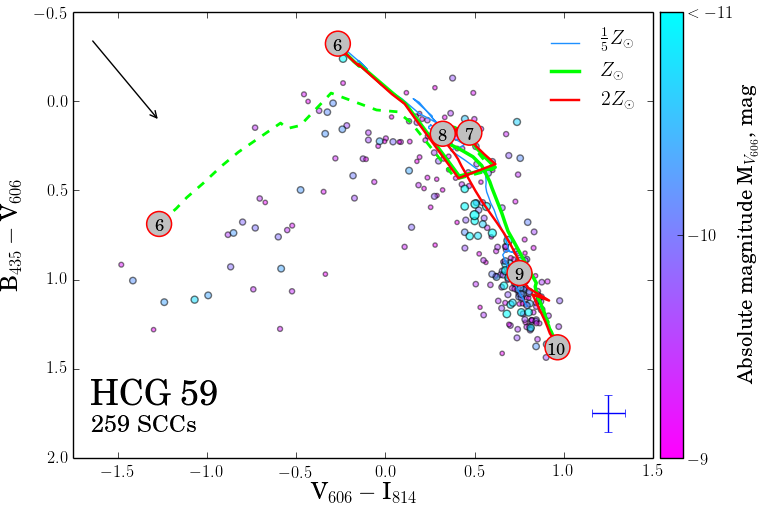}}\\
\caption[Colour-colour plot for all SCCs detected in HCG 59, and for SCCs 
in each galaxy. SCCs spatial distribution and system extent for each galaxy 
in HCG 59]{A colour-colour plot of all star clusters candidates in HCG 
59 (a), including clusters in the intergroup medium, and subplots 
for particular galaxies in that group (c)--(f), continued on the 
next page. For more details see caption for Fig.~\ref{fig:CC07}. 
(A colour version of this figure is available in the online journal.)}
\label{fig:CC59}
\end{figure*}
\clearpage

\begin{figure*}
\ContinuedFloat
\centering
\subfloat[]{\includegraphics[width=0.49\textwidth]{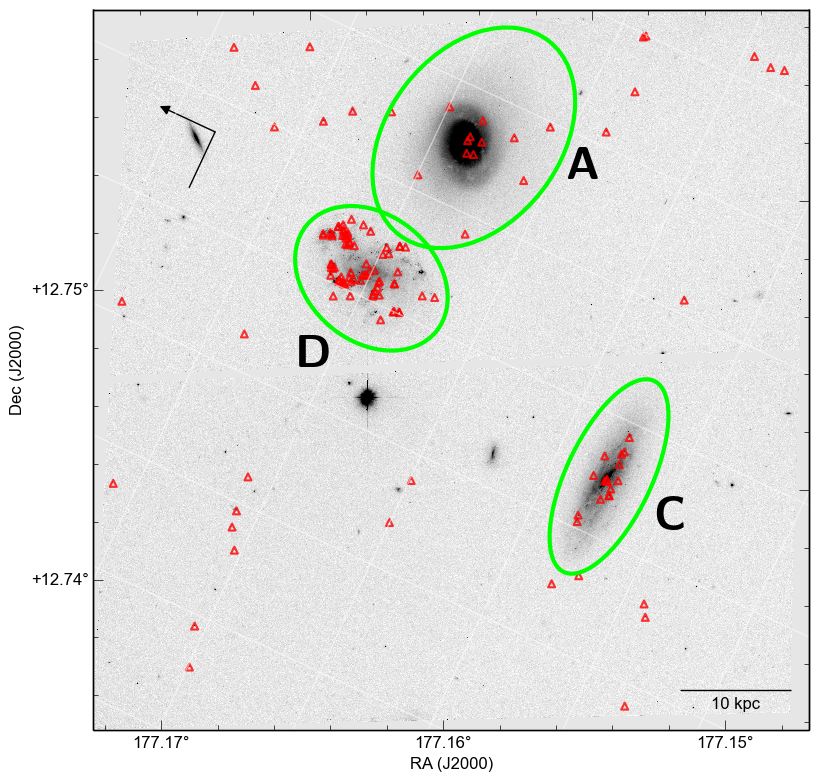}}
\hspace{0.075cm}
\subfloat[]{\includegraphics[width=0.49\textwidth]{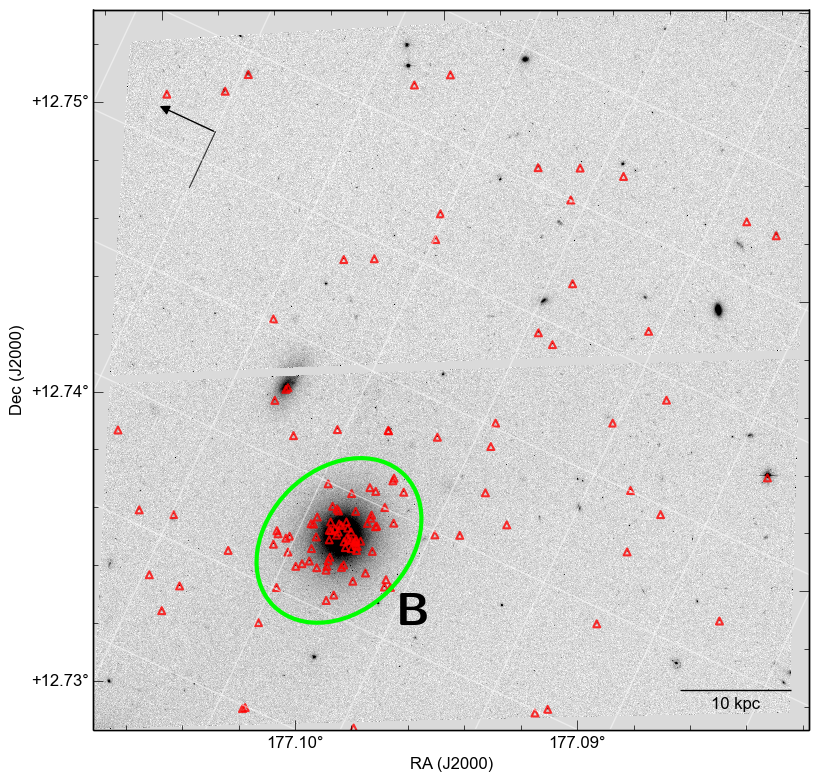}}\\
\subfloat[]{\includegraphics[width = 2.9in]{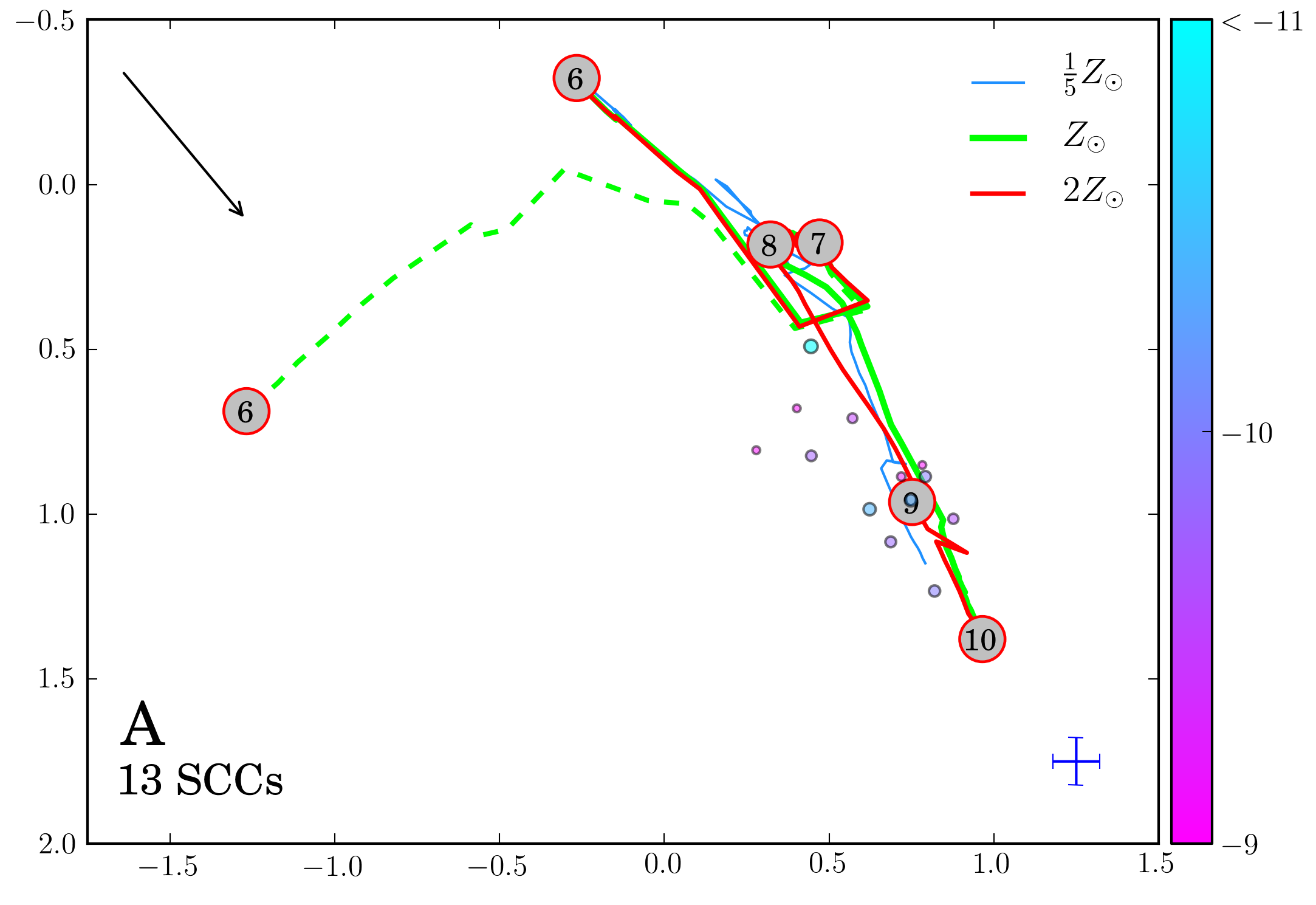}}\hspace{0.4cm}
\subfloat[]{\includegraphics[width = 2.9in]{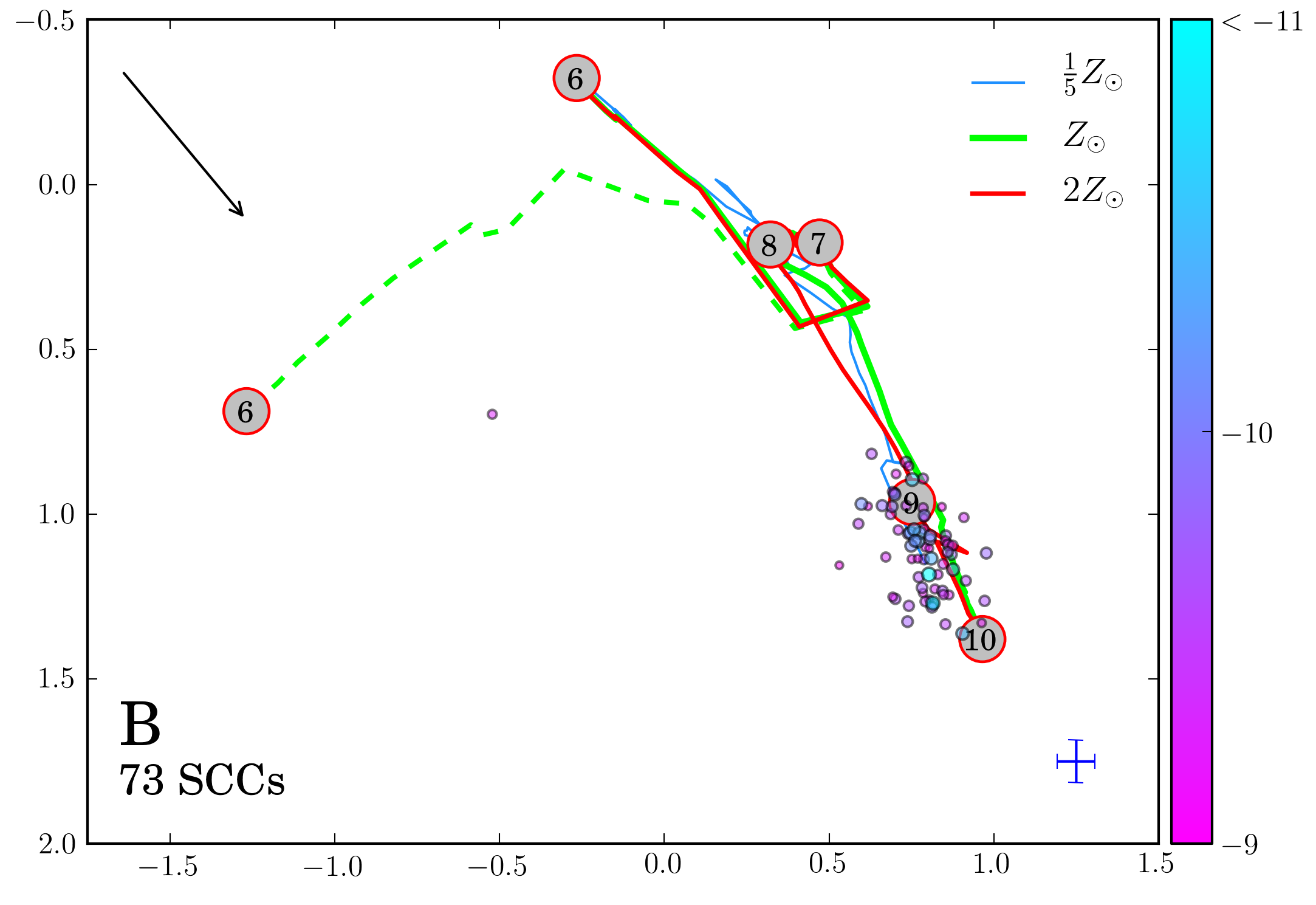}}\\
\subfloat[]{\includegraphics[width = 2.9in]{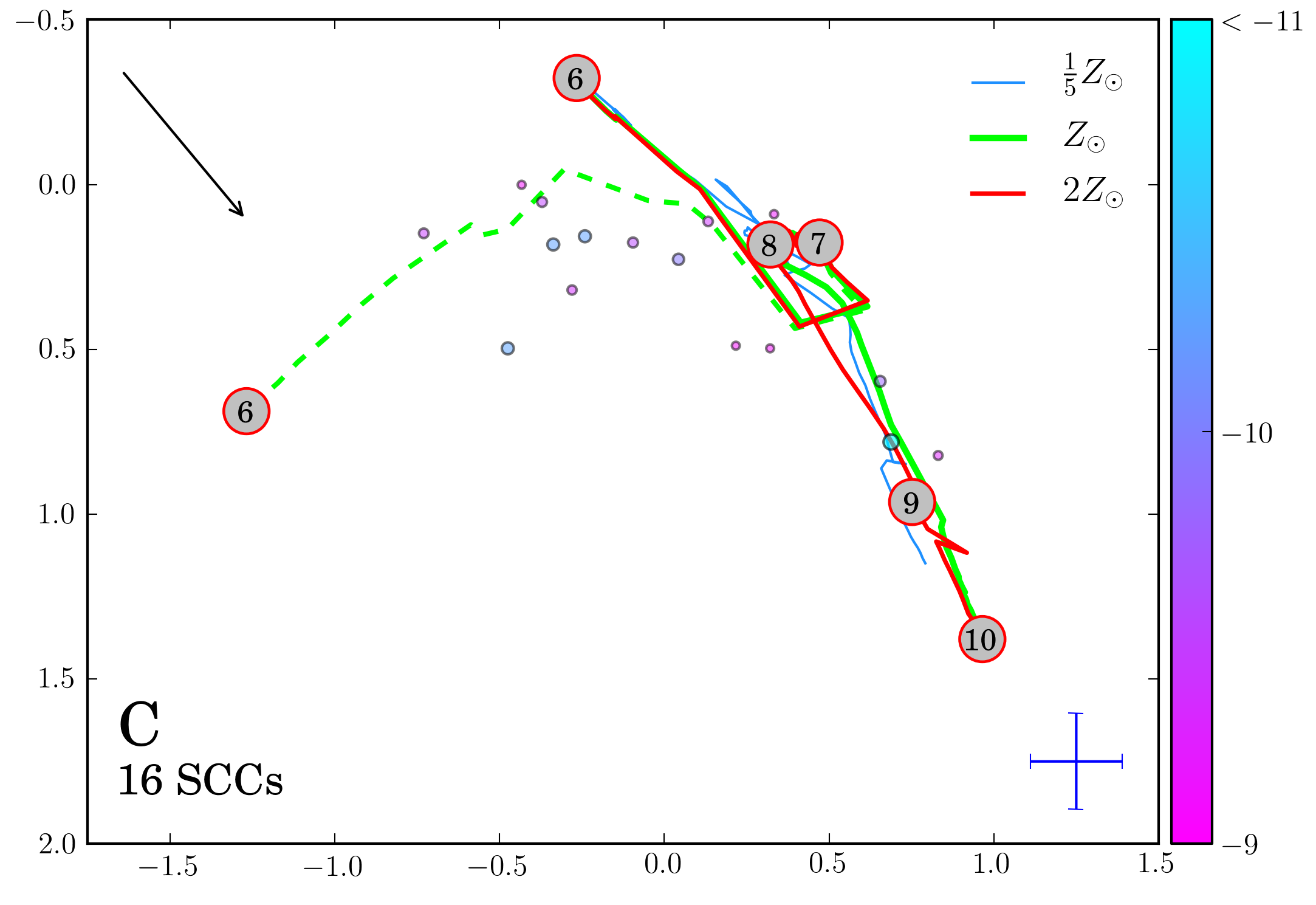}}\hspace{0.4cm}
\subfloat[]{\includegraphics[width = 2.9in]{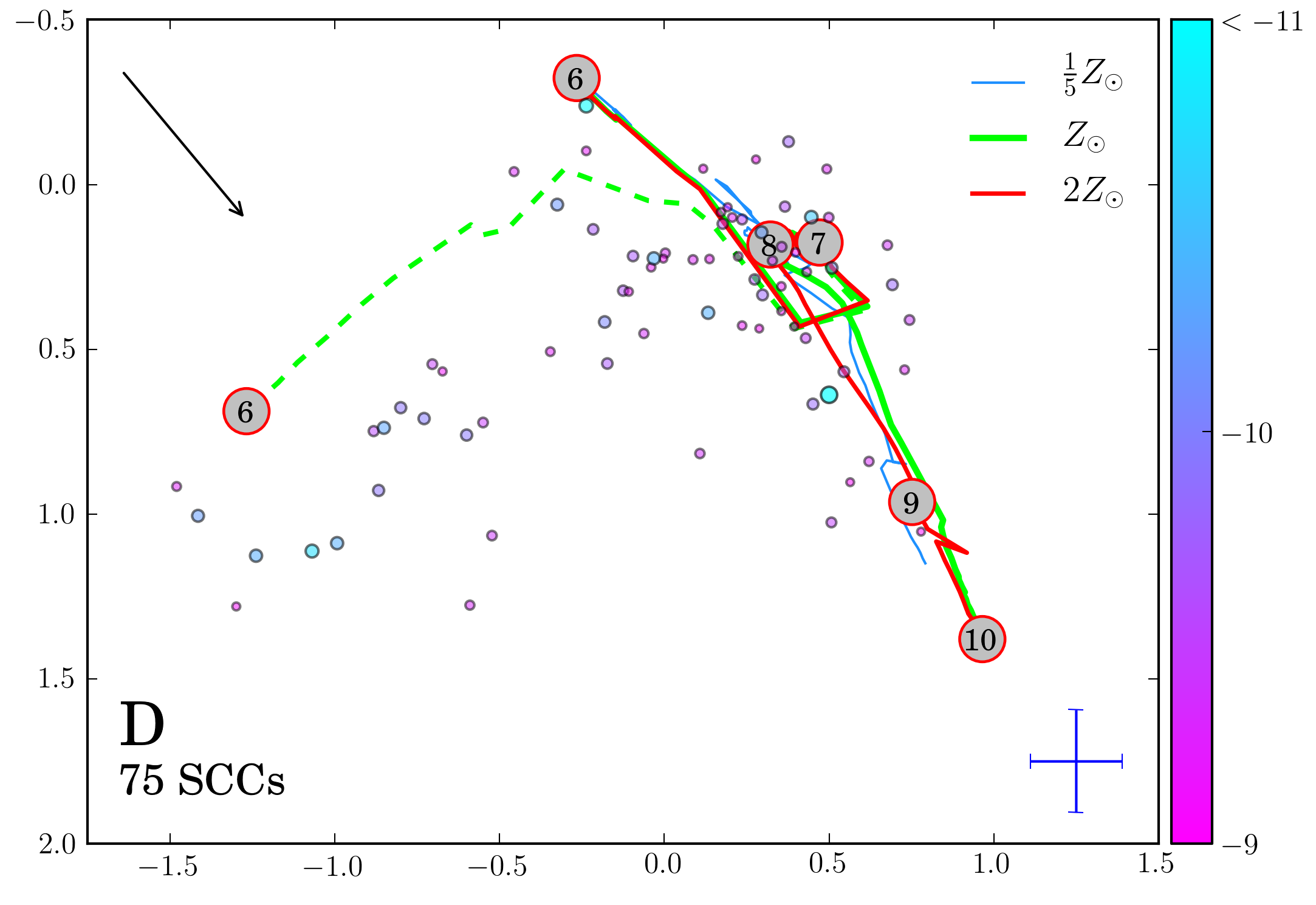}}\\
\caption[continuation of the previous plot]
{$\ldots$ figure continued from the previous page.  Panels 
(b) and (c) are inverted $V_{606}$ images which show the SCC system 
extent as defined by a $V_{606}$ brightness contour of $\sim 1.25 \sigma$ 
above the background level. Panels (c)--(f) are colour-colour 
plots for individual galaxies in the HCG 59 group.  The large 
irregular 59D has a large population of young clusters.  59A hosts 
both old and intermediate-aged clusters, while the elliptical 59B 
has only a globular cluster population. (A colour version of this 
figure is available in the online journal.)}
\end{figure*}
\clearpage

\begin{figure*}
\centering
\subfloat[]{\includegraphics[width=0.95\textwidth]{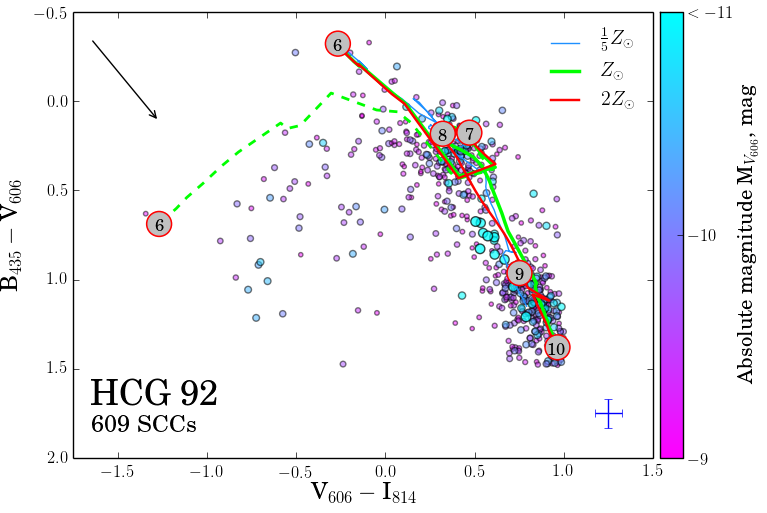}}\\
\caption[Colour-colour plot for all SCCs detected in HCG 92, and for SCCs 
in each galaxy. SCCs spatial distribution and system extent for each galaxy 
in HCG 92]{A colour-colour plot of all SCCs in HCG 92 (a), including 
clusters in the intergroup medium, and subplots for particular 
galaxies in that group (c)--(f) and (h)--(i), continued on the 
next pages. For more details see the caption for Fig.~\ref{fig:CC07}. 
(A colour version of this figure is available in the online journal.)}
\label{fig:CC92}
\end{figure*}
\clearpage

\begin{figure*}
\ContinuedFloat
\centering
\subfloat[]{\includegraphics[width=0.49\textwidth]{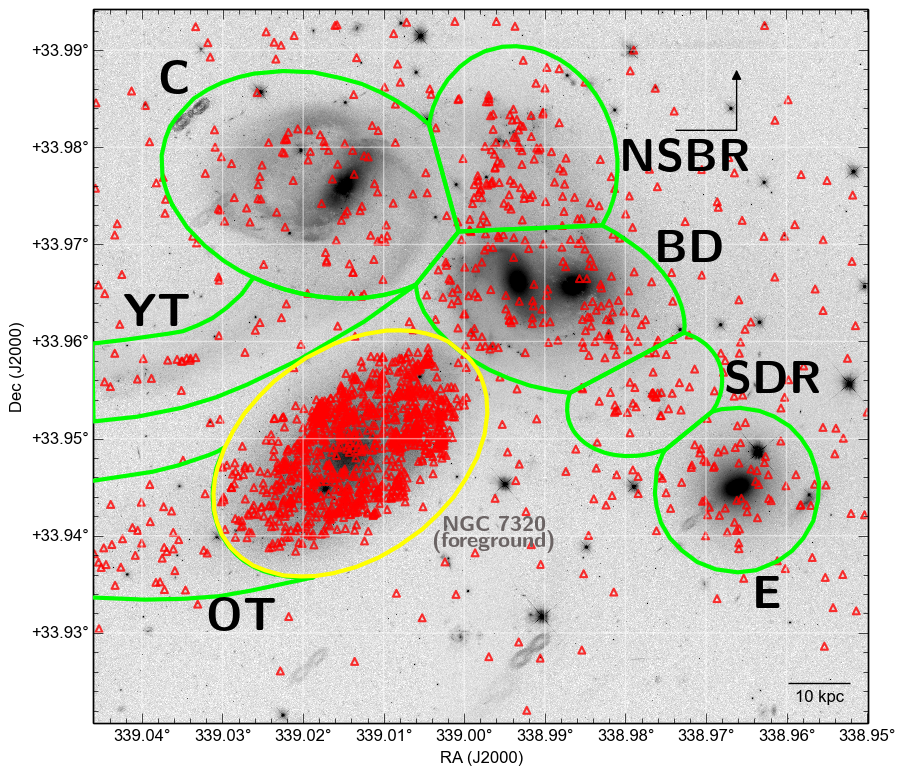}}\\
\subfloat[]{\includegraphics[width = 2.9in]{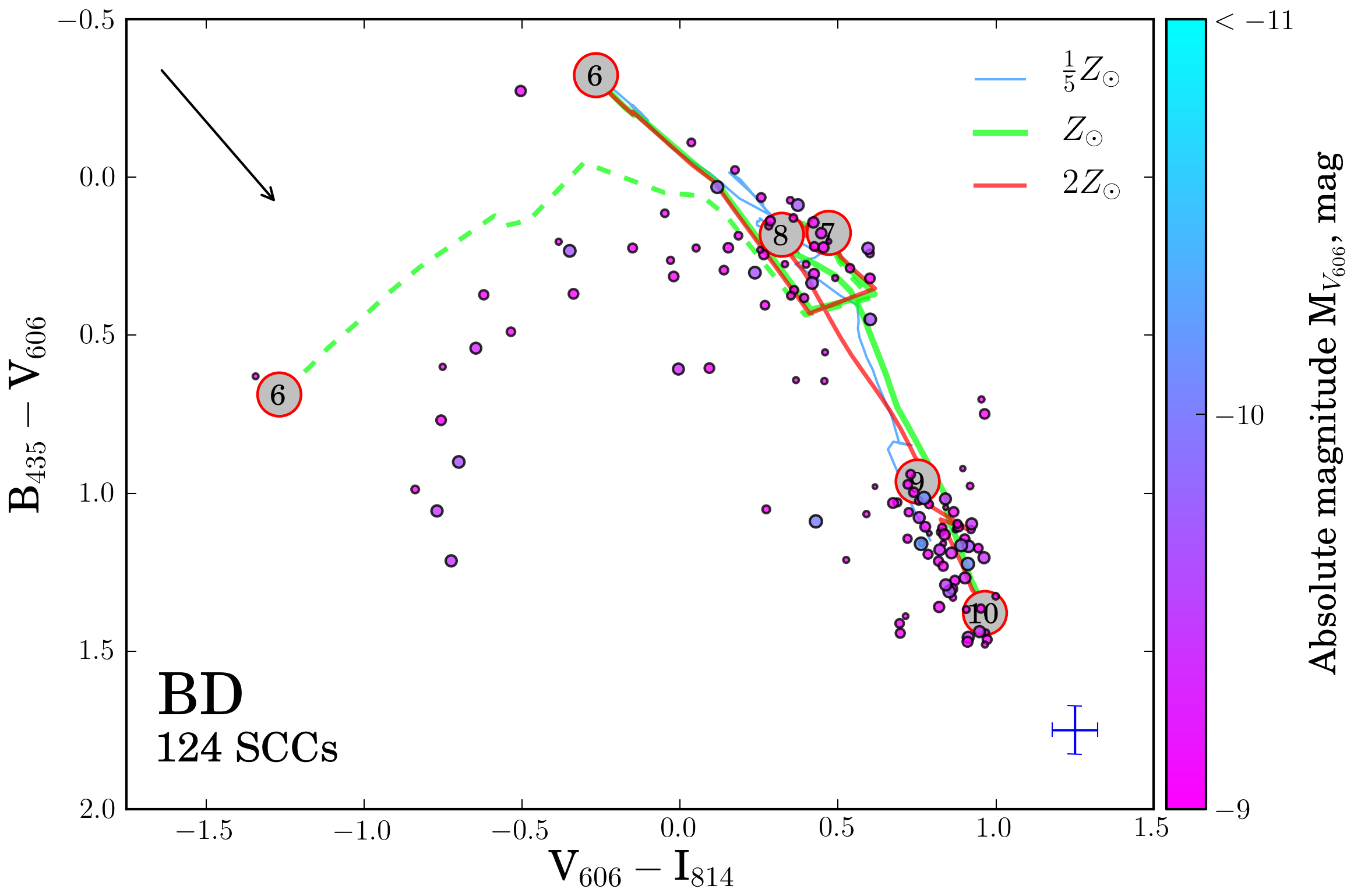}}\hspace{0.4cm}
\subfloat[]{\includegraphics[width = 2.9in]{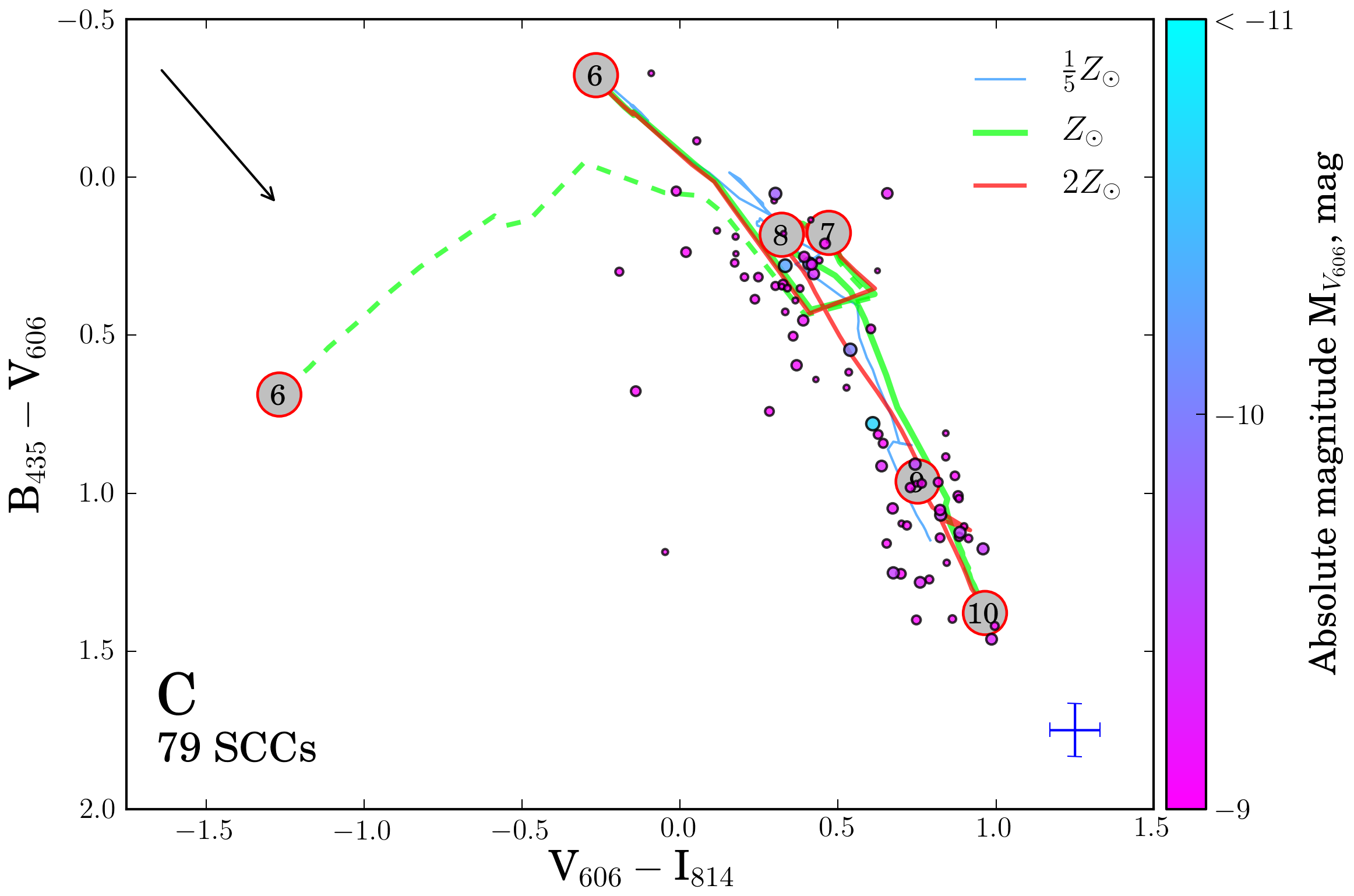}}\\
\subfloat[]{\includegraphics[width = 2.9in]{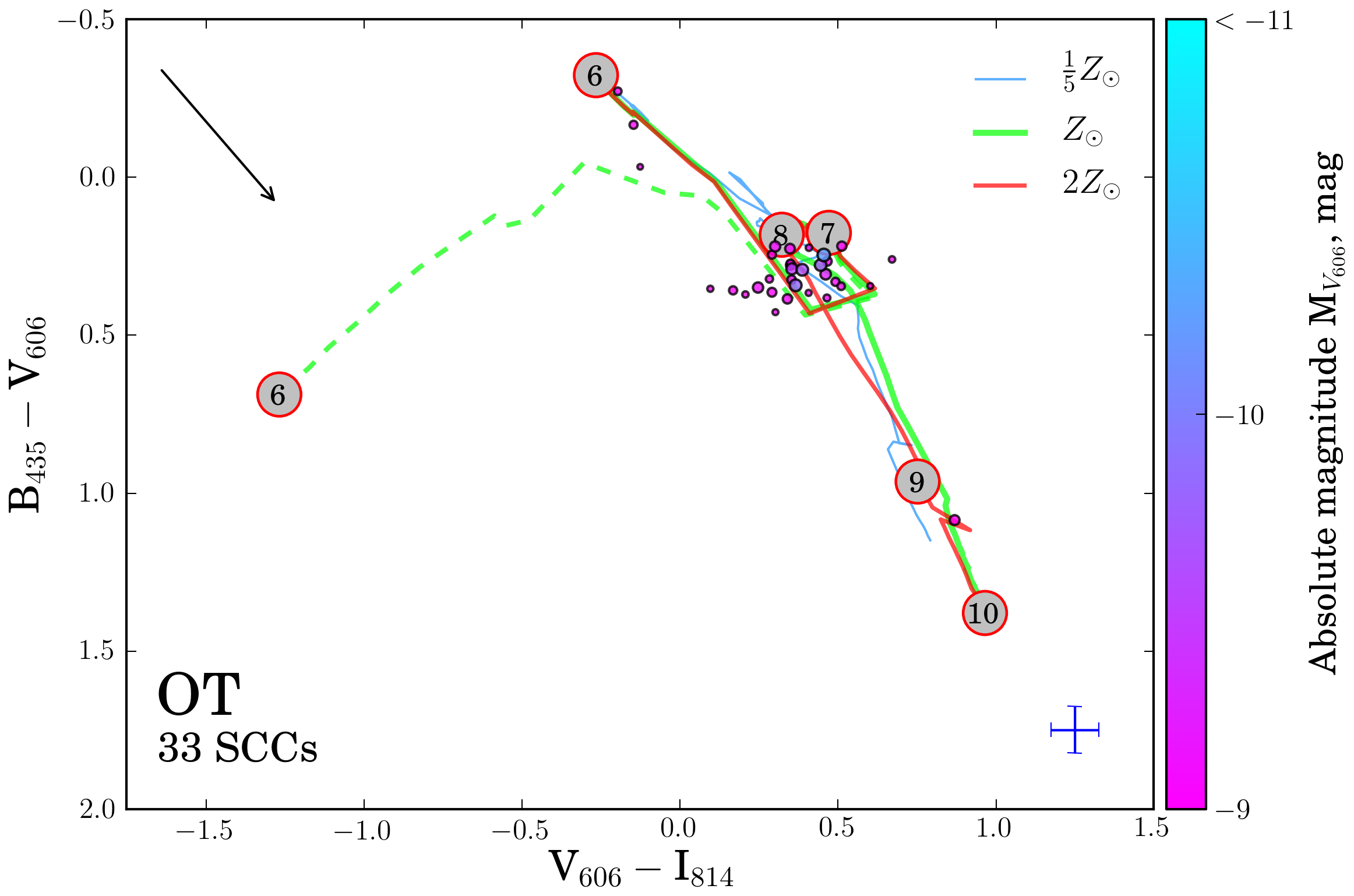}}\hspace{0.4cm}
\subfloat[]{\includegraphics[width = 2.9in]{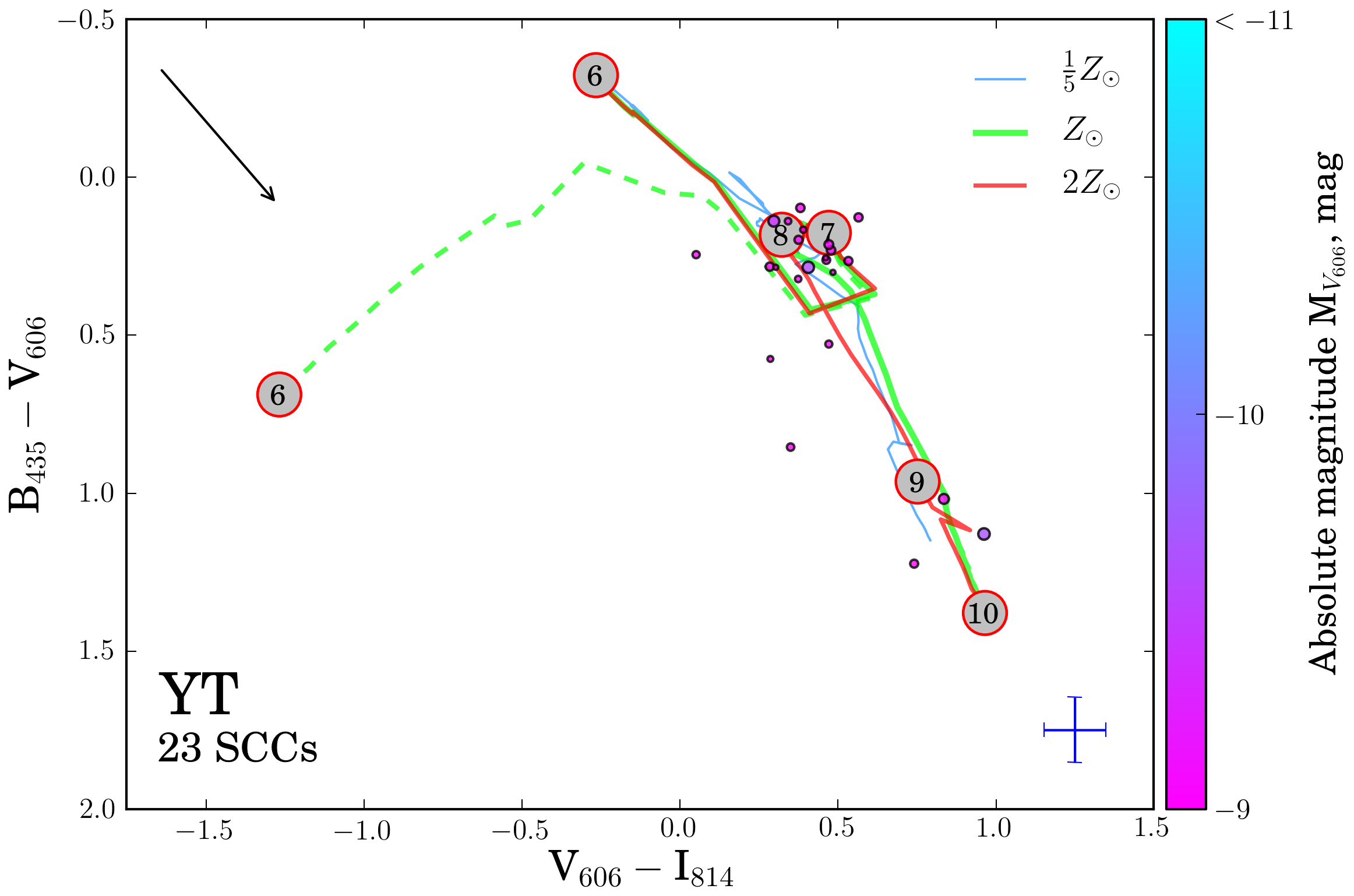}}
\caption[continuation of the previous plot]
{Figure continued from the previous page.  Panel (b) is the inverted 
$V_{606}$ image which shows the SCC system extent as defined by the 
$V_{606}$ brightness contour of $\sim 1.25 \sigma$ above the background 
level.  (The foreground galaxy 92A is outlined in yellow.)  Panels 
(c)--(f) are colour-colour plots for particular galaxies and regions in 
the HCG 92 group.  92BD has a population of young clusters whose formation 
was triggered by the collision of 92B with the cold intra-group medium.  
92C shows evidence for truncated star formation, with an intermediate 
($>100$~Myr) to old SC population.  The Old Tail (OT) and Young Tail (YT) 
tidal features show a small populations of star clusters with well-defined 
colours, tracers of short bursts of star formation in these features.  The 
8-shape objects observed in the panel (b), also in panel (c) in  
Fig.~\ref{fig:GC92}, are ghost images caused by reflections off the CCD 
and return reflections from the CCD housing entrance window in WFC3.  (A 
colour version of this figure is available in the online journal.)}  
\end{figure*}
\clearpage

\begin{figure*}
\centering
\ContinuedFloat
\subfloat[]{\includegraphics[width = 2.9in]{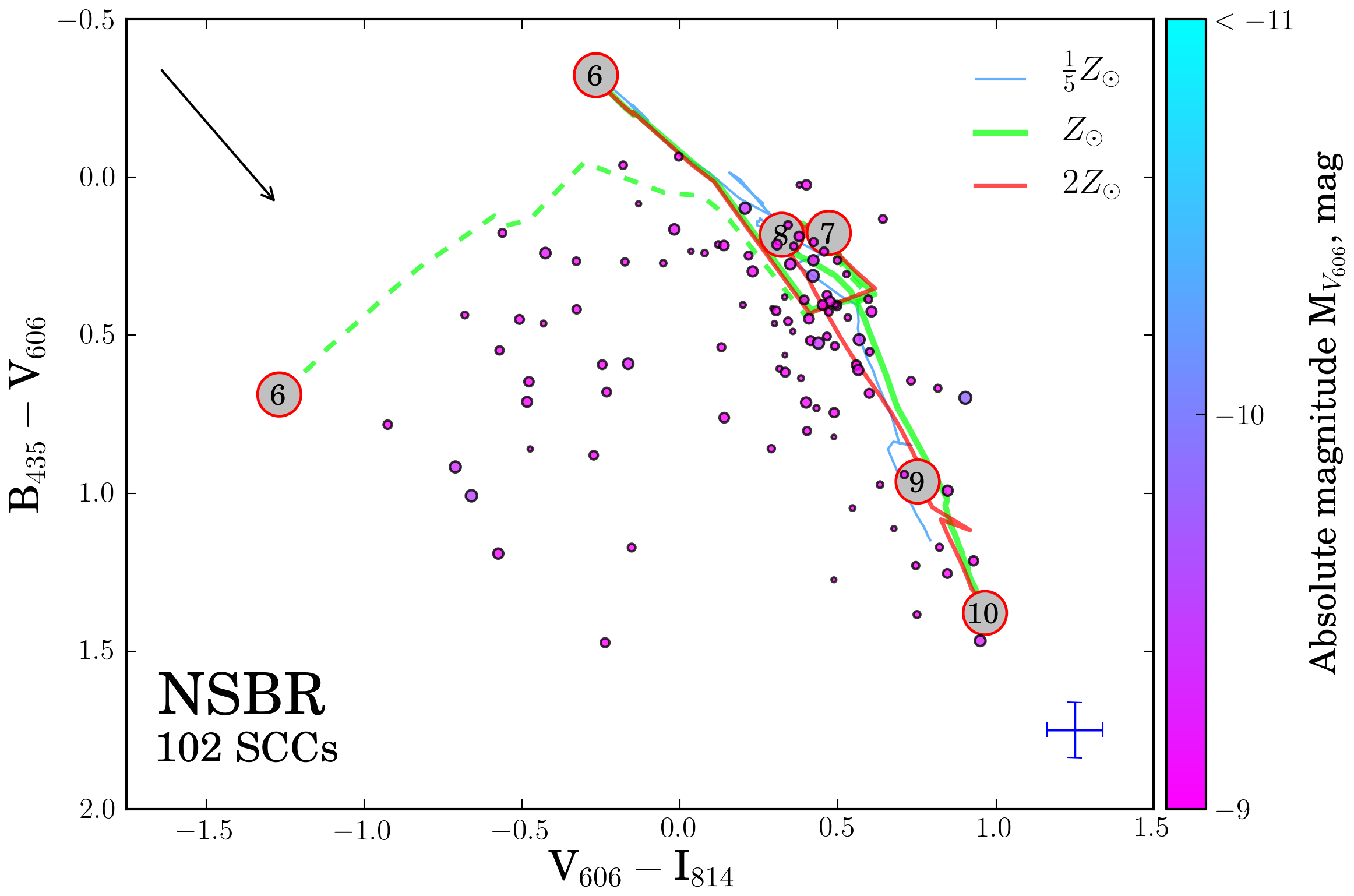}}\hspace{0.4cm}
\subfloat[]{\includegraphics[width = 2.9in]{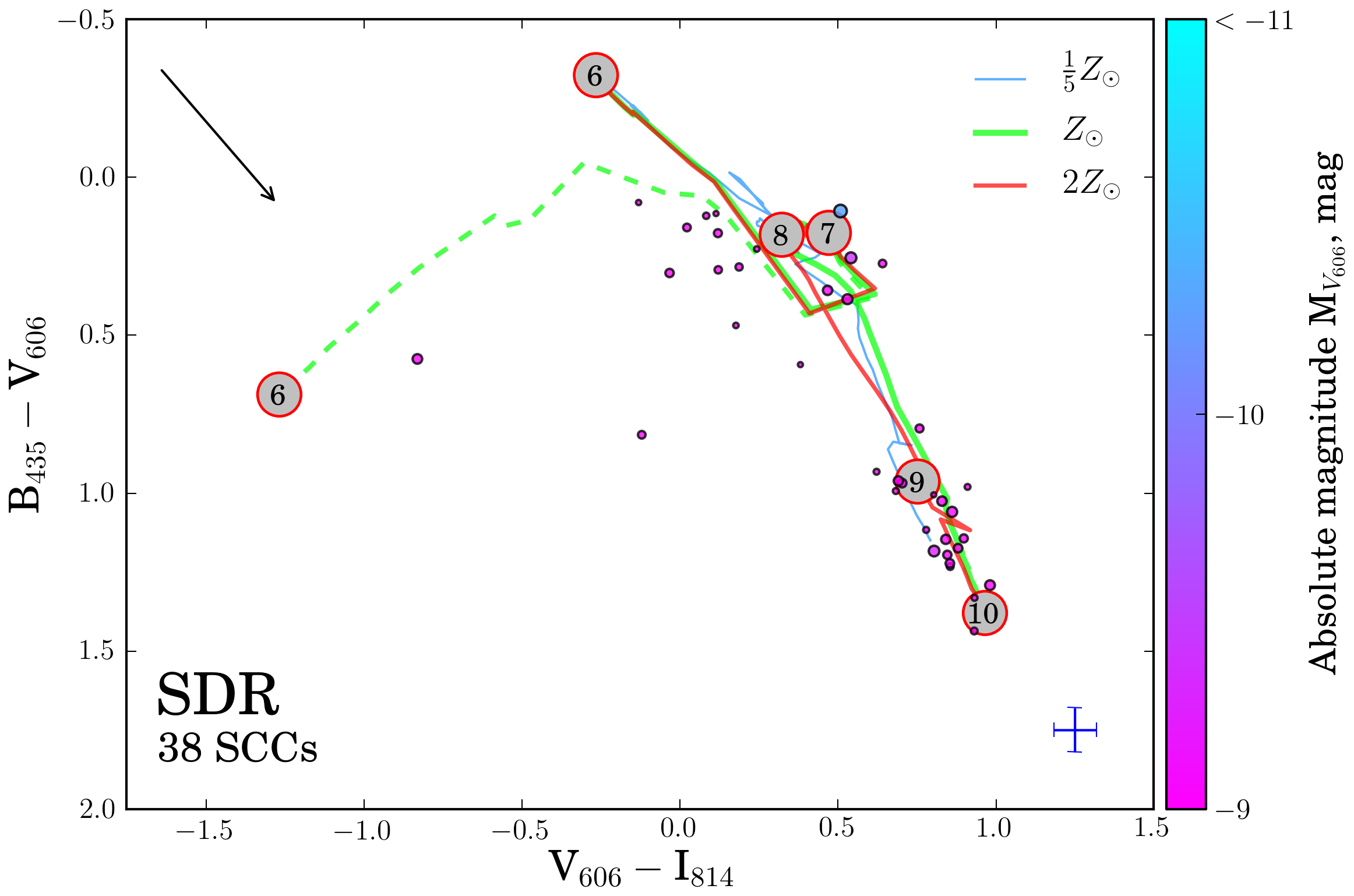}}\\
\subfloat[]{\includegraphics[width = 2.9in]{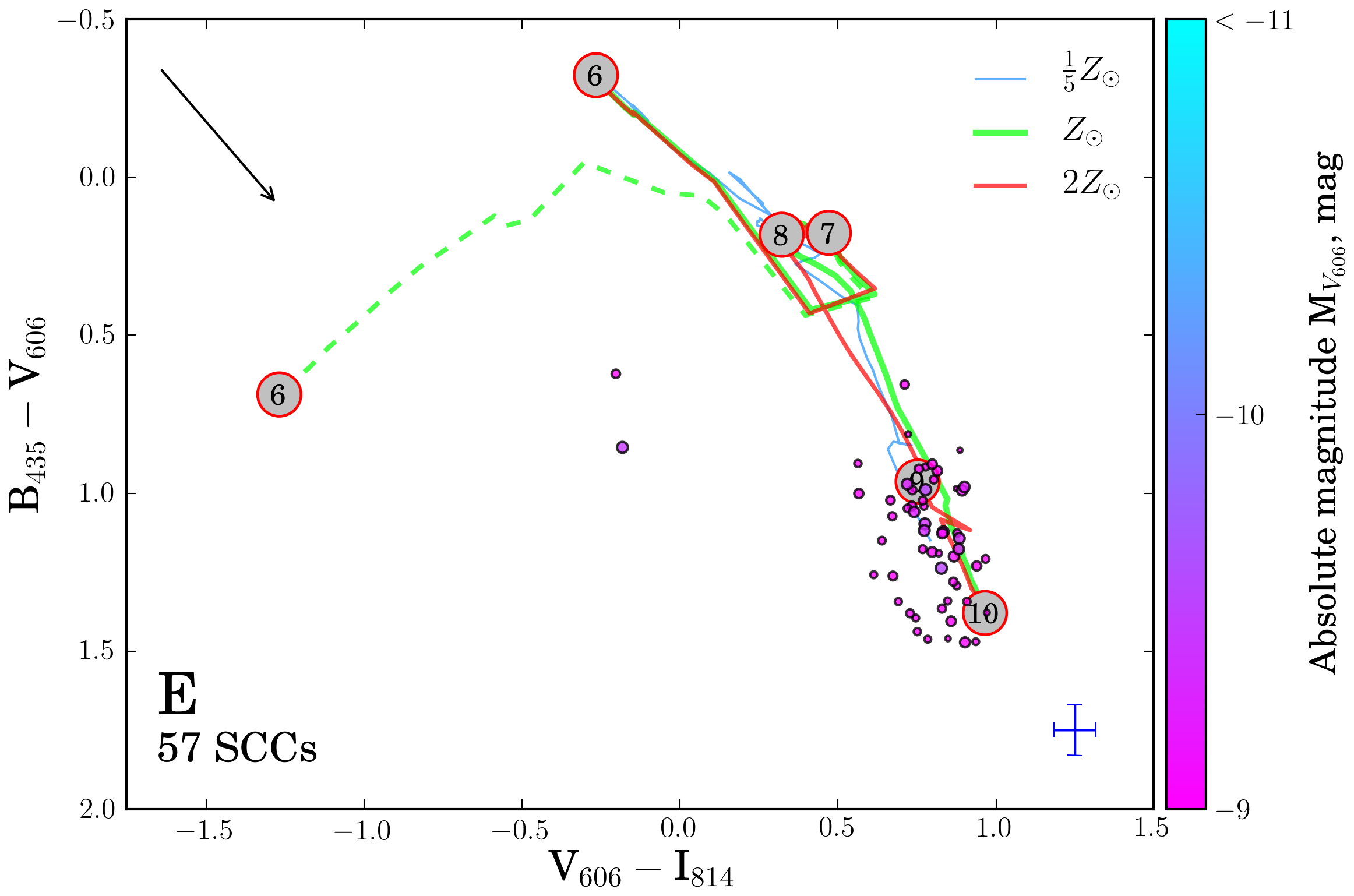}}
\caption[continuation of the previous plot]
{$\ldots$ Figure continued from the previous page.  
Panels (g)--(i) are colour-colour plots for particular 
galaxies and regions in the HCG 92 group.  The Northern 
Starburst Region (NSBR) shows the largest concentration 
of young clusters within the whole group, the Southern 
Debris Region (SDR) has a slightly older population of 
star clusters, and the elliptical 92D has primarily old 
clusters as expected. (A colour version of this figure 
is available in the online journal.)}
\end{figure*}
\clearpage


\begin{figure*}
\begin{center}
\subfloat[]{\includegraphics[width=0.65\textwidth]{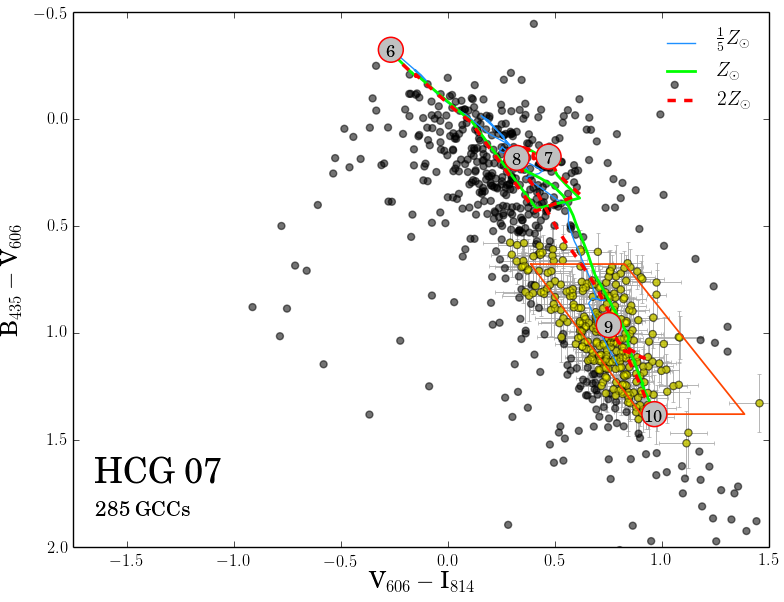}}\\
\vspace{0.1cm}
\subfloat[]{\includegraphics[width=0.65\textwidth]{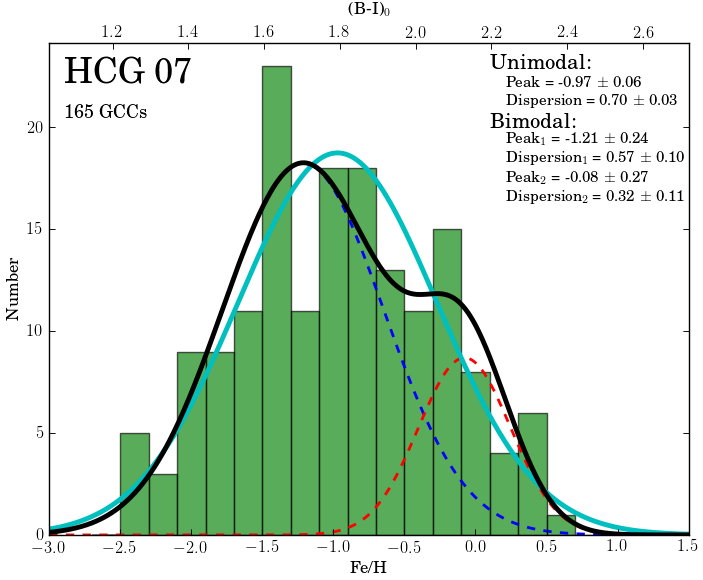}}
\caption[Colour-colour plot of all GCCs detected in HCG 07. Colour-colour 
plots of GCCs for each galaxy in HCG 07. GCCs spatial distribution and GCC 
system extent for each galaxy in HCG 07. GCs metallicities distribution for 
some galaxies in HCG 07]
{A colour-colour diagram of all detected GCCs in HCG 07 
(a), including clusters located in the IGrM, and their metallicity 
distribution (b). The selection parallelogram in (a) is based on 
the colours of Milky Way Globular Clusters \citep{Harris1996}. 
The number of GCCs in the lower left corner is the number of 
clusters that are located inside the selection parallelogram 
or that overlap the selection region with their 1$\sigma$ error 
bars. The thin solid line, solid line, and dashed line trace 
the evolution of SSP models of [0.2, 1.0, 2.0] Z$_{\odot}$ 
\citep{Marigo2008}.  The numbers on the track denote age represented 
in $\log(\text{age}/\text{yr})$.  Note that the measured quantity for the panel (b), 
and for the rest of the plots of the same nature, is the colour index $B-
I$.  It was converted to Fe$/$H values according to prescription in 
\citet{Harris2013} and further analysis were carried out with those 
Fe$/$H values.  Figure continued on the next page. 
(A colour version of this figure is available in the online journal.)}
\label{fig:GC07} 
\end{center}
\end{figure*}
\clearpage

\begin{figure*}
\ContinuedFloat
\centering
\subfloat[]{\includegraphics[width=0.49\textwidth]{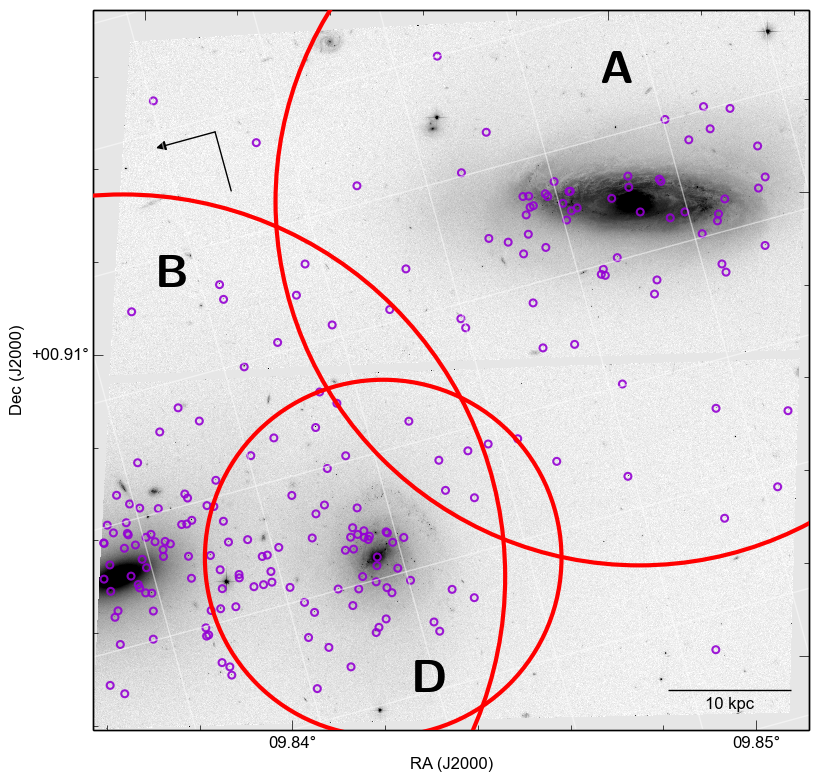}}
\hspace{0.075cm}
\subfloat[]{\includegraphics[width=0.49\textwidth]{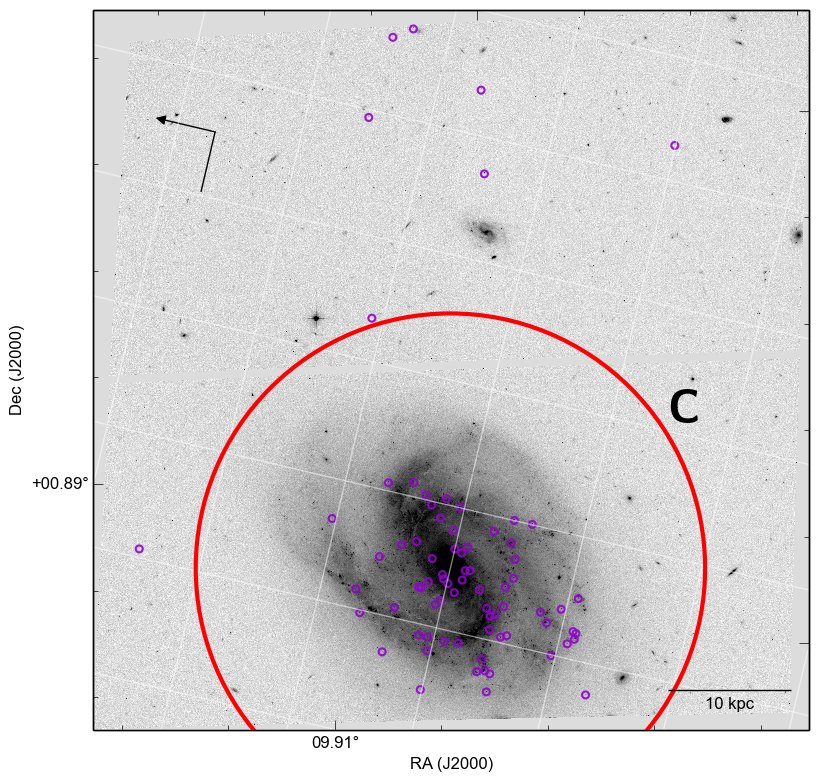}}\\
\subfloat[]{\includegraphics[width = 2.9in]{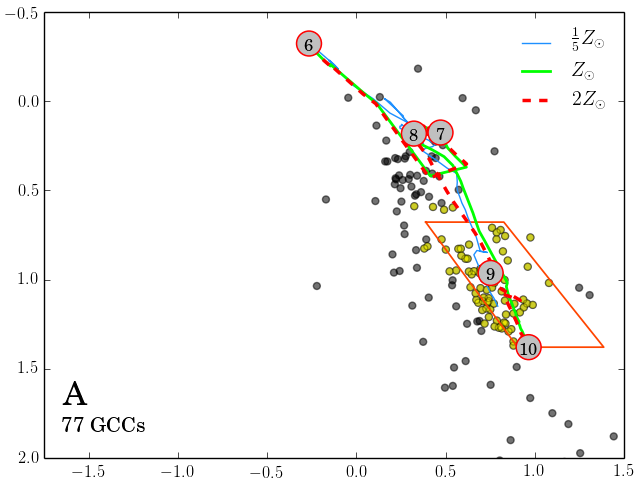}}\hspace{0.4cm}
\subfloat[]{\includegraphics[width = 2.9in]{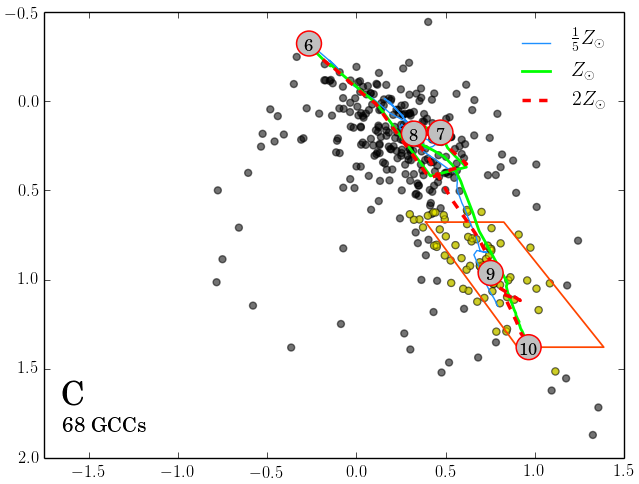}}\\
\subfloat[]{\includegraphics[width = 2.9in]{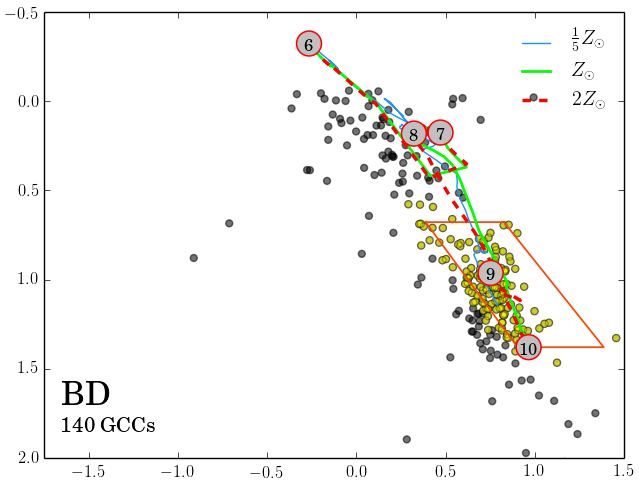}} \hspace{0.4cm}
\subfloat[]{\includegraphics[width = 2.9in]{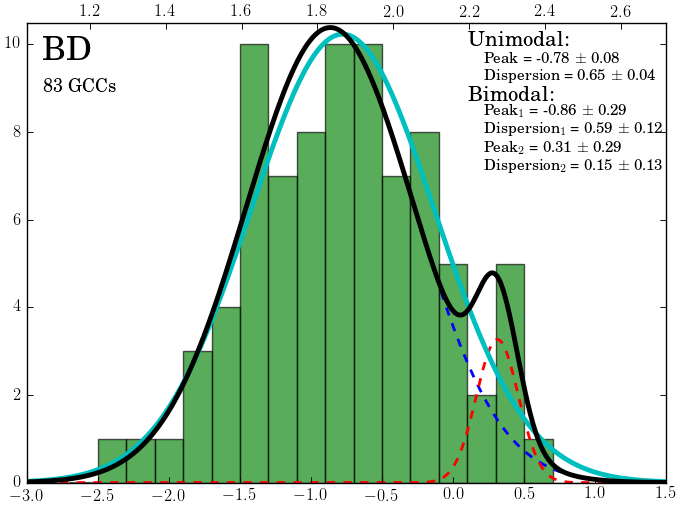}}\\
\caption[continuation of the previous plot]
{$\ldots$  Panels (c) and (d) are inverted $V_{606}$ 
images which show the GCC system extent, with locations of GCCs 
overplotted as circles. GCCs found in the central regions and 
spiral arms of galaxies A and C could potentially be reddened 
young star clusters. Panels (e) and (g) are colour-colour 
plots for particular galaxies in the group.  The systems of 7B 
and D are considered together because of the projected overlap 
of their expected GC system extents.  Their metallicity distribution 
is consistent with a single-peaked Gaussian. (A colour version 
of this figure is available in the online journal.)}
\end{figure*}
\clearpage


\begin{figure*}
\begin{center}
\subfloat[]{\includegraphics[width=0.65\textwidth]{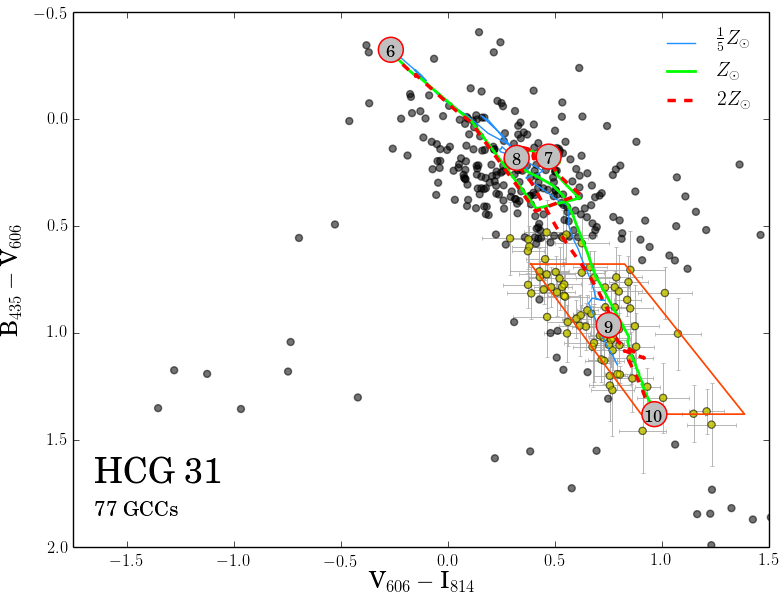}}\\
\vspace{0.1cm}
\subfloat[]{\includegraphics[width=0.65\textwidth]{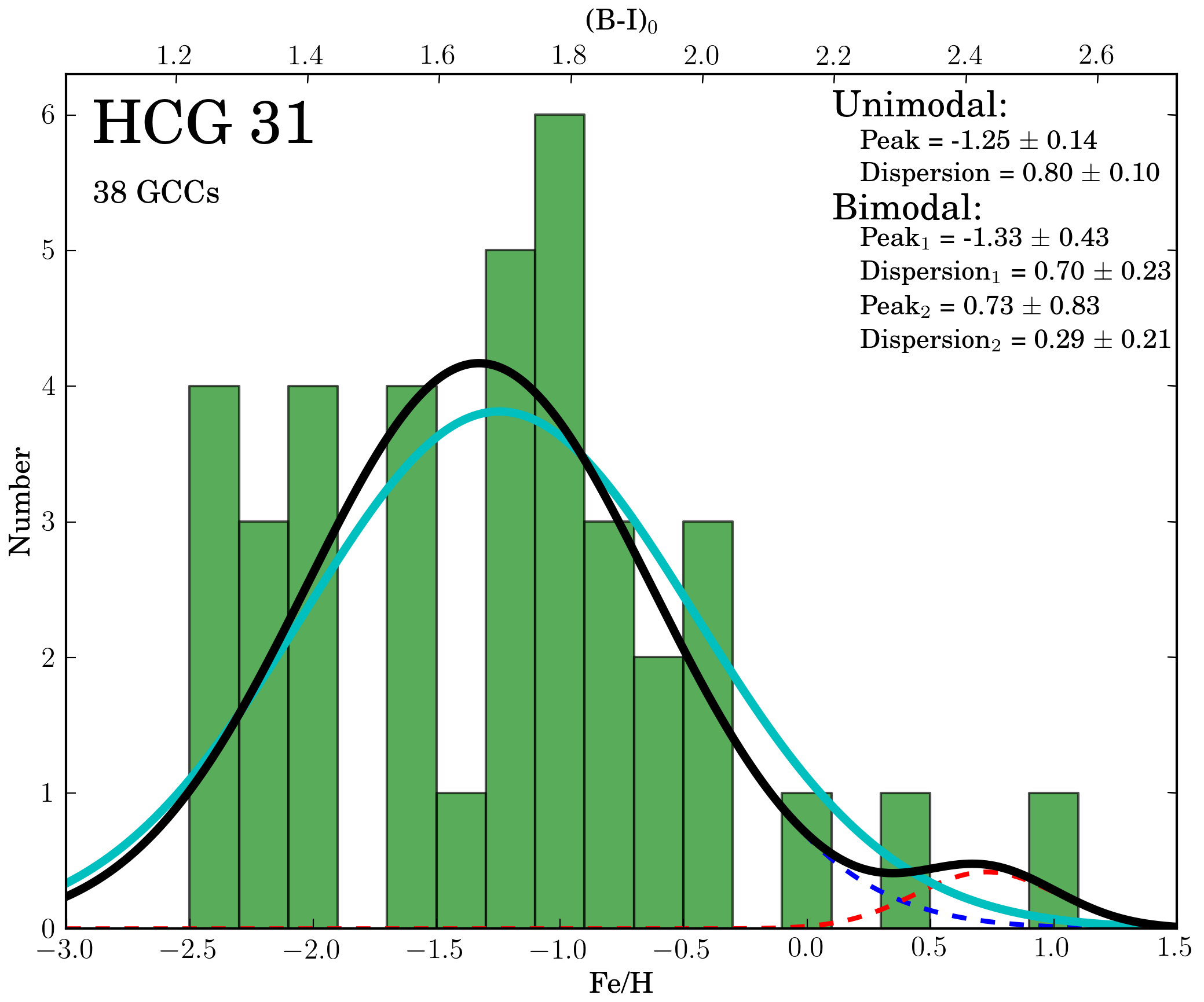}}
\caption[Colour-colour plot of all GCCs detected in HCG 31. Colour-colour 
plots of GCCs for each galaxy in HCG 31. GCCs spatial distribution and GCC 
system extent for each galaxy in HCG 31. GCs metallicities distribution for 
some galaxies in HCG 031]
{The GCC population of HCG 31 (a) and its metallicity distribution (b).  
For more details see caption for Fig.~\ref{fig:GC07}.  Given the high rate 
of star formation in this group, the low masses of the individual 
galaxies, and the spatial distribution of GCCs, it is reasonable to assume 
that the majority of the star clusters labelled as GCCs are in fact 
reddened young clusters. Therefore, the metallicity distribution plot (b) 
should be considered with caution. (A colour version of this figure is 
available in the online journal.)}
\label{fig:GC31}
\end{center}
\end{figure*}
\clearpage

\begin{figure*}
\ContinuedFloat
\centering
\subfloat[]{\includegraphics[width=0.49\textwidth]{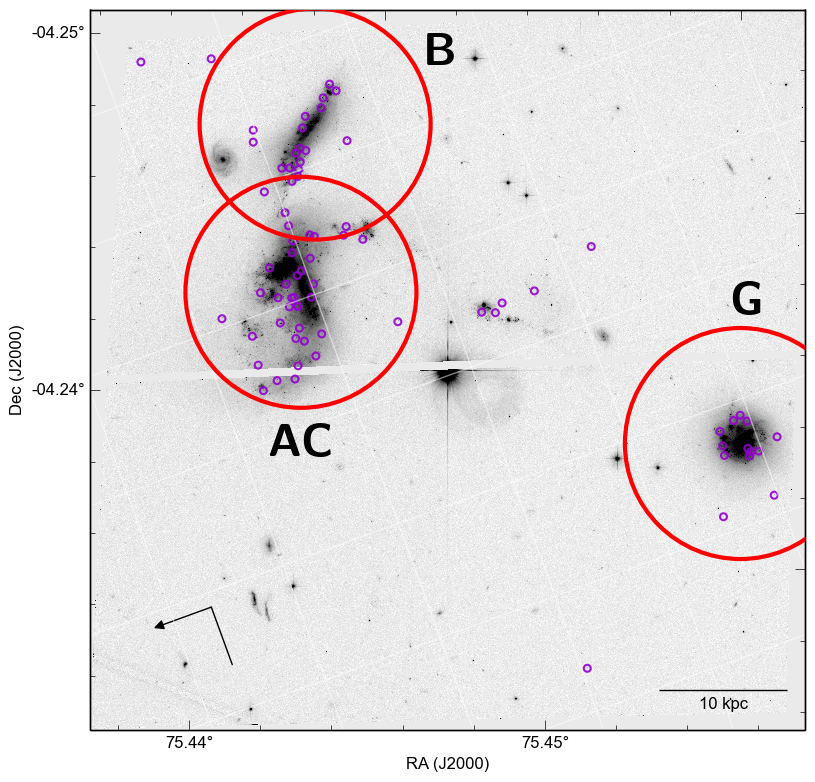}}\\
\subfloat[]{\includegraphics[width = 2.9in]{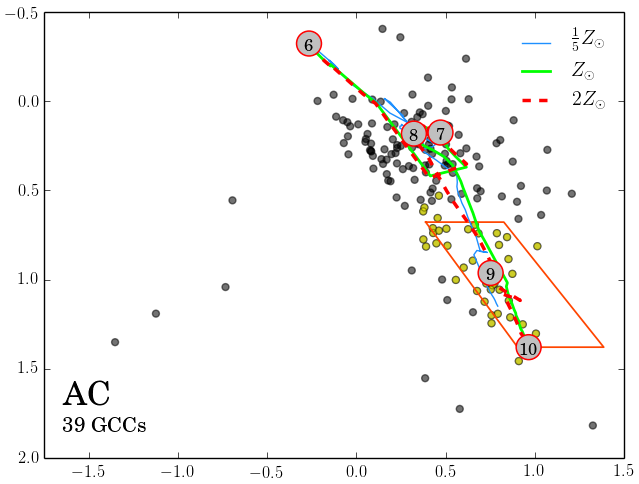}}\hspace{0.4cm}
\subfloat[]{\includegraphics[width = 2.9in]{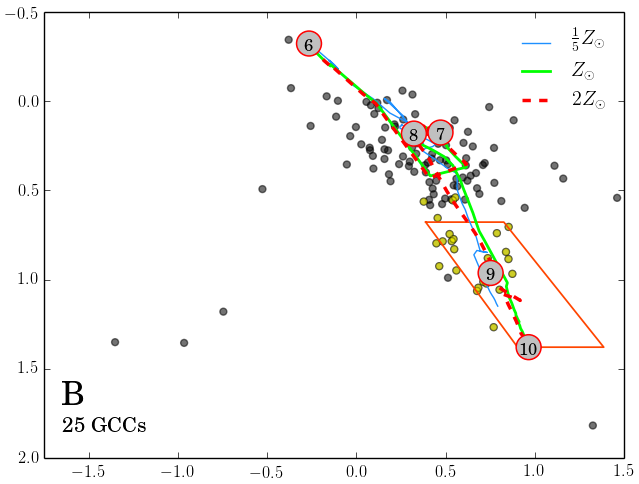}}\\
\subfloat[]{\includegraphics[width = 2.9in]{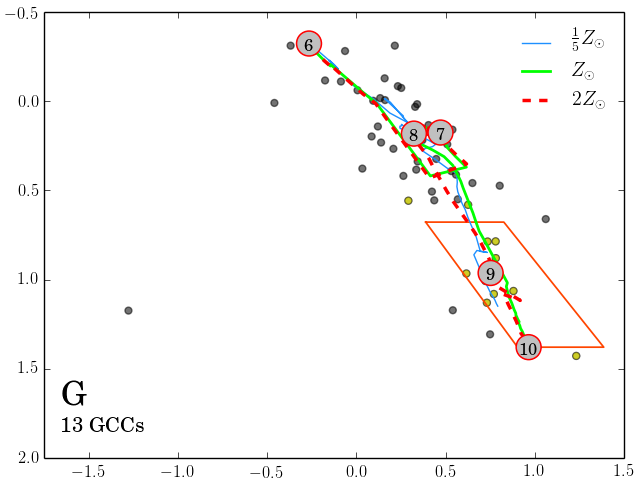}} 
\caption[continuation of the previous plot]
{$\ldots$ Panel (c) is an inverted $V_{606}$ image which 
shows the GCC system extent, with locations of GCCs overplotted 
as circles. Panels (d)--(f) are colour-colour plots for particular 
galaxies/regions in that group. Most likely, the majority of the 
small population of GCCs are reddened young clusters.  (A colour 
version of this figure is available in the online journal.)}
\end{figure*}
\clearpage


\begin{figure*}
\begin{center}
\subfloat[]{\includegraphics[width=0.65\textwidth]{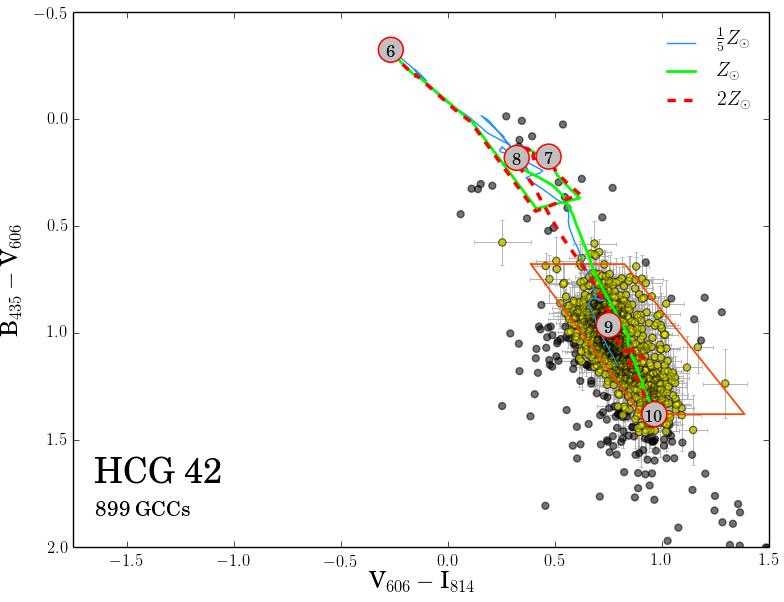}}\\
\vspace{0.1cm}
\subfloat[]{\includegraphics[width=0.65\textwidth]{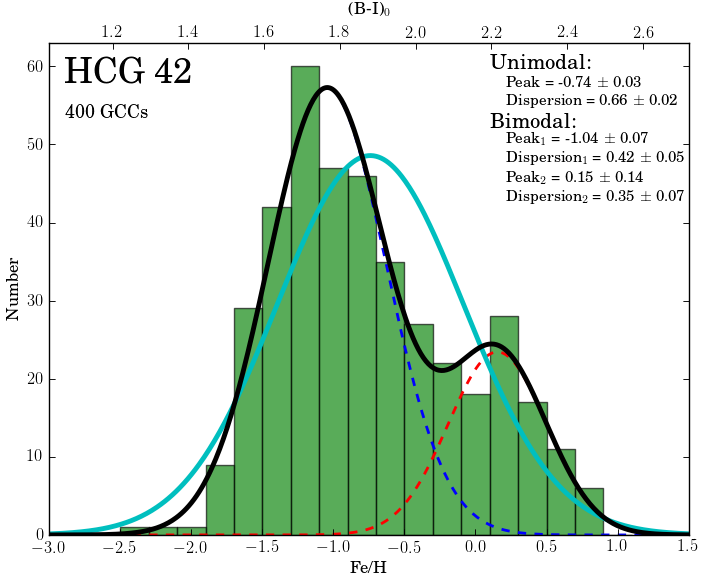}}
\caption[Colour-colour plot of all GCCs detected in HCG 42. Colour-colour 
plot of GCCs for HCG 42A. GCCs spatial distribution and GCC system extent 
for HCG 42A. GCs metallicities distribution for HCG 42A]
{The GCCc population of HCG 42 (a) and its metallicity 
distribution (b).  For more details see caption for Fig.~\ref{fig:GC07}. 
(A colour version of this figure is available in the online journal.)}
\label{fig:GC42}
\end{center}
\end{figure*}
\clearpage

\begin{figure*}
\ContinuedFloat
\centering
\subfloat[]{\includegraphics[width=0.49\textwidth]{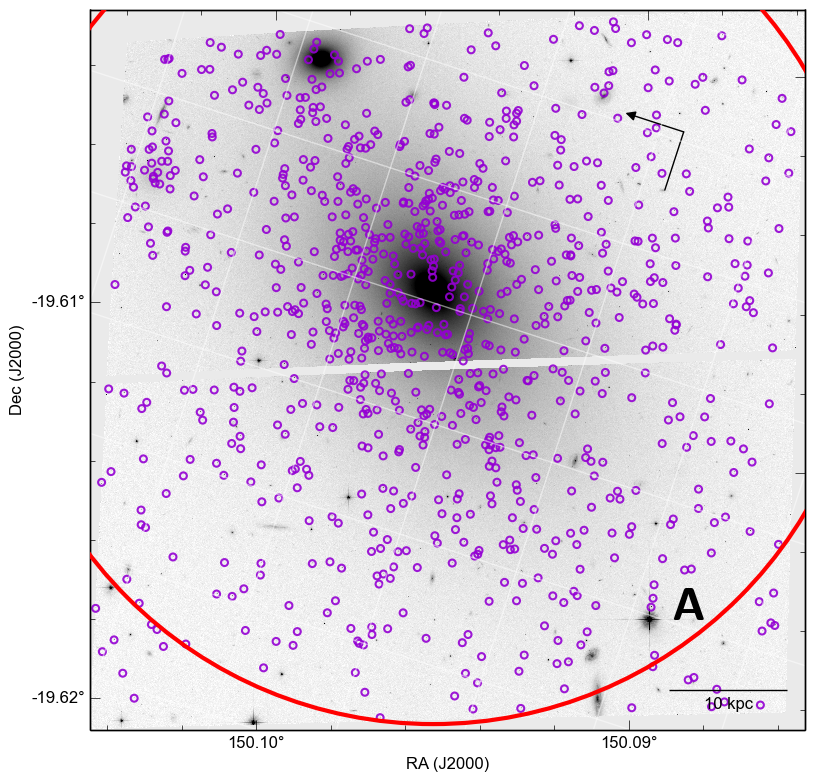}}\\
\subfloat[]{\includegraphics[width = 2.9in]{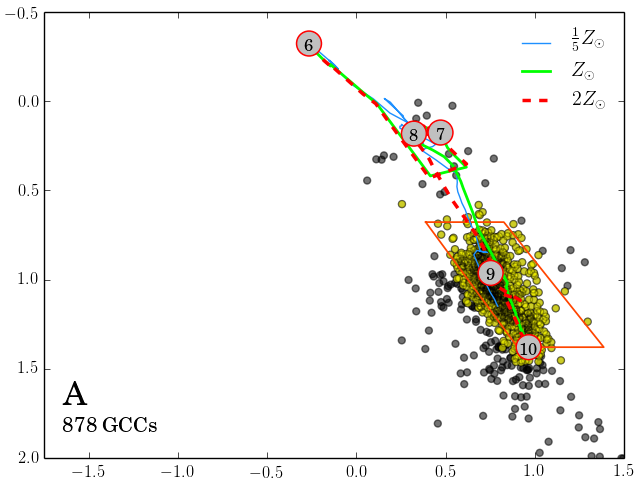}} \hspace{0.4cm}
\subfloat[]{\includegraphics[width = 2.9in]{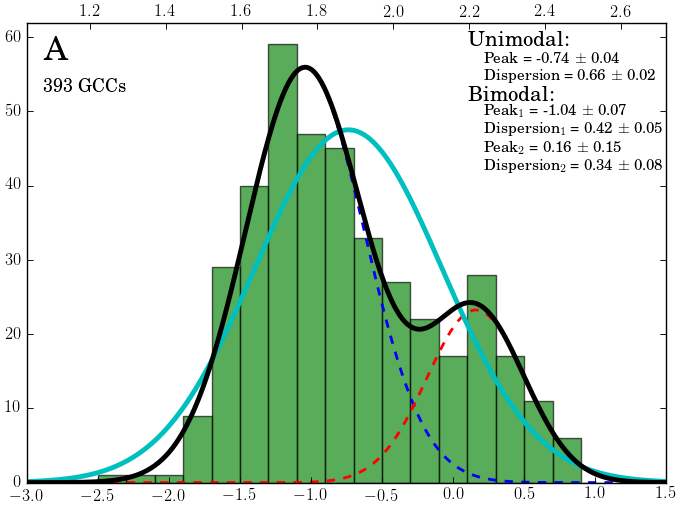}}\\ 
\caption[continuation of the previous plot]
{$\ldots$ Panel (c) is an inverted $V_{606}$ images which 
show the GCC system extent, with locations of detected GCCs 
overplotted as circles. Panel (d) is a colour-colour plot for 
galaxy HCG 42A. A slight overdensity of GCCs close to the left 
upper corner corresponds to the location of a dwarf galaxy, a 
member of HCG 42. Panel (e) is a plot of the metallicity distribution 
of GCCs in HCG 42A. The GMM results favour a bimodal distribution 
with the first peak at [Fe/H] $= -1.04 \pm 0.07$ and the second 
peak at [Fe/H] $= 0.16 \pm 0.15$. (A colour version of this figure 
is available in the online journal.)}
\end{figure*}
\clearpage


\begin{figure*}
\begin{center}
\subfloat[]{\includegraphics[width=0.65\textwidth]{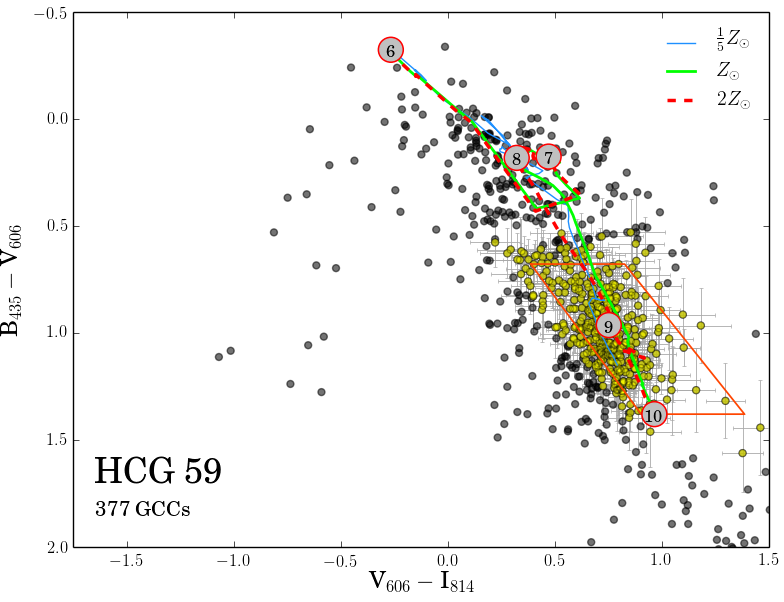}}\\
\vspace{0.1cm}
\subfloat[]{\includegraphics[width=0.65\textwidth]{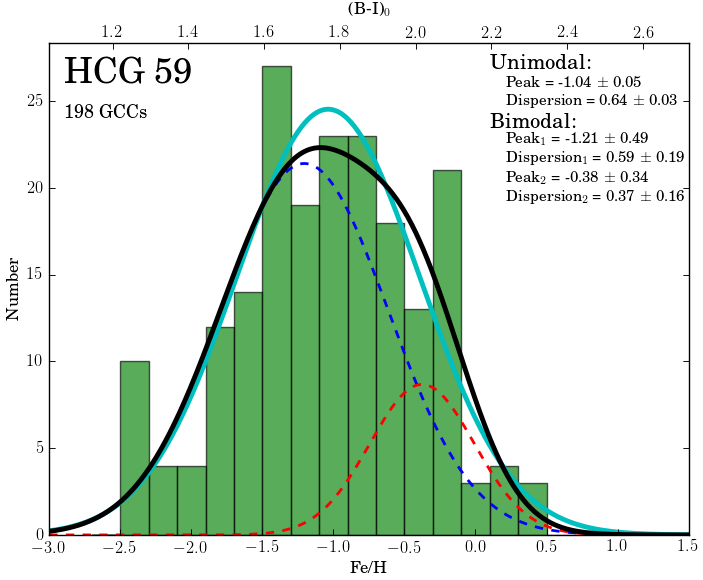}}
\caption[Colour-colour plot of all GCCs detected in HCG 59. Colour-colour 
plots of GCCs for each galaxy in HCG 59. GCCs spatial distribution and GCC 
system extent for each galaxy in HCG 59. GCs metallicities distribution for 
some galaxies in HCG 59]
{The GCC population of HCG 59 (a) and its metallicity 
distribution (b). See caption for Fig.~\ref{fig:GC07}. (A colour 
version of this figure is available in the online journal.)}
\label{fig:GC59}
\end{center}
\end{figure*}
\clearpage

\begin{figure*}
\ContinuedFloat
\centering
\subfloat[]{\includegraphics[width=0.49\textwidth]{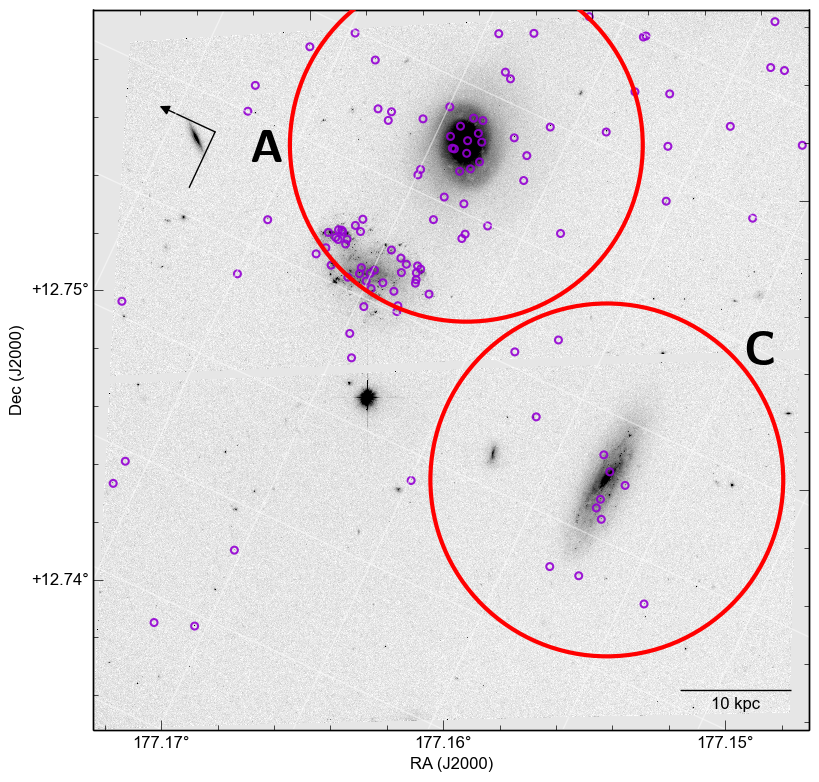}}
\hspace{0.075cm}
\subfloat[]{\includegraphics[width=0.49\textwidth]{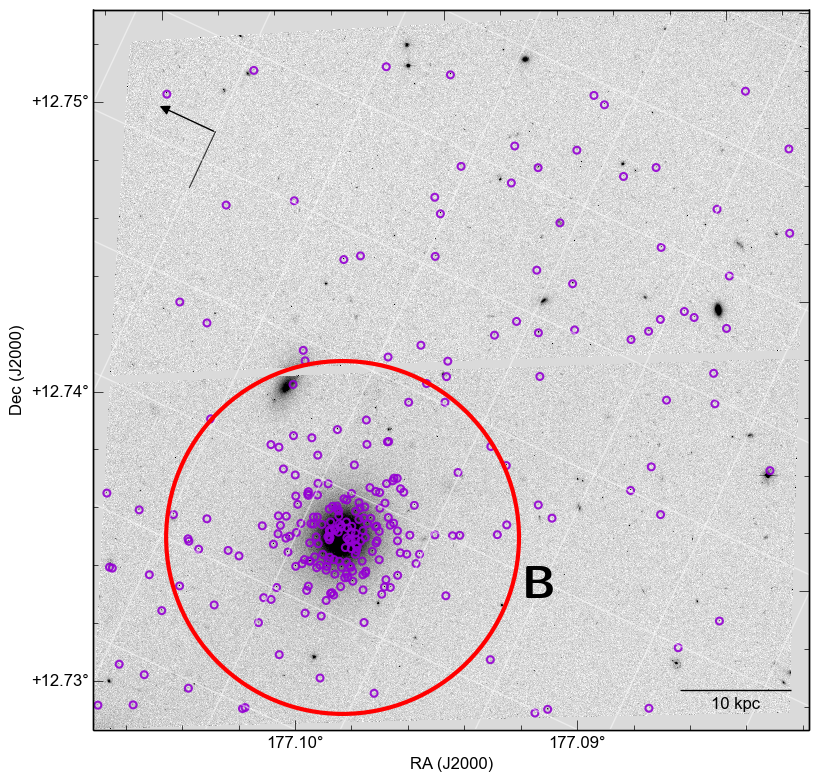}}\\
\subfloat[]{\includegraphics[width = 2.9in]{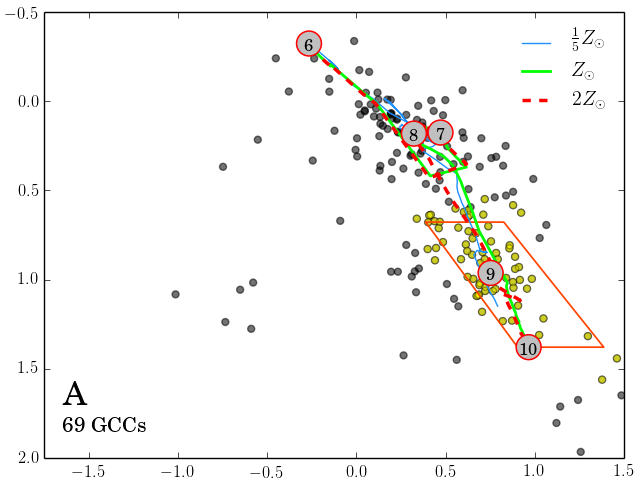}}\hspace{0.4cm}
\subfloat[]{\includegraphics[width = 2.9in]{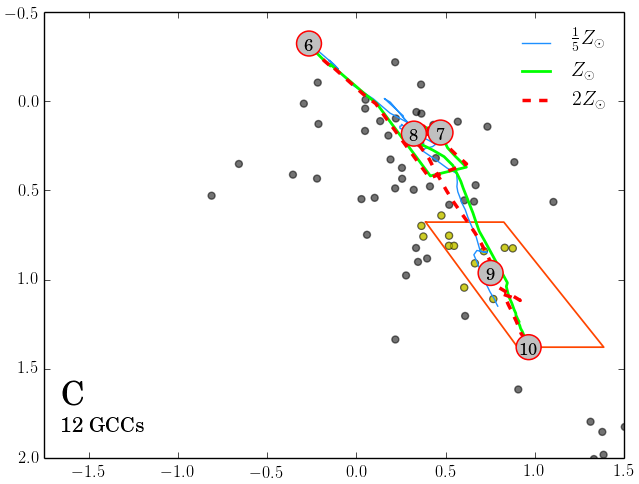}}\\
\subfloat[]{\includegraphics[width = 2.9in]{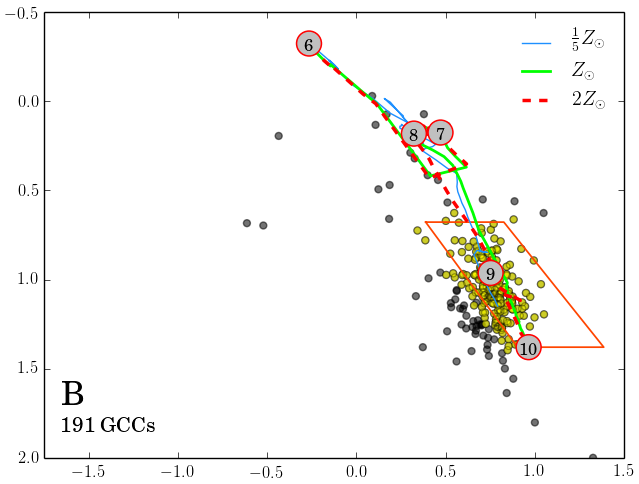}}\hspace{0.4cm}
\subfloat[]{\includegraphics[width = 2.9in]{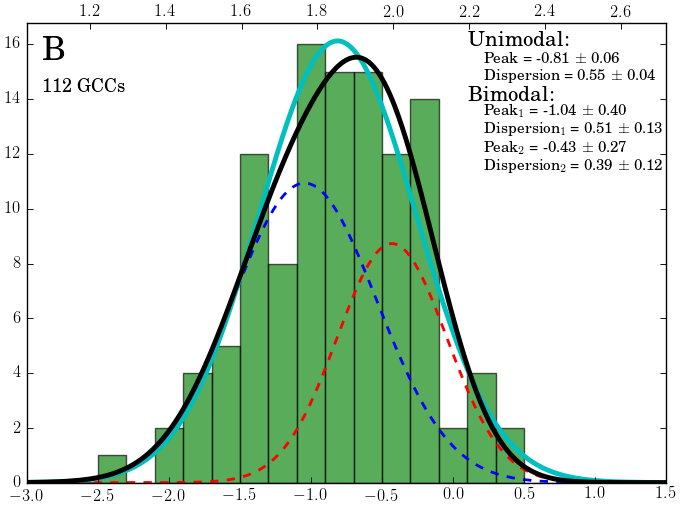}}\\  
\caption[continuation of the previous plot]
{$\ldots$ Panels (c) and (d) are inverted $V_{606}$ images 
which show the GCC system extent, with locations of detected GCCs 
overplotted as circles.  Panels (e)--(g) are colour-colour plots 
for particular galaxies in that group. Panel (h) represent metallicity 
distribution of the large population of GCCs in HCG 59B.  The GMM 
results are inconclusive, and a single-peaked distribution is 
consistent with the data. (A colour version of this figure is 
available in the online journal.)}
\end{figure*}
\clearpage


\begin{figure*}
\begin{center}
\subfloat[]{\includegraphics[width=0.65\textwidth]{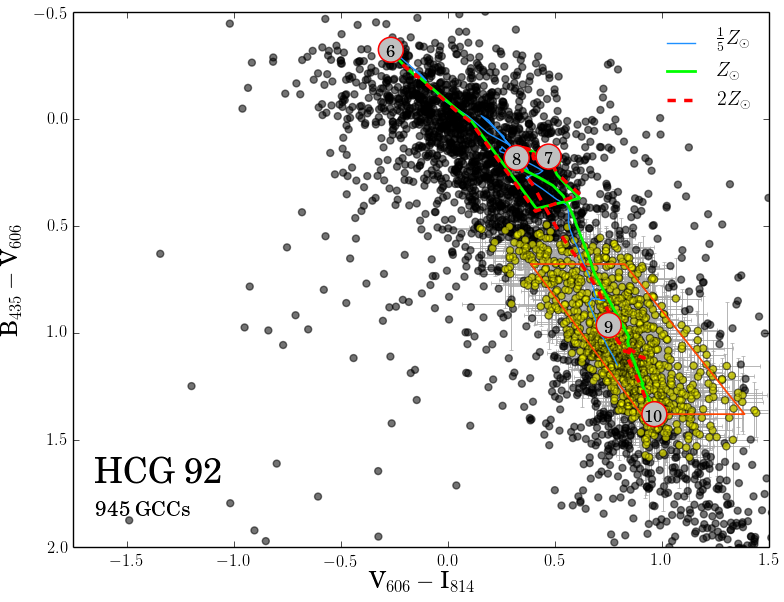}}\\
\vspace{0.1cm}
\subfloat[]{\includegraphics[width=0.65\textwidth]{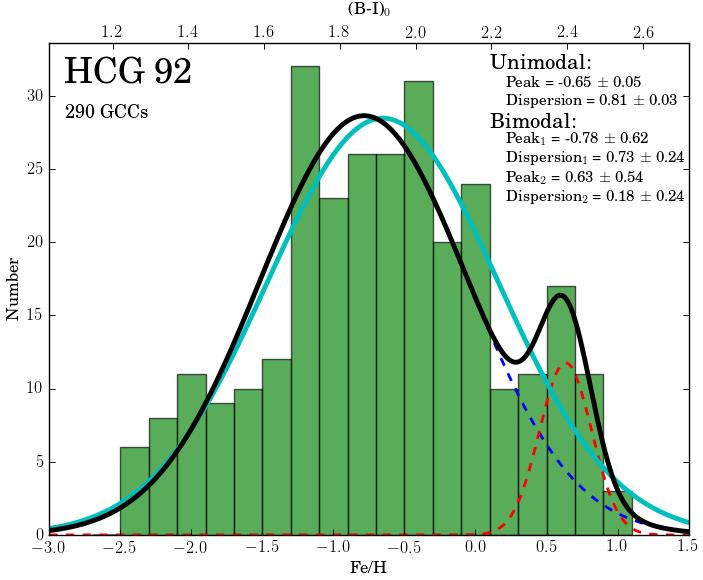}}
\caption[Colour-colour plot of all GCCs detected in HCG 92. Colour-colour 
plots of GCCs for each galaxy in HCG 92. GCCs spatial distribution and GCC 
system extent for each galaxy in HCG 92. GCs metallicities distribution for 
some galaxies in HCG 92]
{The GCCs population of HCG 92 (a) and its metallicity 
distribution (b).  See caption for Fig.~\ref{fig:GC07}. (A 
colour version of this figure is available in the online journal.)}
\label{fig:GC92}
\end{center}
\end{figure*}
\clearpage

\begin{figure*}
\ContinuedFloat
\centering
\subfloat[]{\includegraphics[width=0.49\textwidth]{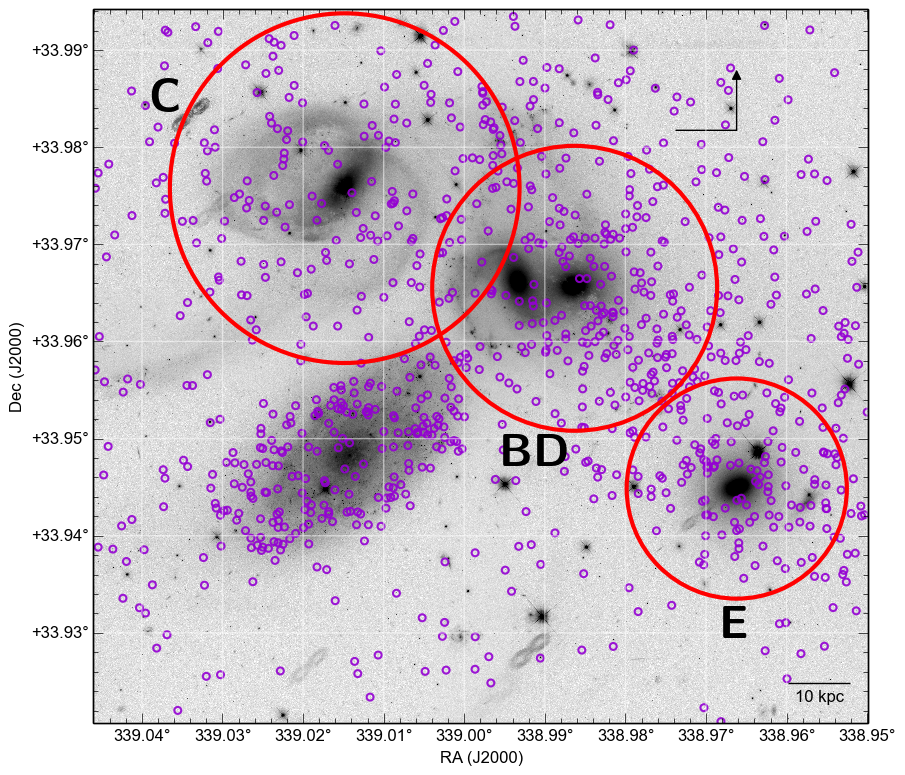}}\\
\subfloat[]{\includegraphics[width = 2.9in]{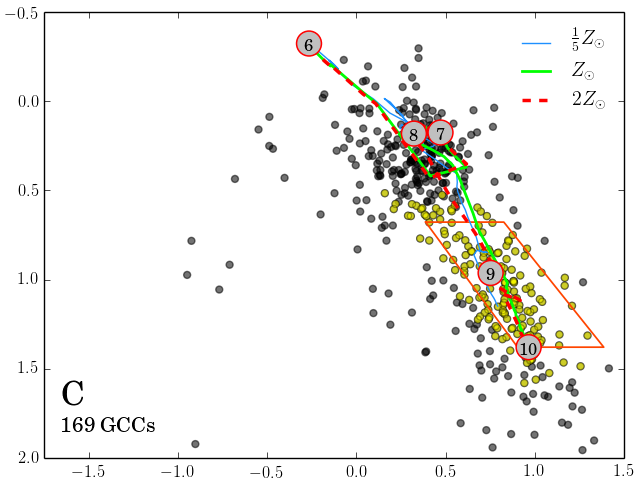}}\hspace{0.4cm}
\subfloat[]{\includegraphics[width = 2.9in]{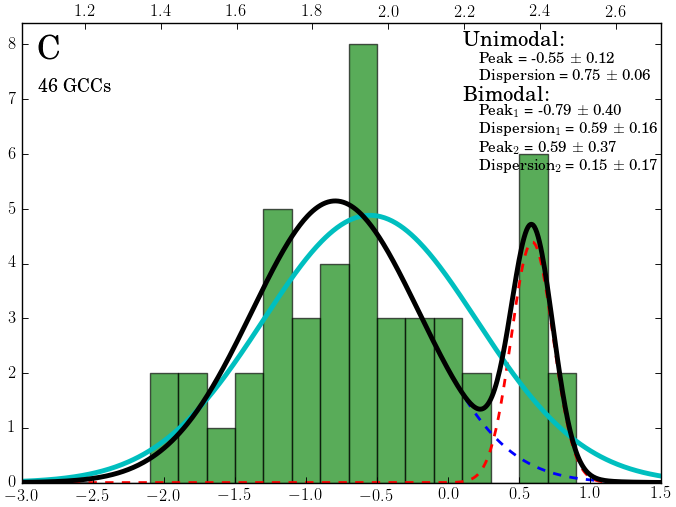}}\\
\subfloat[]{\includegraphics[width = 2.9in]{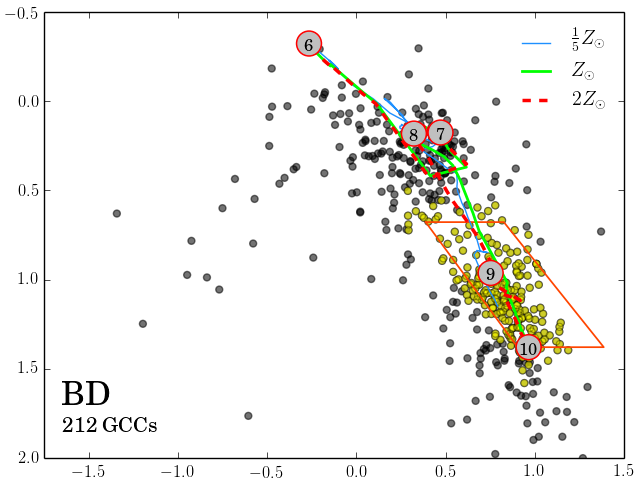}}\hspace{0.4cm}
\subfloat[]{\includegraphics[width = 2.9in]{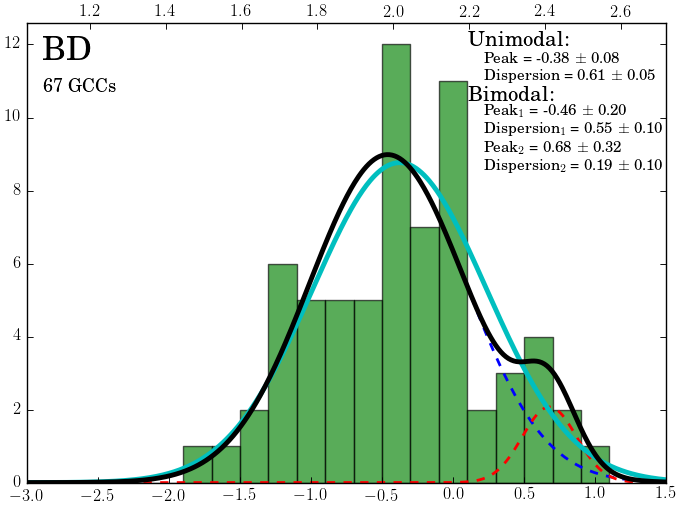}}
\caption[continuation of the previous plot]
{Panel (c) is the inverted $V_{606}$ image which shows the  GCC 
system extents, with locations of detected GCCs overplotted as circles. 
Panels (d) and (f) are colour-colour plots for particular galaxies 
and regions HCG 92. NGC 7319 (HCG 92C) is a face-on spiral and the GCC 
located in the central region, as well as in the spiral arms, could 
potentially be reddened young star clusters. The BD region contains 
the elliptical galaxy NGC 7318A (HCG 92D) and the spiral galaxy NGC 
7318B (HCG 92B) in a field of debris, material left from current and 
previous interactions.  Thus, it is likely that GCCs in the BD region 
are heavily contaminated by reddened young star clusters. Subfigures 
(e) and (g) are  the metallicity distributions for C and BD, respectively. 
(A colour version of this figure is available in the online journal.)}
\end{figure*}
\clearpage

\begin{figure*}
\ContinuedFloat
\centering
\subfloat[]{\includegraphics[width = 2.9in]{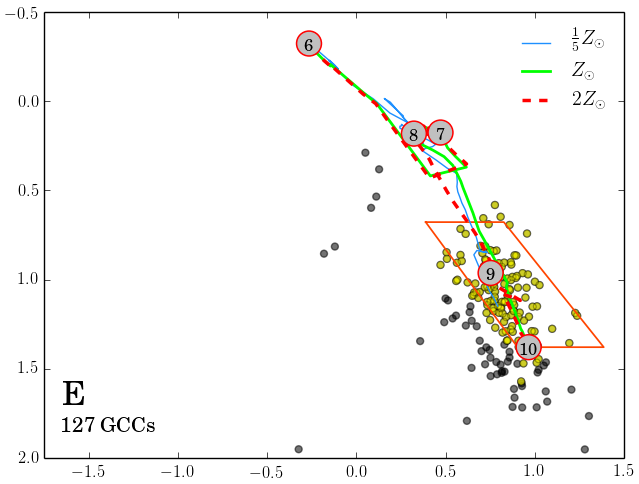}}\hspace{0.4cm}
\subfloat[]{\includegraphics[width = 2.9in]{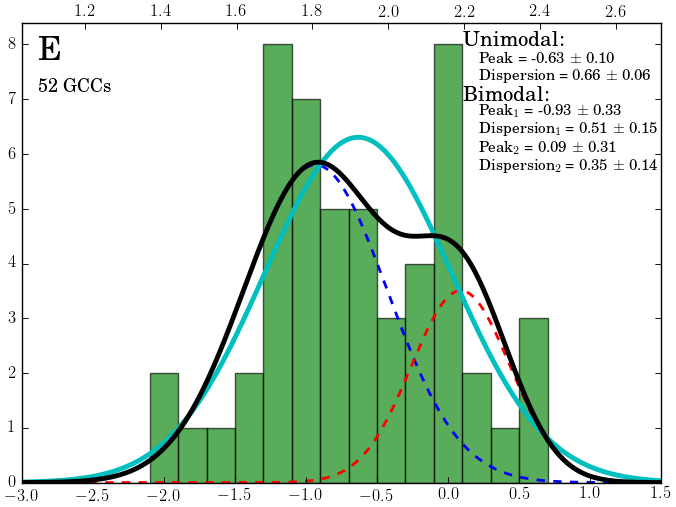}}\\
\caption[continuation of the previous plot]
{\ldots Panel (h) is a colour-colour plot for galaxy E 
and panel (i) is a plot of its metallicity distribution. (A 
colour version of this figure is available in the online journal.)}
\end{figure*}
\clearpage


\begin{figure*}
\begin{center}
\includegraphics[width=0.55\textwidth]{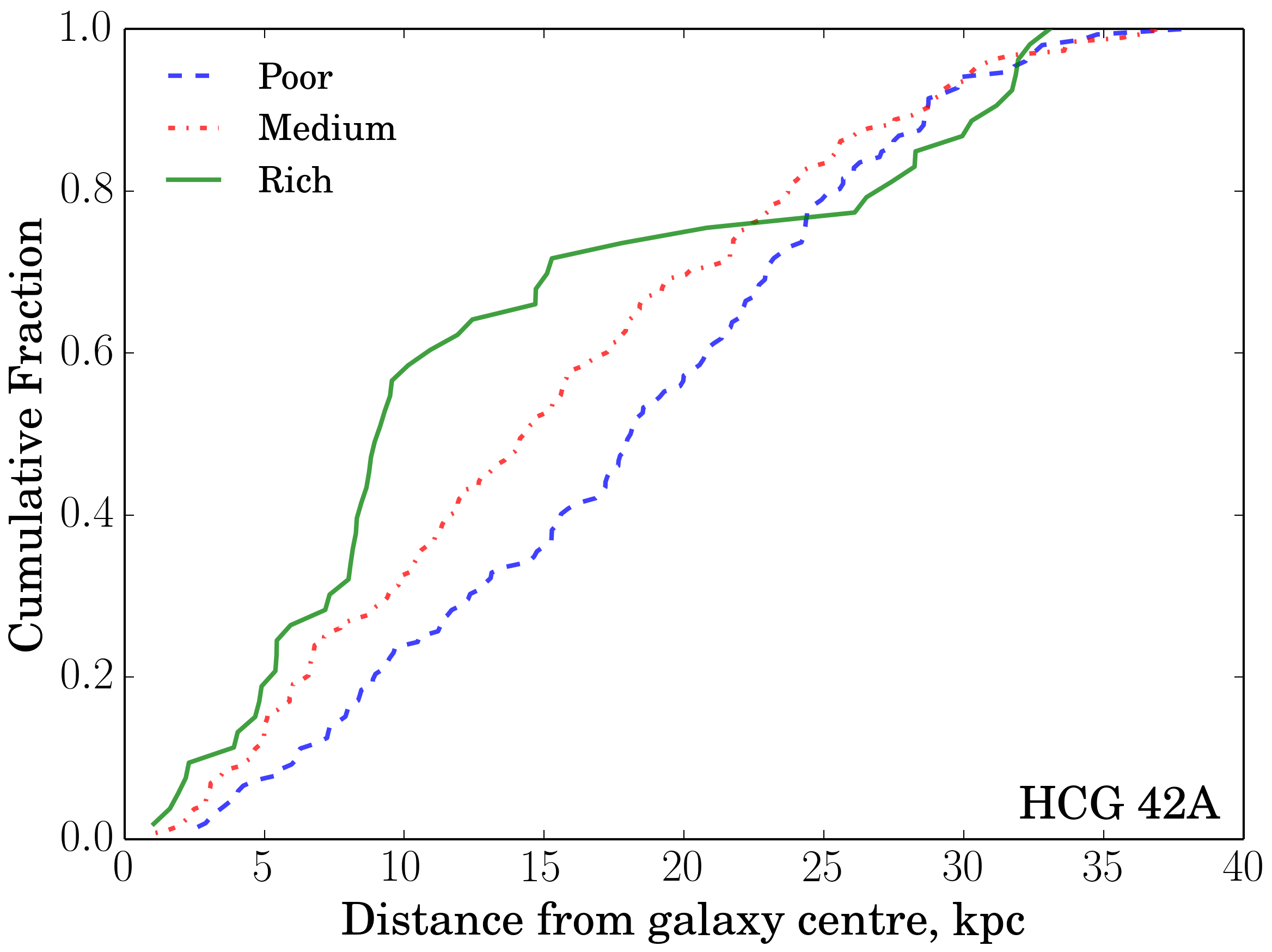}\\
\includegraphics[width=0.55\textwidth]{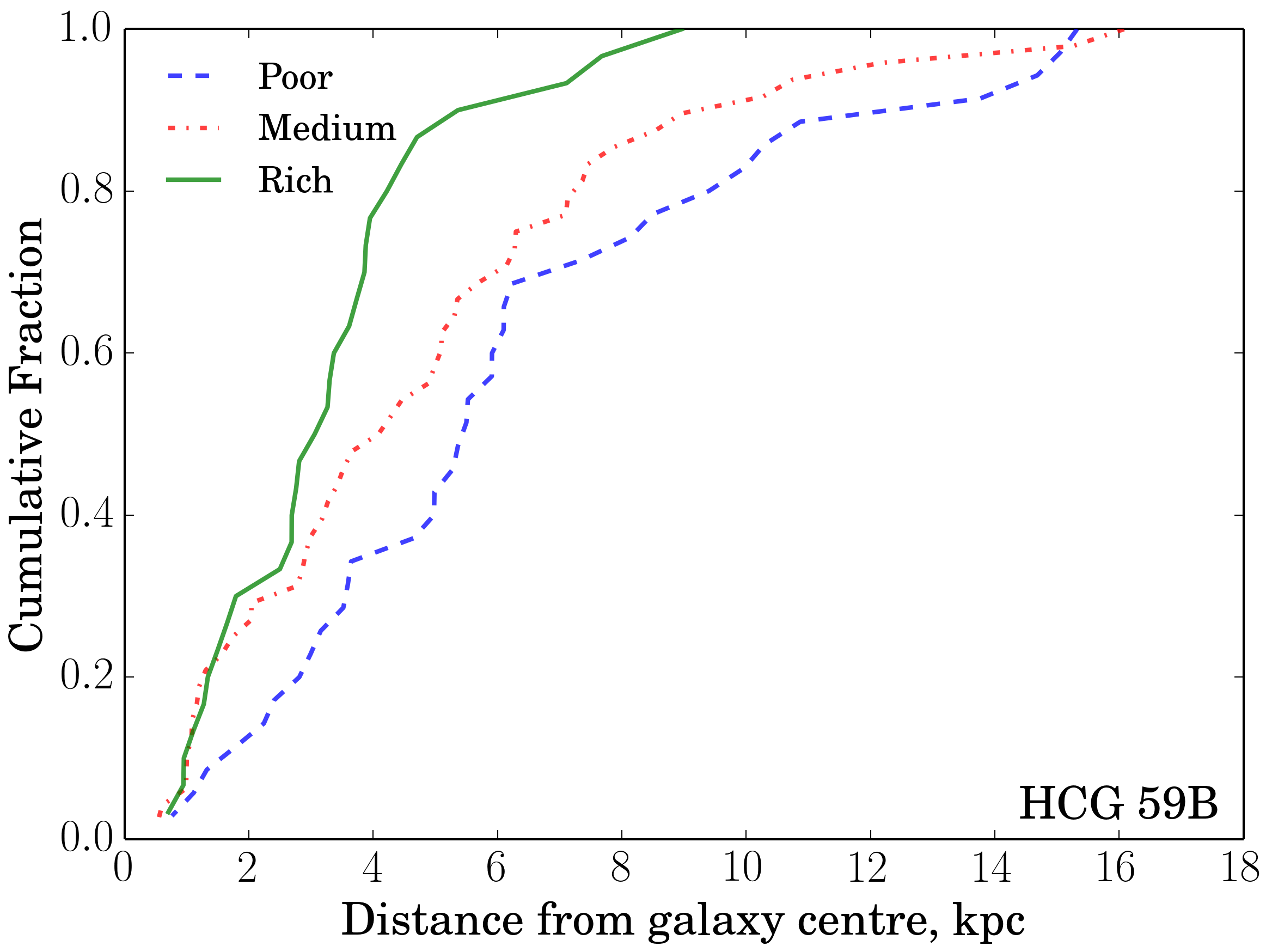}\\
\includegraphics[width=0.55\textwidth]{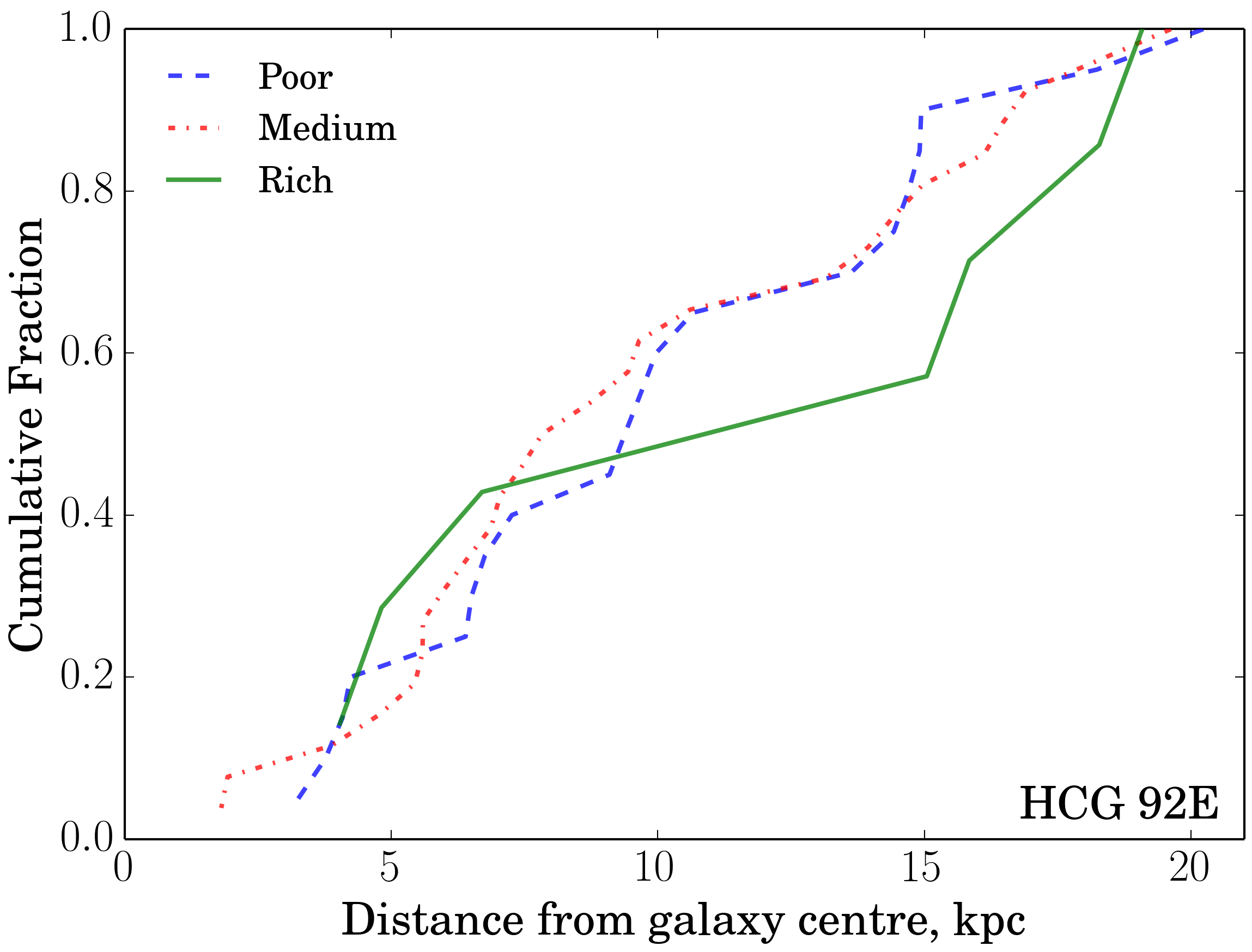}
\caption[Cumulative function of radial distribution of 
clusters with different metallicities for galaxies HCG 
42A, HCG 59B, and HCG 92E]
{Cumulative function of radial distribution of clusters with different 
metallicities for galaxies HCG 42A, HCG 59B, and HCG 92E (top, middle, 
and bottom panels, respectively).  The KS-test has shown that metal rich 
and metal poor populations of 42A and 59B are drawn from different 
distributions (with confidence of $>99\%$).  The GC populations in 92E 
appear to be well mixed throughout the galaxy. (A colour version of this 
figure is available in the online journal.)}
\label{fig:MD}
\end{center}
\end{figure*}
\clearpage

\bibliography{biblio.bib}

\end{document}